\documentclass[12pt,preprint]{aastex}
\usepackage{caption}
\newcommand{\noprint}[1]{}

\shorttitle{TBD}
\shortauthors{Targon et al.}

\begin{document}


\title{Correlating the interstellar magnetic field with  protostellar jets and its sources\footnote{Based on observations made at the Observat\'orio do Pico dos Dias, Brazil, operated by the Laborat\'orio Nacional de Astrof\'\i sica.}}



\author{C.~G. Targon\altaffilmark{2}, C.~V. Rodrigues\altaffilmark{2}, A.~H. Cerqueira\altaffilmark{3}, G.~R. Hickel\altaffilmark{4}}






\altaffiltext{2}{Instituto Nacional de Pesquisas Espaciais/MCT --
Av. dos Astronautas, 1758 -- 12227-010 - S\~ao Jos\'e dos
Campos - SP -- Brazil -- E-mail: claudiavr@das.inpe.br -- CGT now at Universidade de S\~ao Paulo -- Instituto de F\'\i sica de S\~ao Carlos -- Caixa Postal 369
-- 13.560-970 - S\~ao Carlos - SP -- Brazil}

\altaffiltext{3}{LATO-DCET/Universidade Estadual de Santa Cruz --
Rodovia Ilh\'eus-Itabuna, km 16 -- 45662-000 - Ilh\'eus - BA --
Brazil.}

\altaffiltext{4}{DFQ-ICE/Universidade Federal de Itajub\'a -- Av. BPS, 1303 -- 37500-903 - Itajub\'a - MG -- Brazil}


\begin{abstract}

This article combines new CCD polarimetric data with previous information about protostellar objects in a search for correlations involving the interstellar magnetic field. Specifically, we carried out an optical polarimetric study of a sample of 28 fields of 10\arcmin\ $\times$ 10\arcmin\ located in the neighborhood of protostellar jets and randomly spread over the Galaxy. The polarimetry of a large number of field stars is used to estimate both the average and dispersion of the interstellar magnetic field (ISMF) direction in each region. The results of the applied statistical tests are as follows. Concerning the alignment between the jet direction and the interstellar magnetic field, the whole sample does not show alignment. There is, however, a statistically significant alignment for objects of Classes 0 and I. Regarding the interstellar magnetic field dispersion, our sample presents values slightly larger for regions containing T Tauri objects than for those harboring younger protostars. Moreover the ISMF dispersion in regions containing high-mass objects tends to be larger than in those including only low-mass protostars.  In our sample, the mean interstellar polarization as a function of the average interstellar extinction in a region reaches a maximum value around 3\% for A(V) = 5, after which it decreases. Our data also show a clear correlation of the mean value of the interstellar polarization with the dispersion of the interstellar magnetic field: the larger the dispersion, the smaller the polarization. Based on a comparison of our and previous results, we suggest that the dispersion in regions forming stars is larger than in quiescent regions.

\end{abstract}


\section{Introduction}

The role of the magnetic field (MF) in the star formation process is an active area of research. The MF can act in phenomena at large scales, such as the collapse of a molecular cloud \citep[e.g.,][]{Shu1987}, and at smaller scales, such as the formation of protostellar jets and outflows \citep[e.g.,][]{Ferreira2006} and disk viscosity \citep[e.g.,][]{ks2010}. An important present question is if the interstellar magnetic field (ISMF) \citep[e.g.,][]{mou99,mt2010} or the turbulence \citep[e.g.,][]{mo07,crutcher2009} is the main agent of the support of molecular clouds against the gravitational force. 

An indication to the importance of the interstellar magnetic field (ISMF) in the collapse is the alignment between the magnetic field at larger scales and the symmetry axis of the young stellar object (YSO). The ambipolar diffusion model predicts that the collapse of the protostellar object occurs preferentially along the field lines. This results in an alignment of the ISMF with respect to the accretion disk axis \citep{Shu1987}. Simulations of the collapse of magnetized regions that forms the cores also indicate that, in plausible interstellar physical conditions, there will be an alignment. \citet{machida2006} show that the disk axis evolves to a configuration parallel to the ISMF if the rotation rate is small relative to the ISMF strength. The simulations of \citet{Matsumoto2006} result in an alignment of the jet direction with the ISMF, if the ISMF is larger than 80 $\mu$G. Recently, turbulence has been included in those simulations \citep{matsumoto2011}. They have found that the shape of the core depends on its mass: massive cores are prolate and low-mass cores are oblate. And, in any case, the minor axis is parallel to the ISMF.

From an observational perspective, early studies pointed to an alignment between the symmetry axis of the protostellar object and the ISMF. \citet{dl79} and \citet{hz81} obtained near-infrared (NIR) polarimetry of compact IR sources (52 in total) and found a tendency of alignment between the circumstellar polarization and the ISMF direction. \citet{ts89}, also based on NIR polarization, found similar results with a sample of 39 T Tauri objects. \citet{snell80} and \citet{cohen1984} suggested an alignment between YSO outflow and ISMF. These results were later confirmed by \citet{strom1986} using a larger sample, including 38 regions. 

Recently, there has been a revival of this subject. \citet{Menard2004}, using a sample composed of T Tauri stars, did not find any correlation between the YSO disk axis and the ISMF, but suggested that the brighter jets are aligned to the ISMF. Subsequent studies indicate that different interstellar structures are oriented according to the ISMF, independently of presenting or not signs of star formation. The Pipe Nebula and Musca Cloud have their long axis perpendicular to the ISMF (\citealt{fag2010} and \citealt{pereyra2004}, respectively). The same occurs in the Serpens Cloud \citep{sugitani2011}. \citet{li2009} also found that the magnetic field direction in the intercloud medium correlates with the field direction of the dense cores. \citet{anath2008} studied the alignment of outflows and the filaments containing the associated dense core in a sample of 45 objects and found that the directions are approximately orthogonal. \citet{rod2009}, based on a sample of 100 Herbig Ae/Be stars, suggested that the most polarized objects tend to have their polarization position angle aligned with the ISMF direction. In fact, there is evidence that the ISMF is dynamically dominant in interstellar clouds as Pipe and Serpens (\citealt{fag2010} and \citealt{sugitani2010}, respectively).


Besides its direction, another property of the ISMF that can be used to understand the interstellar medium conditions is how ordered it is.  Assuming the equilibrium between kinetic and magnetic forces and isotropy of the motions in the ISM, \citet[][CF53]{Chandrasekhar1953} proposed the following relation between the dispersion of ISMF direction, $\sigma_B$, and  the 3D turbulent velocity, $\sigma(v)$: 

\begin{equation}
\sigma_B=\left(\frac{4}{3}\pi\rho\right)^{\frac{1}{2}}\frac{\sigma(v)}{B},
\label{eq:cf53}
\end{equation}

\noindent where $\rho$ is the mean mass density of the interstellar medium (ISM) and $B$ is the intensity of the interstellar magnetic field. 

But this simple relation has some caveats. Examples are: large fluctuations of the magnetic field amplitude; acting of the nonmagnetic forces on the gas; inhomogeneity of the interstellar material \citep{zwe96}. Even so, numerical simulations of polarimetric maps of molecular clouds indicate that the CF53's relation is valid at least as an order of magnitude estimate \citep{ostriker2001,pad01,hei01}. 

As ISMF dispersion and turbulence are connected, we cite some recent studies on the origin of the interstellar turbulence, which are not yet conclusive. Some authors, based on simulations, suggested that stellar outflows can replenish the turbulent motion in dense cores \citep{nl2007,carroll2009,carroll2010}. From an observational perspective, this scenario seems to be true in the Serpens cloud, in which the outflows have enough energy to power the observed turbulence \citep{sugitani2010}. On the other hand, others authors, based on simulations and observations, show that the turbulence is injected in the molecular clouds by a large-scale process \citep{brunt2009,padoan2009}. A third different view is that in which the dispersion of the velocity is the result of gravitational collapse \citep{heitsch2009}. 

In this paper, we present a search for connections between the properties of the interstellar magnetic field and the stellar formation to constrain the magnetic field role in the star formation process. The ISMF is probed through optical polarimetry. Our observational technique allows the measurement of the polarization of a large number of objects in a region. It improves the statistical significance of our results, compared with previous studies based on photoelectric measurements, and enables the estimate of the ISMF dispersion, a quantity that is poorly explored previously due to the lack of enough data. The regions in which we sample the ISMF are spread over many molecular clouds throughout the Galaxy. This should result in unbiased results valid in the context of star formation in the Galaxy, not specific for a given star formation complex. Moreover, our fields of view are small, so more plausibly associated with the interstellar medium nearby the YSO.

This article is organized as follows. Section \ref{sec_observations} describes the procedure of data acquisition and reduction as well as how both the average and the dispersion of the interstellar magnetic field direction are estimated. The YSO information is compiled from the literature as explained in Section \ref{sec:compilation}. The statistical analysis and discussion are presented in Section \ref{sec_stat}. Our conclusions are summarized in the last section. This paper is the result of the Master dissertation of C. \citet{Targon2008}. 

\section{Polarization measurements and estimation of the properties of the interstellar magnetic field}
\label{sec_observations} 

We obtained the optical polarimetry of 28 southern regions close to Herbig-Haro (HH) objects from the Reipurth's catalog \citep{Reipurth1994}. We randomly selected among the richer stellar fields, since the richer the field the better the mapping of the ISMF. Each region covers a sky area of 10\farcm5 $\times$ 10\farcm5. The total number of HH objects found in a radius of 20\arcmin\ from the center of each field is 82, considering all fields. However, a jet can be associated with more than one HH object. Furthermore, a YSO can have a jet and its counterjet. Hence, we considered as one object the group of HHs associated with a given protostellar source (jet and counterjet). In doing this, the final number of objects (= jets, hereafter) in our data is 60. To help the reader to relate the fields to the sites of star formation, Table \ref{tab:localization} displays a non-uniform identification of the region, cloud, or globule in the direction of each line-of-sight.

The observations were carried out in 1998 December 18, between 2005 February 11 and 17, and on 2007 May 7, with the 0.60-m Boller \& Chivens telescope at the Observat\'orio do Pico dos Dias, Brazil, operated by the Laborat\'orio Nacional de Astrof\'\i sica, Brazil. A CCD camera modified by the polarimetric module described in \citet{mag96} has been used. The CCD array used was a SITe back-illuminated, $1024 \times 1024$ pixels. This results in a field-of-view of 10\farcm5 $\times$ 10\farcm5 (1 pixel = 0\farcs62). All the fields were observed with a $R_C$ filter. The employed technique automatically eliminates the sky polarization \citep{Piirola1973}.  Polarimetric standards stars \citep{Serkowski1975, Bastien1988, Turnshek1990} were observed to calibrate the system and to estimate the instrumental polarization. The measured values of the unpolarized standard stars are consistent with zero within the errors. Table \ref{tab:dados} presents a log of the observations.

The reduction was carried out using the IRAF\footnote{IRAF is distributed by National Optical Astronomy Observatories, which is operated by the Association of Universities for Research in Astronomy, Inc., under contract with the National Science Foundation.} facility. The first step in the reduction process consists of CCD imaging correction: bias and flat-field. Then, aperture photometry of the ordinary and extraordinary images of each object is performed. The resulting counts are used to calculate the polarization using the method described in \citet{mag84}. For the polarimetric analysis, the package {\sc PCCDPACK} \citep{per00} was used.

Our aim is to obtain the interstellar polarization of as many objects in the field as possible. 
The position angle of the polarization, $\theta$, is assumed to be the direction of the ISMF \textit{as projected in the plane of sky}. The following analysis is based on the objects with good signal-to-noise ratio. Specifically, we have selected those with $P/\sigma_P \geq 3$ and this provides a maximum error on the position angle of 10\degr. The only exception is Field 19, including HH~160, for which we used $P/\sigma_P \geq 2.5$. The number of objects considered in each field, $N_f$, can be found in Table \ref{tab:reducao}: it runs from 14 to 559. A polarimetric catalog of each field is made available as a online-only table.
Table \ref{tab:polarimetriccatalog} shows an example of this catalog.

\begin{figure}
\includegraphics[width=13cm,angle=0]{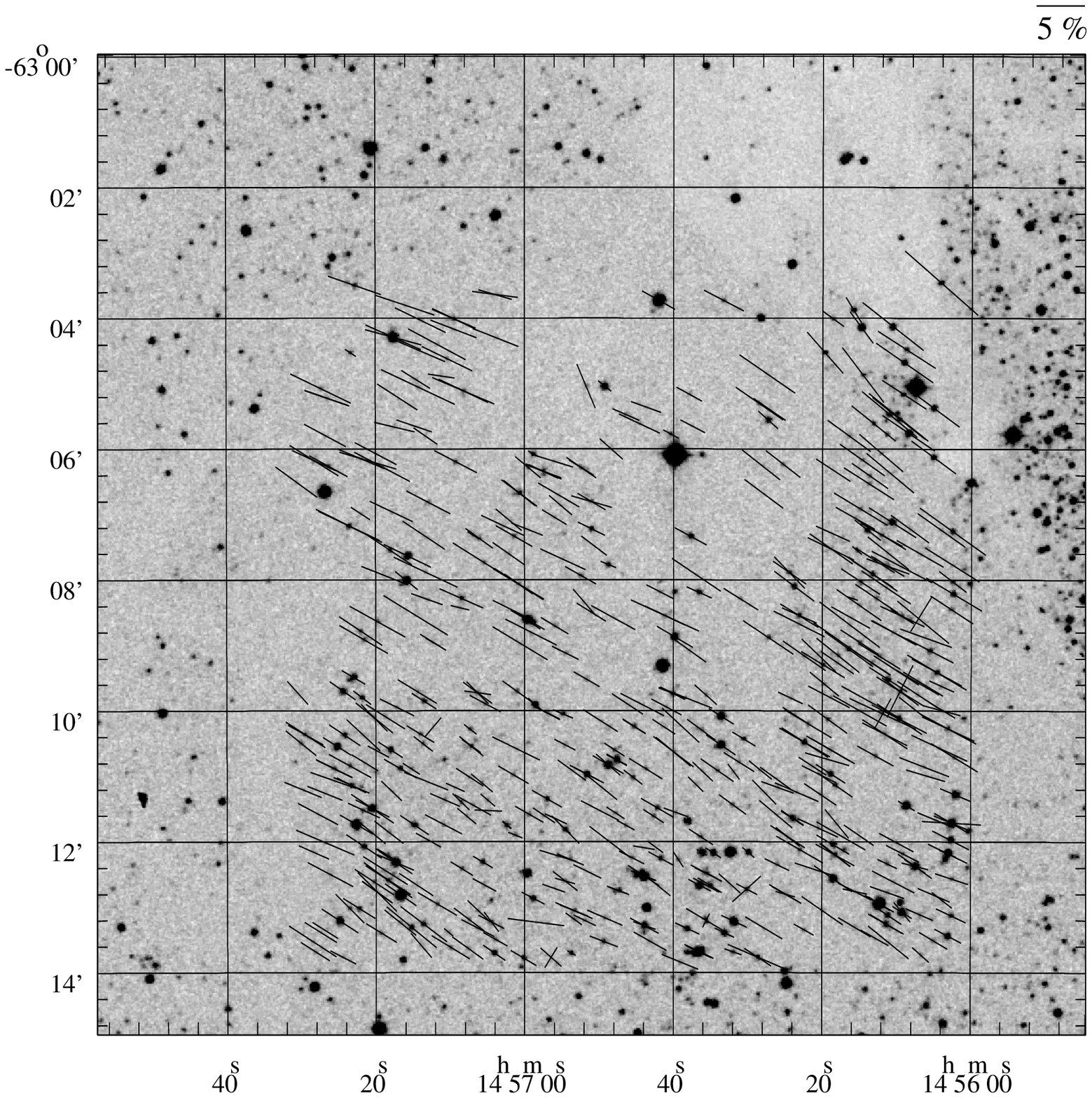}

\vspace{-5cm}
\includegraphics[width=6cm,angle=-90.]{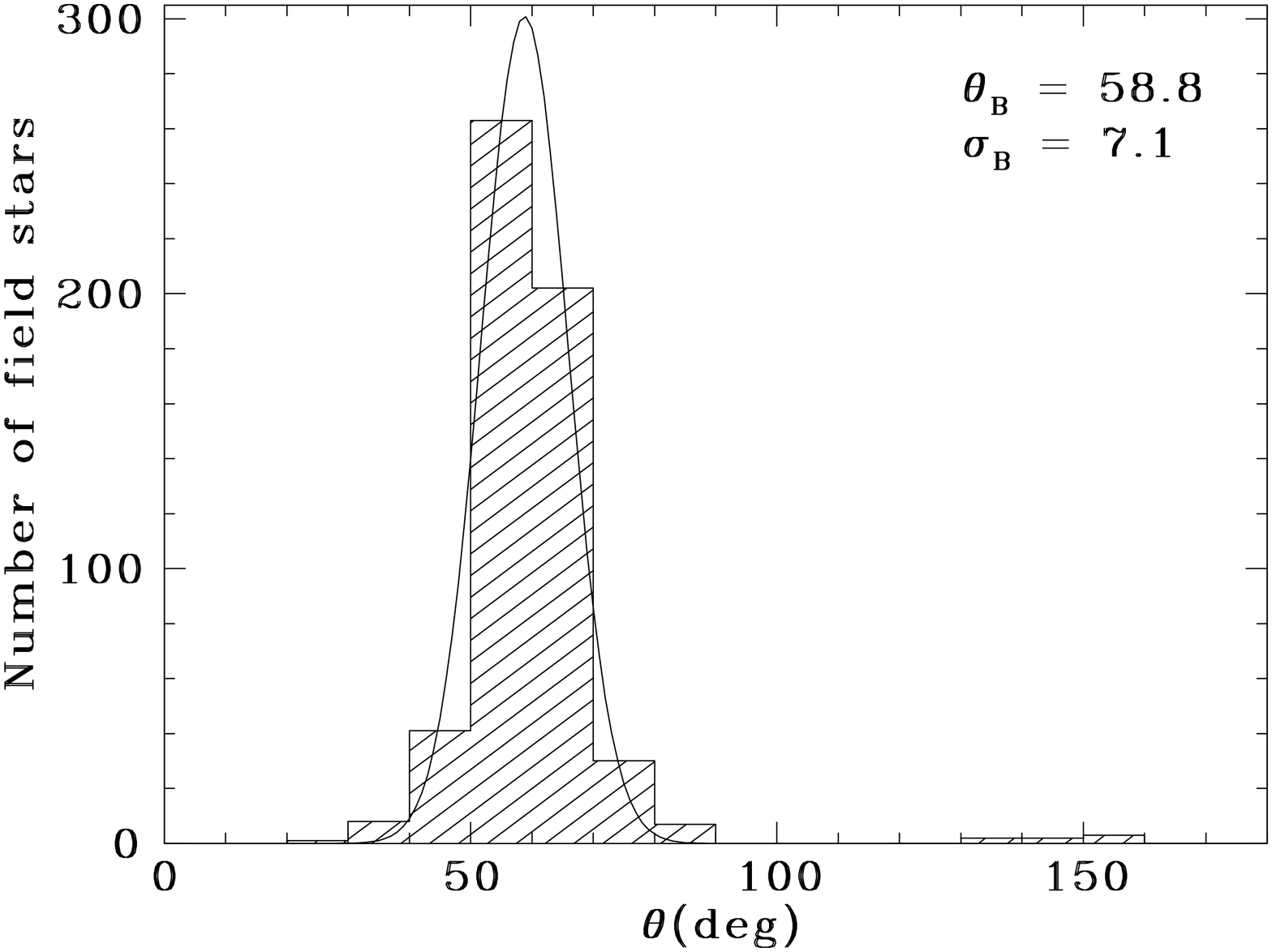}
\caption{Polarimetry of the Field~17 which contains HH~76-77 and HH~139. Upper panel: The observed polarization vectors overplotted on a DSS2 red image. The coordinates are B1950. Lower panel: The histogram of the corresponding position angles of the polarization, $\theta$. In the upper right corner, it is shown the average and the dispersion of the interstellar magnetic field used in the analysis. A Gaussian curve using these values is also depicted. }
\label{fig:hh139}
\end{figure}

For each field, we constructed histograms of $\theta$. For most fields (23/28), this histogram has a very well defined Gaussian shape. In these cases, it is straightforward to estimate the average and dispersion of the direction of the ISMF from a Gaussian fit. Figure~\ref{fig:hh139} shows our polarization vectors superposed on a DSS image in the line-of-sight of HH~139 (Field 17) and the corresponding histogram of the position angle of the polarization. The graphs for the other fields can be found as online-only material. Exceptions are the graphs for Fields 02 and 15, which are presented in \citet{Hickel2002} and \citet{rod2007}, respectively. Table \ref{tab:reducao} summarizes the interstellar polarimetric characteristics of each field. It lists, per field, the number of objects ($N_f$), the position of the peaks in the histogram of $\theta$ ($\theta_P$), the value adopted for the direction of the ISMF ($\theta_B$) and its dispersion ($\sigma_B$), and the average of the polarization modulus ($P_{ISM}$). The dispersion is unbiased from the mean error of the position angle as suggested by \citet{Pereyra2007}. The Gaussian curves depicted in the figures use the \textit{unbiased} value of the dispersion.

Five fields do not show a histogram with a clear Gaussian shape: they have two peaks. In these cases, more elaborated procedures were performed to estimate the properties of ISMF direction, which are described in the following paragraphs.

Field~15 (HH135/HH136) was studied by \citet{rod2007}. Here we summarize their procedure to estimate the ISMF properties in the star forming region because it was used in other fields of this study. In this region, there are two peaks in the histogram of $\theta$. Each peak is associated with objects having different mean values of polarization. In addition, the most polarized objects are fainter than those having the smallest polarizations. These two facts can be explained if the stars in this field-of-view, and hence in the histogram, come from two populations. Furthermore, we can assume that the faintest objects are also the furthest ones. Consequently one population, associated with the nearest stars, is polarized by a single interstellar cloud. The second population, including the furthest objects, is polarized by two interstellar clouds, one of which is responsible for the polarization of the first population. As the HH objects in this line-of-sight are around 3~kpc, we assumed that the more distant dust cloud is associated with the star forming region. Thus, to obtain the direction of the polarization produced in that region, the average polarization of the population which has the brighter stars was calculated and then this value was vectorially subtracted from the polarization of the population including the fainter objects (associated with the star forming region).

Fields~16 and 28, similarly to the Field~15, shows two peaks in the histogram of  $\theta$ which are associated with different values of polarization. One of the peaks is composed of a small number of objects which are also the less polarized. We assumed that these objects constitute a foreground polarization component; therefore, we have adopted the same procedure as for Field~15.

In Field~13 (HH~120), the two peaks do not have different values of polarization. They are, however, spatially segregated. To represent the direction of the ISMF, we chose the value of $\theta_P$ corresponding to objects near HH~120.

Field 19 (HH~160) has two peaks and a small number of objects. Therefore it is difficult to perform an appropriated statistical analysis. We used the value of the dominant (in number) component.

Three fields (5, 13, and 18) have been observed with two different exposure times, $t_{exp}$, to check if the observed distribution of $\theta$ depends on how deep the image is. For the three fields, the displacement in the peak position, $\theta_P$, between the polarimetry using different $t_{exp}$ is negligible ($<$ 2\degr). For the analysis, we adopted the values obtained with the larger  $t_{exp}$, which are presented in Table \ref{tab:reducao}.

There are two measurements of Fields~12 and 24, which were carried out in two adjacent lines of sight (denoted by A and B) to verify if the distribution of the ISMF direction can change in the scales of the projected angular distance of an image. Notice that the sizes of the jets in these two fields are very different. While Field~24A contains HH~271-272 and the associated YSO (hence the entire jet extension), Fields~12A and 12B map the region of HH~90-92 and 597-598, a giant jet \citep{Bally2002}, which is much larger than both fields together. The results for these two cases are discussed below.

Fields~24A and 24B have similar distributions of $\theta$. Therefore we have assumed that they both trace the same ISMF distribution, and hence we have combined the two fields to obtain the values of $\theta_B$ and $\sigma_B$ used in the analysis. These values are presented in Table \ref{tab:reducao} (24AB). This table also presents the values for the individual fields for comparison.

The jet associated with HH~90-92 and HH~597-598 extends for approximately 25\arcmin\ \citep{Bally2002}. An image has a side of 10\arcmin. Field~12A includes the mid portion of the jet (HH~90) and Field~12B is displaced in a direction perpendicular to the jet axis. The distribution of $\theta$ in each field is composed of an single component, but  with $\theta_P$ differing by 70\degr . For the following analysis, we have adopted the values from field~12A, because it includes the jet, but we have kept the values of field~12B in Table \ref{tab:reducao} for comparison. Jets as large as the one associated with HH~90 do not dominate our sample: only 8 jets (out of 54) have extensions larger than 5\arcmin. The extension is the distance from the YSO to the end of the jet. The number 54 corresponds to the jets which has an estimate for their extension. See Section~\ref{sec:compilation} and Table~\ref{tab:jets} for the extension of jets used here.

The distances of the HHs in our sample range from 130 to 4300~pc (see Section \ref{sec:compilation} and Table \ref{tab:jets}). Consequently, our images map regions on the plane of sky extending from 0.38~pc (slightly larger than a dense core) to 12.5~pc (dimension associated with molecular clouds). The average HH distance in our sample is 750~pc, which matches to an image size of 2.2~pc. We remind the reader that the objects included in an image are from a 3D conical structure and, therefore, sample regions of different sizes depending on the distance. 

\section{Compiled information about the Herbig-Haro objects and its sources}
\label{sec:compilation}

We compiled some properties of the 60 jets of this work from the literature. We collected the distance, extension and position angle of the jet (Table \ref{tab:jets}) and the name, luminosity, mass, and class of the associated YSO (Table \ref{tab:ysos}). We could not find estimates for all the above properties for some objects. We did not include the jets associated with HH~273, HH~731 and HH~735-736 in the analysis due to the lack of information in the literature.  Consequently, our analysis is restricted to 57 jets. In the following paragraphs, some remarks about the compilation are given.

We used the position angle of the jet (PA) as projected in the plane of sky and in the range 0--180\degr. For some jets, this value can be directly read from previous articles. For the others, we estimated it using the coordinates of the HH knots and YSO or the image itself. When there is more than one value of PA (for jet and counter-jet, for instance), we use the average.  Whenever possible, we crosschecked information found in the literature. In doing so, we found an inconsistency between the jet directions cited by \citet{Ray1994} for HH~140-143 and those calculated by us using the knots coordinates presented in the same article. Our analysis indicates that there is probably an error in the knots coordinates. Hence we propose new values, which are given in Table \ref{tab_hh140}.

The \textit{projected} value of the jet extensions were directly collected from the literature or estimated using available data. We define the jet extension as the distance from the YSO to the jet extremity. Consequently if the literature quotes the distance between the jet and counterjet extremities, we use half this number. In some cases, the extension was derived using the angular size of the jet and the object distance. In cases in which there is more than one estimate for the distance, we used the smallest value. This is justified by the fact that the sample is biased towards smaller extensions, since (1) we use projected lengths and (2) the farthest jet point can be missed. Among the extensions collected in the literature, the only inconsistency found is for HH~444-447. The extensions presented by \citet{rei1998} are correct, and not those from \citet{Mader1999}. If the angle with the line of sight is known, what is true for some jets, we could derive the real jet extensions. However, only a part of our sample has this information. Hence, to keep a homogeneous procedure for all jets, we considered the length as seen projected in the plane of sky. 
 
Concerning mass classification, we simply adopt the same qualitative classification found in the literature among low, intermediate, and high mass. HH~160 (M = 2-3 M$_\odot$) and HH-217 (Sp: F0-G0) were both considered as intermediate mass. The objects classified as T Tauri were considered as low-mass objects.

Some remarks about the adopted classes should be done. The objects classified in previous articles just as T Tauri stars were considered as Class II. The source of the jet associated with HH~55 seems to be a very evolved object \citep{Graham1994}. Hence we have considered this object as a Class II/III. In fact, it is the only possible Class III object in the sample. Our sample includes some FU Ori variables, which are probably in a transition phase from Class I to Class II \citep{Hartmann1996}.

\section{Statistical analysis and discussion}
\label{sec_stat}

This section presents the analysis of the ISMF direction and its dispersion estimated from our optical polarimetric data. Our main aim is to verify if there are statistically significant correlations between the properties of the ISMF and those of the HH and its associated YSO. This section also contains the discussion of our results.

Our statistical analysis is performed using non-parametric tests of cumulative histograms. We have used the Kolmogorov-Smirnov test \citep[e.g.,][]{Press1986} and the Kuiper test \citep[e.g.,][]{Press1986,pal04}. From hereafter, these tests will be called  KS and Kp, respectively. Both tests compare two cumulative distributions, but each test uses a distinct quantity (statistic) to measure how different the distributions are. The statistic of the KS test is the maximum value of the absolute difference between the two distributions. The Kp test uses the sum of the absolute values of negative and positive differences and, hence, is more sensitive to differences in the entire abscissa range. Additionally, the Kp test is more appropriate for cyclic quantities. We will call the tests between a given data distribution and a uniform one KS1 and Kp1. The tests between two data distributions will be denoted as KS2 and Kp2.

The results relative to the ISMF direction are presented in Section \ref{sec:deltatheta}. Section \ref{sec:dispersao} contains the study of the ISMF dispersion. Our findings about the average polarization in the fields are shown in Section \ref{sec:polarizacaomedia}.

\subsection{The alignment between the jet and the interstellar magnetic field}\label{sec:deltatheta}

To measure the alignment of a jet with the ISMF, we define $\Delta \theta$: the difference between the position angle of the jet and the direction of the interstellar magnetic field. Its value is defined in the range $0- 90$\degr\ because we compare directions, not senses. To study $\Delta \theta$, the Kp test is more appropriate than the KS test because $\Delta \theta$ is a cyclic quantity. Hence, in the graphs only the probability associated with the Kp test is shown. In spite of that, we have quoted, throughout the text, the probabilities associated with both tests. This enables the reader to verify the dependence of the results on the statistical method. 

\begin{figure}
\includegraphics[height=6.5in,angle=270]{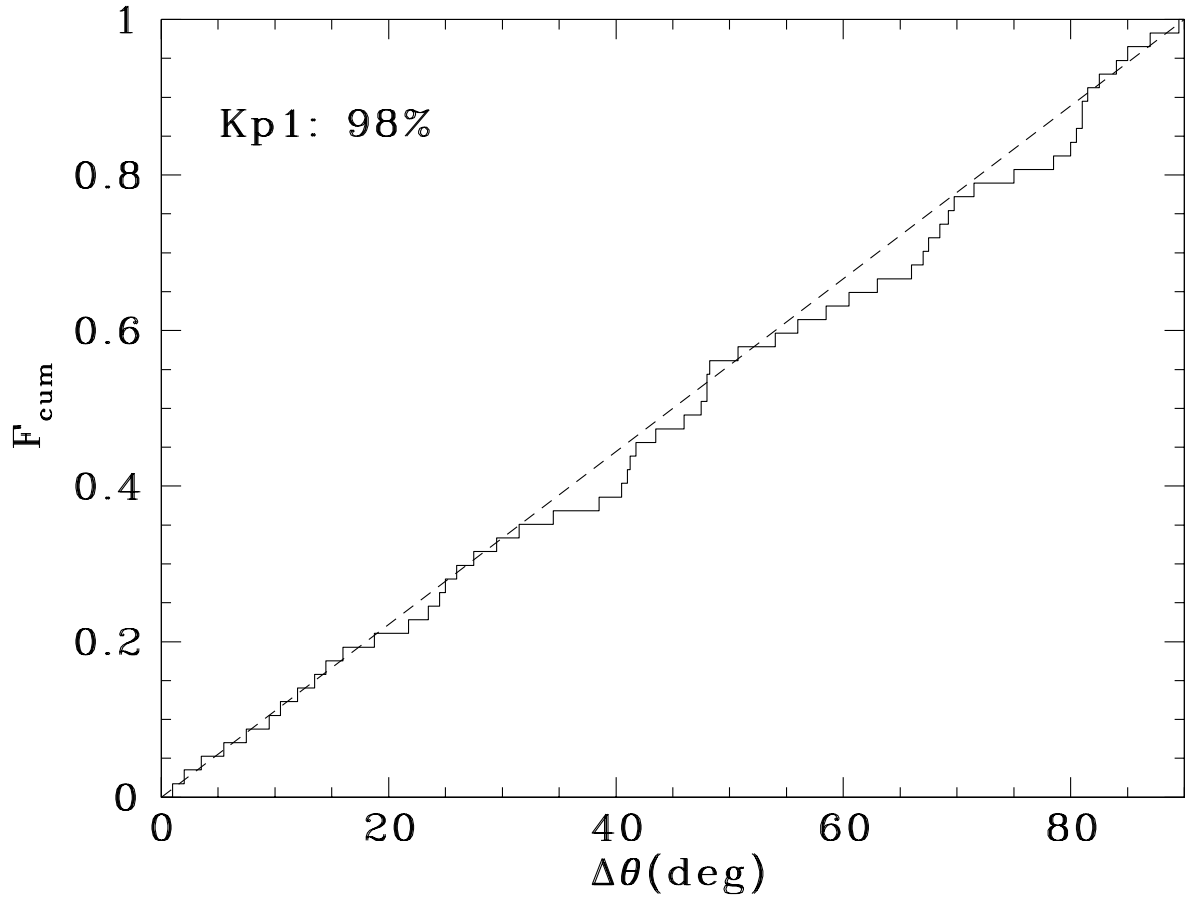}
\caption{Cumulative histogram of the difference between the position angle of the jet and the direction of the ISMF ($\Delta\theta$). $\rm{F_{cum}}$ represents the cumulative fraction of objects with $\Delta\theta$ smaller than a given value. The solid line is the observed distribution and the dashed line represents a uniform distribution for comparison. In the upper left corner, it is shown the probability that the observed $\Delta\theta$ is drawn from a uniform distribution according to the Kuiper test (Kp1).}
\label{fig_dtheta_all}
\end{figure}

Figure \ref{fig_dtheta_all} shows the cumulative distribution of $\Delta \theta$ for all jets in our sample (57 objects). The dashed line corresponds to a uniform cumulative distribution. The observed distribution of $\Delta \theta$ is very similar to the uniform one. Indeed, the statistical tests result in large probabilities for a random distribution of the jet position relative to the ISMF: Kp1 = $98\%$;  KS1 = $82\%$. Hence, our sample, as a whole, does not show evidence of  alignment between the jet axis and interstellar magnetic field. 

\begin{figure}
\includegraphics[height=6.5in,angle=270]{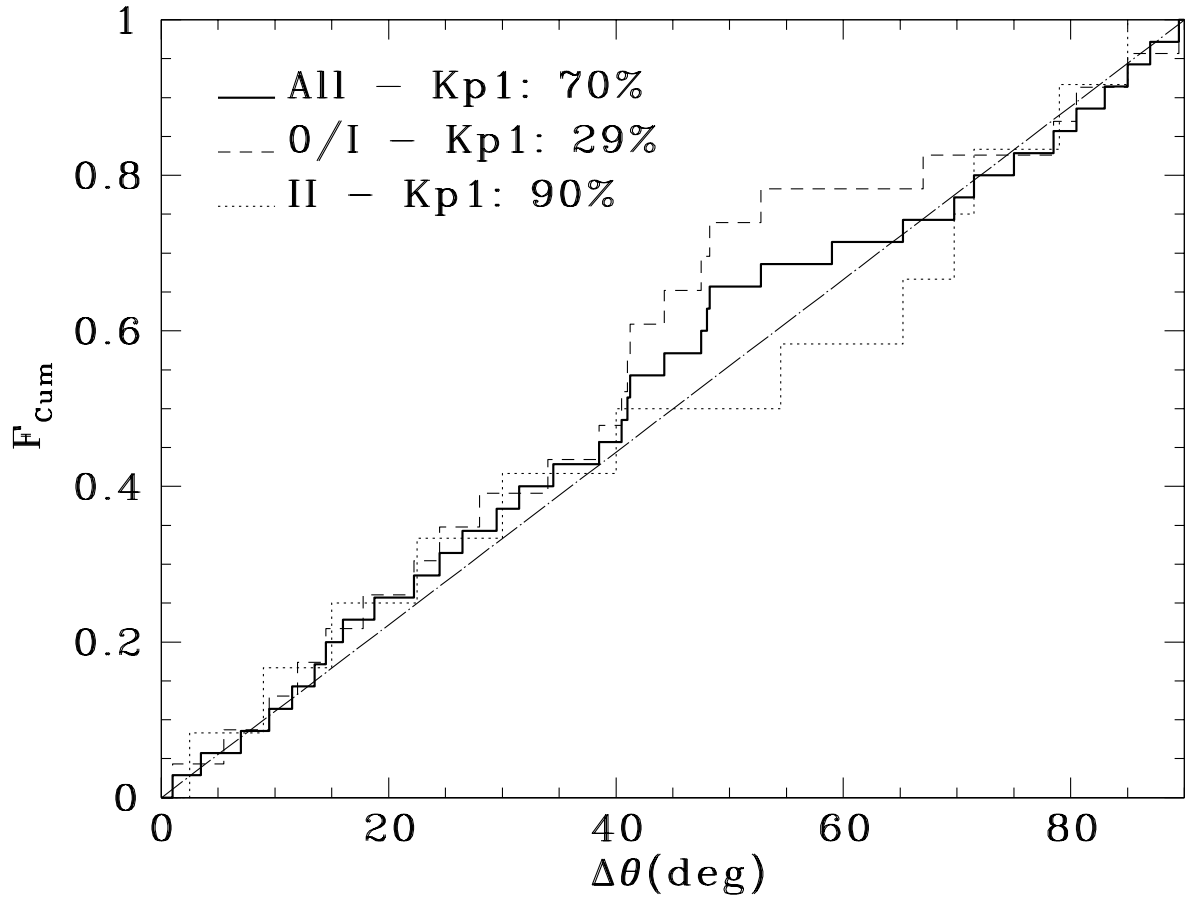} 
\caption{Cumulative histograms of $\Delta\theta$ for all objects having an estimate for their Class (solid line). The histogram for objects of Classes 0 and I is shown as a dashed line and that for objects of Class II as a dotted line. The dot-dash-dot line corresponds to a uniform distribution. It is also shown the probability that each distribution comes from a uniformly distributed population according to the Kuiper test.} 
\label{fig_dtheta_classes}
\end{figure}

We checked if this randomness remains if the objects are grouped by age. Figure \ref{fig_dtheta_classes} presents the objects separated in two groups: (1) Class 0 or I (23 objects, dashed line); (2) Class II or III (12 objects, dotted line).  This figure also presents the observed distribution for all objects in which the class has been determined (35 objects, solid line). The objects whose classification is uncertain between Class I or II, 3 objects, were included in the initial stages group, together with those in Classes 0 or I. Nevertheless, the statistical tests do not exhibit significant changes if they are classified as Class II. 

The position angles of the jets associated with T Tauri stars (Class II/III) have an observed distribution consistent with a random distribution relative to the ISMF: Kp1 = 90\% and KS1 = 74\%. This is not true, however, for the embedded phases whose distribution is not random: Kp1 = 29\% and KS1 = 28\%. The probabilities that the distributions from early and evolved objects are drawn from the same population are small: Kp2 = 65\% and KS2 = 48\%. The histogram for Class 0 and I objects shows an excess of objects \textit{above} the uniform distribution, indicating that these objects tend to have low values of $\Delta\theta$ with respect to a uniform distribution. A possible interpretation is that younger YSOs have jets preferentially aligned to the ISMF, while those nearer to the main-sequence have jets randomly distributed relative to the ISMF.

\begin{figure}
\includegraphics[height=6.5in,angle=270]{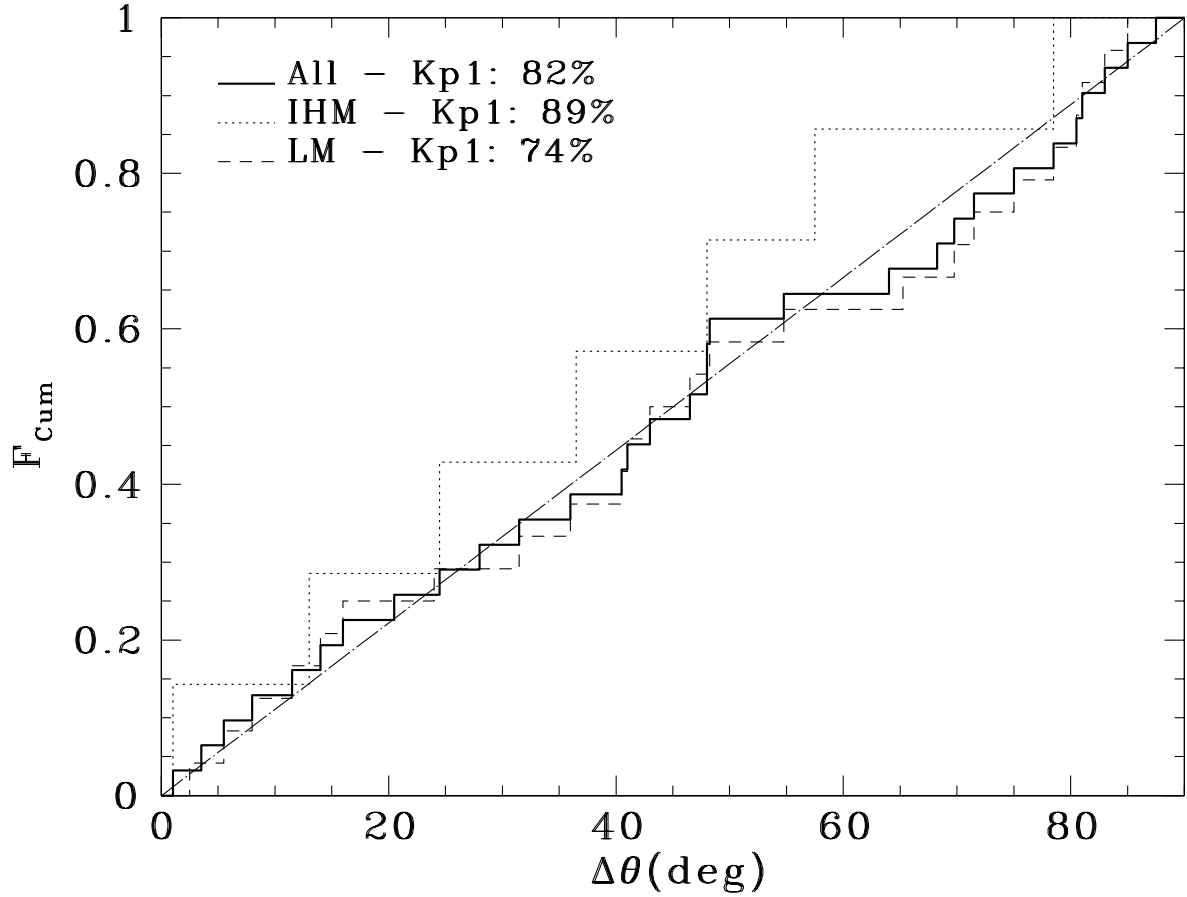}
\caption{Cumulative histograms of $\Delta\theta$ for all the objects of our sample that have the mass determined (solid line), for the low mass objects (dashed line) and for intermediate and high mass objects (dotted line). It is also shown the resulting probability from a Kuiper test comparing each observed distribution with the uniform one.
\label{fig_dtheta_masses}}
\end{figure}

Our sample has 31 objects with estimated masses. Only 7 objects have intermediate or high mass (IHM). The Kp1 test gives us a probability of 89\% (KS1: 96\%) for this distribution be random (Figure \ref{fig_dtheta_masses}), but the  number of objects is in the limit for a reliable result. Using Kp1 in the remaining 24 low-mass (LM) objects, we have obtained a probability of 74\% (KS1: 64\%) for a random distribution, which is a relatively small value. However, using KS2 and Kp2 to compare the low-mass and high-mass samples, we obtain that the two distributions come from the same population with probabilities of 89\% and 86\%, respectively. It is probably the result of the small number of IHM objects. We conclude that our data is not enough to reveal any evidence that the alignment between jet and ISMF depends on the mass of YSO.


In short, the only sub-sample in which a possible non-random behavior of $\Delta \theta$ is found is that of younger objects. This result is consistent with previous studies that have found an alignment between the ISMF and the axis or outflow of YSOs in early evolutionary phases \citep[e.g.,][]{dl79,snell80,hz81,cohen1984,ts89}. Even \citet{Menard2004} also suggest a possible alignment in T Tauri's with brighter jets, which may also be associated with less evolved objects \citep{myers98}. The embedded outflows from Classes 0 and I protostars are also correlated with the filament direction \citep{anath2008}. Recently, arguments in favor of a magnetic field dynamically important in the ISM and in the star formation process have been summarized by \citet{li2011}. Besides the alignments of the magnetic field in interstellar structures of different size scales and of the dense cores minor axis with the surrounding ISMF, another fact pointing to the importance of the ISMF is the turbulence anisotropy observed by these authors. From a modeling perspective, \citet{matsumoto2011} have found that the rotation axis, bipolar outflows, and the magnetic field  become aligned during the process of collapse. All these results suggest that the interstellar magnetic field has a role in shaping protostars. 

\subsection{The dispersion of the interstellar magnetic field}
\label{sec:dispersao}

The degree of alignment of the interstellar magnetic field direction on each field-of-view can be quantified by its dispersion, $\sigma_B$. See Section \ref{sec_observations} for details on its estimation from our polarimetric data and Table \ref{tab:reducao} to see the values. In the following analysis, the entire sample is composed of 28 objects because there is a single value of $\sigma_B$ for each field. 

As done in the previous section, we use the Kolmogorov-Smirnov and the Kuiper tests. However, we do not compare the observed distributions of $\sigma_B$ with a uniform distribution, because there is no expectation that $\sigma_B$ is a random distribution. Therefore, we used the tests to compare two groups: KS2 and Kp2. Another issue is the mass and class associated to a given field-of-view. For the scope of this article, the field is considered of low-mass (or of intermediate/high-mass) if the YSOs associated with the HHs in the field have low-mass (intermediate/high mass) central sources. The same is assumed for the classes. Concerning masses, there is no ambiguity: our regions have only low-mass HHs or only intermediate/high mass objects. For classes, there are only two fields with objects in different evolutionary phases. But, even in these cases, there is a clear dominant class: in Field~1, Classes 0 and I objects; in Field~10, Class II objects.

\begin{figure}
\includegraphics[height=6.5in,angle=270]{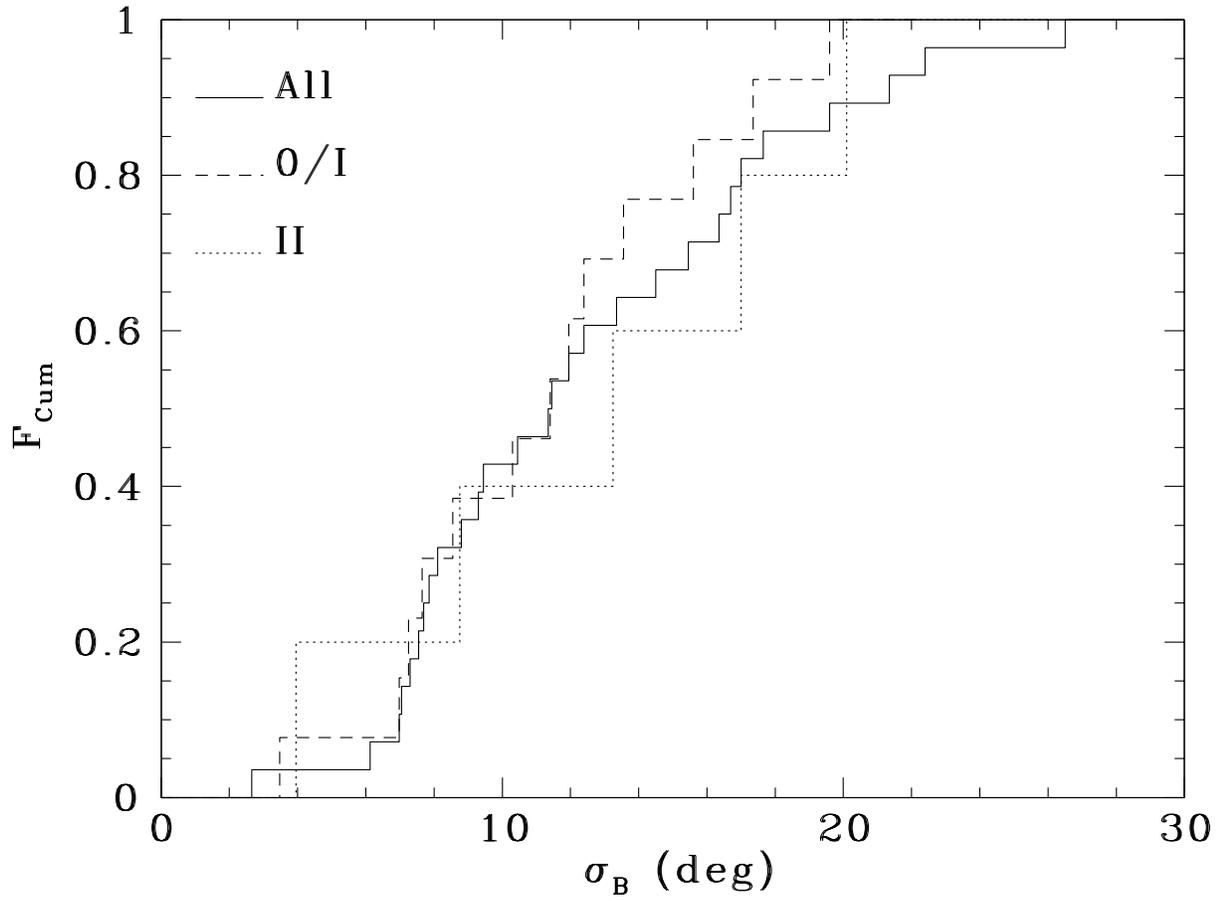} 
\caption{Cumulative histogram of the dispersion of the interstellar magnetic field, $\sigma_B$, for all fields, for fields having objects classified as Classes 0 or I and for regions with objects of Class II.}
\label{fig_sigma_classes}
\end{figure}

Figure \ref{fig_sigma_classes} shows the observed cumulative distribution of $\sigma_B$ for: the entire sample (solid line); regions with Classes 0 or I objects (13 regions - dashed line); and regions with Class II or III objects (5 objects - dotted line). The KS2 probability that the distributions of regions containing objects with different ages come from the same population is only 36\%. In contrast, a different result is found using the Kp2 test: 81\%. This is the only case in which the two tests give inconsistent results and should be caused by the small sample of Class II or III regions. We did not find in the literature the smallest sample for which the Kp test can be applied. The KS test is trustful for samples larger than 4 objects \citep{Press1986}, but, even so, the result should be take with caution. An alternative approach is to look for a possible difference between the two distributions using their mean values. The regions with less evolved objects have $\langle \sigma_B \rangle = 12.0 \pm 1.3 $, while those including more evolved objects have $\langle \sigma_B \rangle = 14.9 \pm 2.8 $. These values are barely consistent, so there might be a difference in the interstellar magnetic field dispersion of regions containing younger or older YSOs.

This possible increase of the dispersion of the ISMF from regions having less evolved objects to those closer to the main sequence can be the result of the injection of energy in the ISM by the outflows, which had already enough time to occur in the T Tauri objects (around ${\rm 10^6}$ years). In an observational study of high-mass star-formation regions, \citet{pillai06} have found an increase in line widths from less to more evolved objects, specifically from infrared dark cloud to high-mass protostellar objects and then to ultracompact HII regions. Other observations point to the same conclusion. For instance, \citet{bm89} show that the width of ammonia lines in dense cores tends to be smaller in cores without star formation than in cores harboring protostars. And the dispersion of the polarization direction may increase with the turbulence, according to \citet{sen2000}, who measured the polarization in a sample of Bok globules of different line widths. The above results may be an indication that the dispersion of the interstellar magnetic fields is caused by the onset of stellar formation. 


Regarding the polarization dispersion, it is worth to cite the results of \citet{mg1991} who studied the interstellar polarization in clouds, clusters, and complexes. They calculated the dispersion in regions whose size is tipically much larger that those of this work. They found that the regions with embedded clusters have greater dispersion than regions without clusters. They suggested that it is the consequence of the enhancement of the gas density and not of the higher stellar content.

\begin{figure}
\includegraphics[height=6.5in,angle=270]{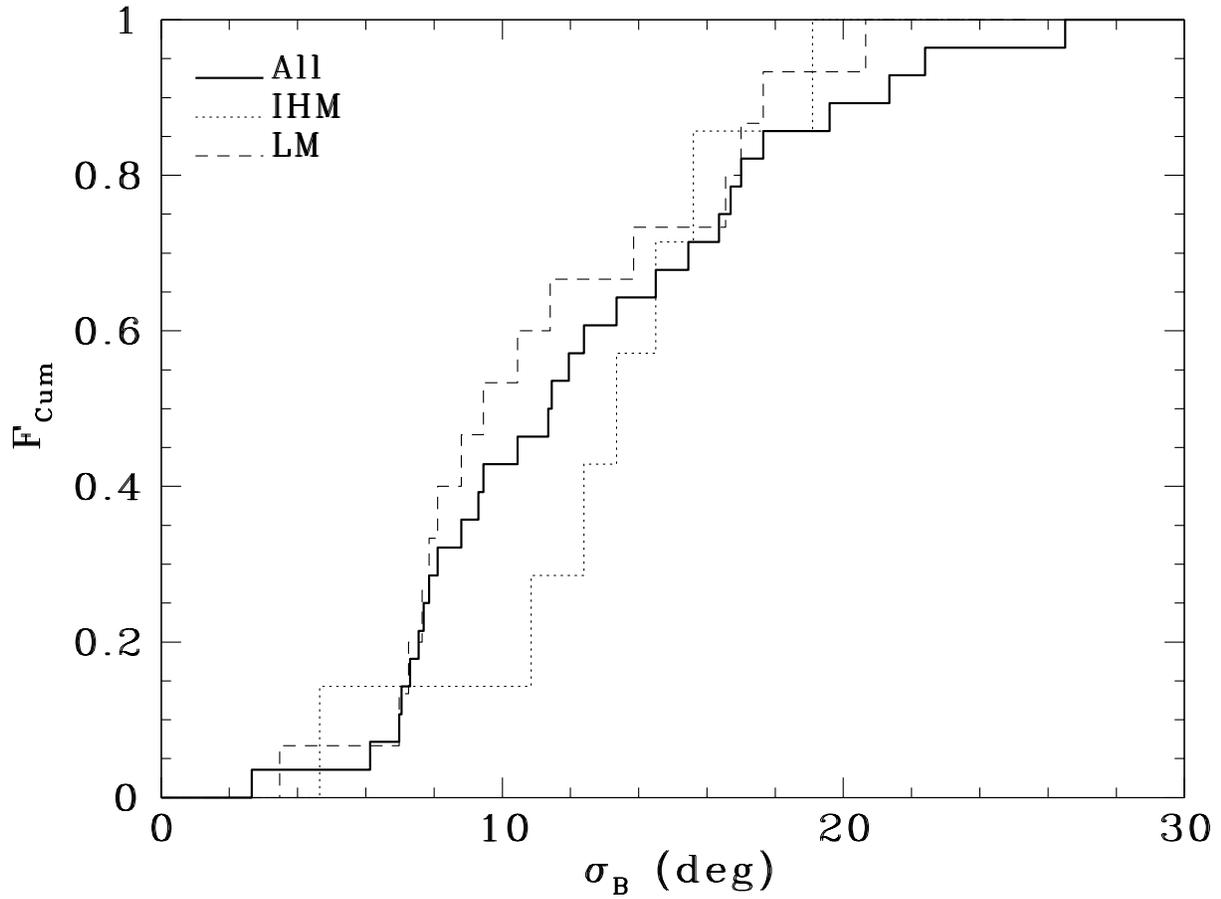} %
\caption{Cumulative histogram of the dispersion of the interstellar magnetic field, $\sigma_B$, for all fields, for fields having low mass objects (LM) and for regions with objects of intermediate or high mass (IHM).}
\label{fig:histocumtodasdisp}
\end{figure}

Figure \ref{fig:histocumtodasdisp} shows the cumulative histograms for the dispersion of all the 28 fields (solid line), for the fields having low-mass objects (LM - 15 fields - dashed line) and for those having intermediate or high-mass objects (IHM - 7 fields - dotted line). Comparing the high and low mass distributions, we obtained a probability of only 10\% (KS2 - the Kp2 test provides a value of 12\%) that the distributions are drawn from a same parent population. The regions with IHM objects have higher values of $\sigma_B$: $\langle\sigma_B\rangle_{IHM}=14.5 \pm 1.5$ and $\langle\sigma_B\rangle_{LM}=11.9 \pm 1.3$. This high dispersion in IHM regions can reflect (i) different properties of these regions prior to star formation or (ii) the injection of energy in the medium by the outflow of the IHM YSO. 

Previous studies indicate that the turbulence is larger in regions that form high-mass stars. \citet{plume1997} have shown that cores associated with massive star formation have larger densities and line widths than those of low-mass formation. Considering quiescent regions, \citet{pillai06} has shown that infrared dark clouds with high masses, hence probably sites of high mass star formation, have line widths larger than the low-mass analogues. Moreover, using observations of $\rm{C^{18}O}$ lines in clumps, \citet{saito2006} have found that the more massive cores are also the more turbulent (and dense), independently if the core has signs of star formation or not. In a complementing article, \citet{saito2007} present a similar trend: the maximum mass of the star formed in a clump increases with the line width. 

\subsection{The average polarization of the interstellar medium}
\label{sec:polarizacaomedia}

In the previous sections, we have studied the direction of the polarization as a tracer of the ISMF. However, we expect that also the value of the polarization carries information on the interstellar medium properties. In this section, we present some results involving the average interstellar polarization, $P_{ISM}$, of the observed fields (column 6 of Table \ref{tab:reducao}). 

\begin{figure}
\includegraphics[height=6.5in,angle=270]{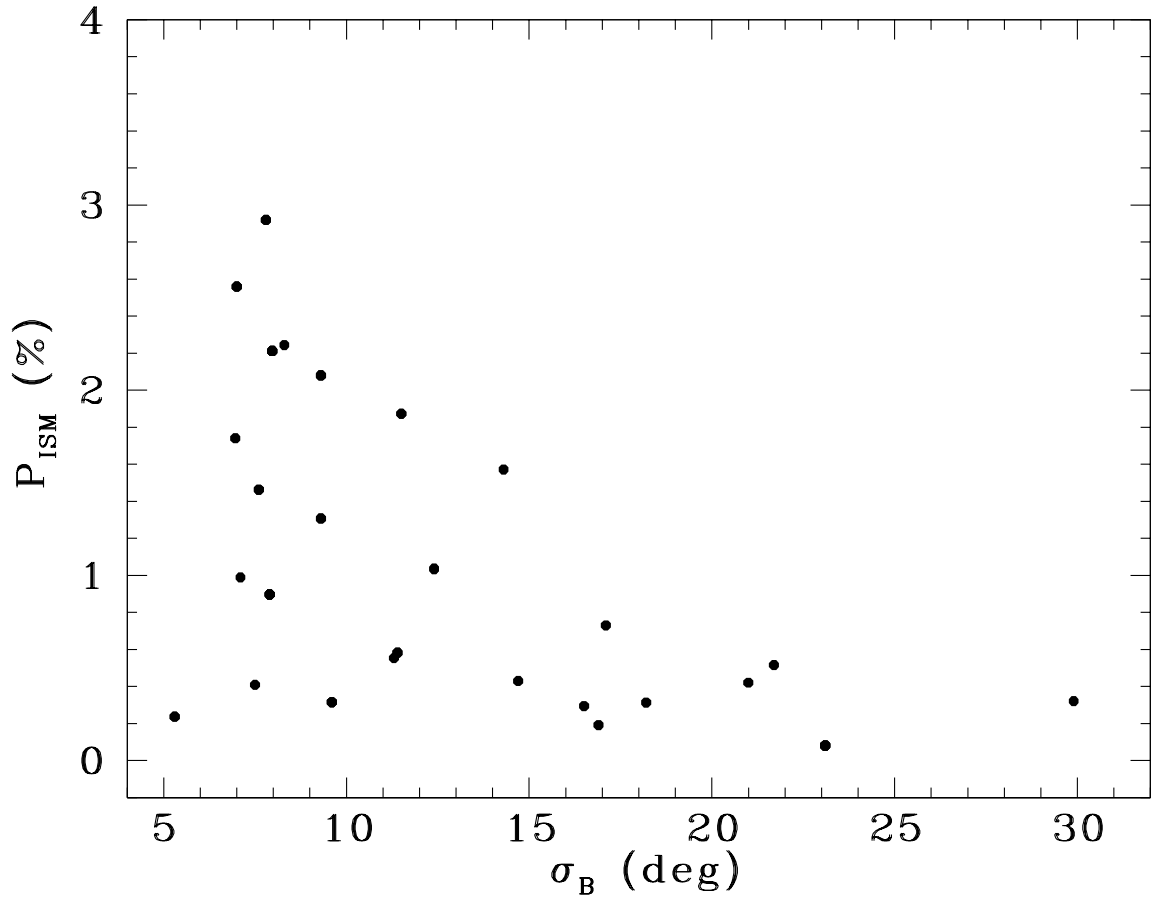}
\caption{Average interstellar polarization of a field plotted as a function of the dispersion of the interstellar magnetic field, $\sigma_B$.}
\label{fig:graficopmediadisp}
\end{figure}

Figure \ref{fig:graficopmediadisp} shows a plot of $P_{ISM}$ against $\sigma_B$. This graph shows clearly that the $P_{ISM}$ has a maximum value for a given $\sigma_B$. This result can be compared with that of \cite{Alves2008} for the \textit{Pipe Nebula}. In this region, $P_{ISM}$ is also correlated with $\sigma_B$: the higher the dispersion, the smaller the polarization. Interestingly, this dark cloud presents regions with and without star formation: they are characterized by different polarimetric properties. In the quiescent region, the dispersions are tipically less than $5^{\rm o}$, with a polarization modulus higher than 5\%. In the region where there is a new-born star identified, the polarization vectors show the higher dispersions (larger than $5^{\rm o}$) and lower degrees of polarization (lower than 4\%). Our sample is not limited to an specific molecular cloud (see Table \ref{tab:localization}). Even so, our results are completely consistent with the range of values associated to the star forming portion of the Pipe Nebula. In Musca dark cloud, the smallest values of polarization (2-3\%) and largest values of dispersion are located near the only site of star formation \citep{per00,pereyra2004}. These results may be an evidence that the average and dispersion of the polarization of a given interstellar region is intrinsically connected to the presence of stellar formation.

\begin{figure}
\includegraphics[height=6.5in,angle=270]{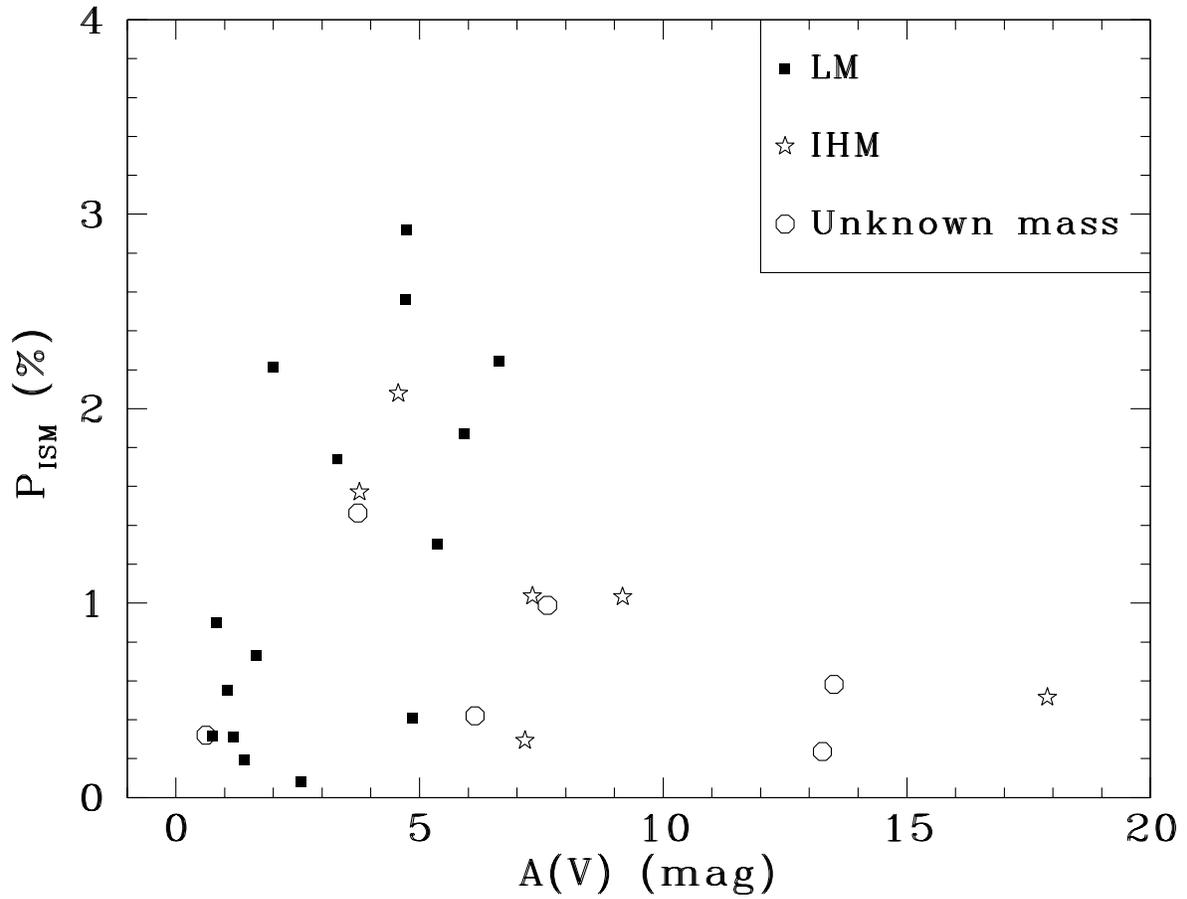} 
\caption{The average polarization of the ISM as a function of the interstellar extinction. Different symbols are used to represent regions harboring objects with different masses according to the legend.} 
\label{fig:graficopmavmassa}
\end{figure}

Figure \ref{fig:graficopmavmassa} depicts the average polarization of a field as a function of the interstellar extinction in that direction. The reddening in band V, A(V), is presented in Table \ref{tab:reducao} and is estimated from dust maps of \citet{sfd1998}. The different symbols are used to represent regions having star formation of low mass (LM, filled square), intermediate and high mass (IHM, open star) and unknown mass (open circle). Figure \ref{fig:graficopmavmassa} shows a clear tendency of smaller polarization for larger extinction. 
Being more specific, the polarization seems to have an initial increase with A(V), to reach a maximum value of around 3\% at A(V)~=~5~mag, and then to decrease. It is not clear if our result is function of the mass of the star being formed in the region. But it is possible that the initial increase is limited to regions forming low-mass stars.

The above result can have a parallel with the decrease of polarization observed towards the center of submillimeter cores \citep[e.g.,][and references therein]{curran2007,Matthews2000}: in both cases, the polarization decreases in regions of large extinction. The initial increase in the polarization with the reddening is expected: more matter, more reddening and polarization. The decrease of the polarization with A(V) could be explained in different ways. We present three possible scenarios. In regions of high density (and hence extinction) the efficiency in the grain alignment can decrease due to microscopic details of the alignment mechanism. For instance, the collisions between particles can increase. Another possible effect of higher density is the increase in the size of grains which occurs together with the change in the grain shape making them more spherically symmetric. Another possibility is an integrated effect in the line-of-sight: with the increase of the column density there is a dispersion of the alignment \textit{along} the line of sight. Simulations of the dense cores, more suitable to the context of sub-mm emission, can reproduce this decrease \citep{flk2008,pelkonen2007}.


\section{Summary} 
\label{summary}

Recent results on star formation process suggest that the magnetic field is important at large scales and at the early phase of star formation process, since the ambipolar diffusion mechanism seems to play a dominant role to determine the collapse phase of molecular clouds \citep{Girart2006,Alves2008}. On the other hand, a  study of a sample of classical T Tauri stars (CTTs) in Taurus-Auriga suggests that there is no correlation at all between the CTTs disk orientation and the ISMF mean direction \citep{Menard2004}. Taking these pieces of evidences, we choose to conduct a study on the orientation of YSOs with respect to interstellar magnetic field for a sample of YSO that is spatially randomly distributed, which means that our sample is not confined to a given molecular cloud complex. Our sample is composed of jets randomly chosen from the \citet{Reipurth1994}'s Herbig-Haro catalog and includes low-mass and high-mass YSOs from Class 0 to Class II.  The direction and dispersion of the interstellar magnetic field  of 28 fields is estimated using optical CCD polarimetry. The analysis considers 57 protostellar jets in those lines of sight. The large number of polarization measurements in each field allows us to also study the magnetic field dispersion. Our main results are summarized below.

\begin{itemize}

\item The sample as a whole does not present an alignment between the ISMF and the jet direction.

\item There is a statistical evidence that the alignment between the jet and the ISMF is a function of the age (Class) of the YSO: there is a tendency of alignment for jets of Classes 0 and I objects, which is not observed for jets associated with T Tauri objects. This suggests that the interstellar magnetic field affects the initial phases of the star formation process and, later on in the star formation process, such a memory can be lost.

\item The cumulative distributions of the dispersion of the ISMF direction seems to be different for younger and evolved objects. Specifically, our sample suggests that the dispersion is slightly larger for objects nearer to the main sequence. Considering this is really the case, a possible interpretation is that the star formation process, probably through mass outflows, can efficiently transfer momentum to the ISM. 

\item The dispersion of the direction of the ISMF is higher in regions having intermediate and high-mass YSOs than in those having low-mass star formation. The same trend is observed in previous works that measured the turbulence.

\item The average interstellar polarization, $P_{ISM}$, decreases for higher values of the dispersion of the ISMF, $\sigma_B$. All the values of $P_{ISM}$ and $\sigma_B$ obtained in our sample are consistent with the ones found by \cite{Alves2008} in the portion of \textit{Pipe Nebula} having star formation. Are the values of $P_{ISM}$ and $\sigma_B$ related to the presence of star formation in a given region? None of the regions studied in this work - all sample star forming regions - has a dispersion as small as those of quiescent regions in the Pipe Nebula. Hence we suggest that $P_{ISM}$ and $\sigma_B$ have different values in regions that form or not form stars. However, measurements of the dispersion in other regions with no star formation should be done to confirm that.

\item The maximum value of $P_{ISM}$ grows with extinction till a reddening of about A(V)~=~5~mag and decreases for higher column densities. A(V)~$\approx$~5~mag is also the inferior reddening in which we see, in our sample, intermediate and high-mass star formation.  

\end{itemize}





\acknowledgments

This study was partially supported by CAPES (CGT), CNPq (CVR: 308005/2009-0; AHC: 307036/2009-0 and 471254/2008-8) and Fapesp (CVR: 2001/12589-1). CGT and CVR are indebted to Germ\'an A. Racca for helpful discussions. We acknowledge the use of: the SIMBAD database, operated at CDS, Strasbourg, France; the NASA's Astrophysics Data System Service; the NASA's {\it SkyView} facility  (http://skyview.gsfc.nasa.gov) located at NASA Goddard Space Flight Center; and the NASA/IPAC Extragalactic Database (NED) which is operated by the Jet Propulsion Laboratory, California Institute of Technology, under contract with the National Aeronautics and Space Administration. 



{\it Facilities:} \facility{OPD}.

\clearpage



\clearpage

\clearpage

\clearpage

\clearpage

\clearpage

\clearpage

\clearpage

\clearpage

\begin{deluxetable}{cl}
\tablecolumns{2}
\tablewidth{0pc}
\tablecaption{Star forming region or cloud in the direction of each field
\label{tab:localization}}
\tablehead{
\colhead{Field}    &   \colhead{Region} \\
 &  \\
}
\startdata
1 & L1630 - South of NGC2068 - Orion\\ 
2 &  Gum Nebula - Bok globule Sa 111 \\ 
3 &  Chamaeleon II dark cloud \\ 
4 &  Lupus 2 \\ 
5 &  Norma 1 \\ 
6 &  L1641 - Orion \\ 
7  & L1660 - Vela \\ 
8 &  Near the small nebula Re 6  - Vela Molecular Ridge \\ 
9 &  Southern edge of the small cloud S114 - Gum Nebula \\ 
10 &   R Coronae Australis Molecular Cloud \\ 
11 & L1641 - Orion \\ 
12 &  Orion B \\ 
13 &  Cometary globule CG30 - Gum Nebula \\ 
14 &  S114 or DC 268.0+1.0 - central part of Vela R2 \\ 
15 &  DC~290.4+01.9, near the bright-rimmed HII region BBW 47 - eastern Carina  \\ 
16 & DC~291.4-02 - Globule No. 103 - Sandqvist No. 127 (S127)  \\ 
17 & G317-4 - Western end of the Circinus molecular cloud complex\\ 
18 & Main Circinus core, western part of the Circinus complex \\ 
19 & S296  \\ 
20 & DC~278.6-0.9 - Vela  \\ 
21 & S109 - Gum Nebula  \\ 
22 & Puppis/Vela \\ 
23 & L1634 - Orion \\ 
24 & GGD 17 - Monoceros R2 \\ 
25 & Small cometary globule Ori I-2 - Orion \\ 
26 & BHR 71 - a Bok Globule  \\ 
27 & Trifid Nebula \\ 
28 & Orion 1b association \\ 
\enddata
\end{deluxetable}

\begin{deluxetable}{lccccccl}
\tablecolumns{8}
\tablewidth{0pc}
\tablecaption{Journal of observations\label{tab:dados}}
\tablehead{
\colhead{Field}    &  \colhead{ HHs } & \colhead{ AR }& \colhead{ Dec } & \colhead{ Filter} & \colhead{$n_{im}$} & \colhead{$t_{exp}$} & \colhead{Date Obs.}  \\
& \colhead{in the field} & \multicolumn{2}{c}{B1950} & & & (s)\\
}
\startdata
1&19, 20, 21, 22, & 05 42 30 & 00 00 00 &$R_C$&12&200& 2005 Feb 15\\
 & 23, 24, 25, 26, & & & & &\\
 & 27, 37, 70 &&& & & &\\
2\tablenotemark{1} & 46, 47& 08 24 17 & -50 50 34 & $R_C$& 9 & 600 & 1998 Dec 18\\
3& 52, 53, 54& 12 50 20 & -76 42 30 & $R_C$& 12& 200 & 2005 Feb 16\\
4 & 55 & 15 52 50 & -37 37 37 & $R_C$ & 12 & 200 & 2005 Feb 16 \\
5A & 56, 57 & 16 28 45 & -44 48 20 & $R_C$ & 8 & 10 & 2005 Feb 12 \\
5B& 56, 57 & 16 28 45 & -44 48 20 & $R_C$& 8& 300& 2005 Feb 12\\
6 & 68, 69 & 05 39 10 & -06 27 00 & $R_C$ & 9 & 200 & 2005 Feb 17 \\
7 & 72 & 07 18 00 & -23 59 00 & $R_C$ & 16 & 300 & 2005 Feb 13 \\
8 & 73, 74 & 09 01 15 & -44 39 00 & $R_C$ & 8 & 60 & 2007 May 07 \\
9 & 75 & 09 09 35 & -45 34 00 & $R_C$ & 12 & 180 & 2005 Feb 17 \\
10 & 82, 96, 97,  & 18 57 42 & -37 14 00 &$R_C$&8&100& 2007 May 07\\
   & 98, 99, 100, &&&&&& \\
  &  101, 104, 729, &&&&&&\\
  &  730, 731, 732, &&&&&&\\
  &  733,734, 735, &&&&&&\\
  &   736, 860 &&&&&&\\
11 & 59, 60, 83 & 05 31 00 & -06 32 00 & $R_C$ & 16 & 300 & 2005 Feb 14\\
12A& 90, 91, 92, 93,  & 05 38 57 & -01 08 00 &$R_C$&8&200&2005 Feb 16\\
   & 597, 598 &&&&&&\\
12B& 90, 91, 92, 93, & 05 39 10 & -00 57 30 &$R_C$&8&250&2005 Feb 16\\
   &  597, 598 &&&&&&\\
13A & 120 & 08 07 47 & -35 59 48 & $R_C$ & 8& 20& 2005 Feb 14\\
13B &120 & 08 07 47 & -35 59 48 &$R_C$ &8 &300 &2005 Feb 14 \\
14 & 133 & 09 08 35 & -45 14 00 & $R_C$& 12& 250& 2005 Feb 17\\
15 &135, 136& 11 10 00 & -58 30 00 &$R_C$&16&300&2005 Feb 12\\
16 & 137, 138 & 11 11 30 & -60 37 00 & $R_C$ & 16 & 120 & 2005 Feb 16 \\
17 &76, 77, 139& 14 56 51 & -63 07 28 &$R_C$&16&300&2005 Feb 13\\
18A &  140, 141, 142, & 14 59 00 & -63 12 00 & $R_C$ & 4 & 10 &2005 Feb 14\\
    &  143  &&&&&&\\
18B &140, 141, 142, & 14 59 00 & -63 12 00 &$R_C$&6&300&2005 Feb 14\\
    &  143 &&&&&&\\
19 & 160 & 07 01 37 & -11 26 00 & $R_C$ & 8 & 40 & 2005 Feb 17 \\
20 & 171 & 09 46 45 & -54 42 30 & $R_C$ & 16 & 60 & 2005 Feb 15 \\
21 & 188, 246 & 08 19 05 & -49 29 30 & $R_C$ & 16 &300 &2005 Feb 13 \\
22 & 217 & 08 15 39 & -35 43 43 & $R_C$ & 16 & 150 & 2005 Feb 17 \\
23 & 240, 241 & 05 17 27 & -06 00 06 & $R_C$ & 8 & 600 & 2005 Feb 11 \\
24A & 271, 272,  & 06 10 25 & -06 11 00 & $R_C$&8&250&2005 Feb 16\\
    &  273  &&&&&&\\
24B & 271, 272, & 06 10 07 & -06 23 00 & $R_C$ & 8 & 100 & 2005 Feb 16\\
    &  273  &&&&&&\\
25 & 289 & 05 35 40 & -01 46 00 & $R_C$ & 8 & 300 & 2005 Feb 15\\
26 & 320, 321 & 11 58 40 & -64 59 00 & $R_C$ & 16 & 120 & 2005 Feb 16 \\
27 & 399 & 17 59 20 & -23 10 00 & $R_C$ & 8 & 100 & 2007 May 07\\
28 & 444, 445, & 05 37 15 & -02 31 00 & $R_C$ & 16 & 250 & 2005 Feb 15\\
   &  446, 447  &&&&&&\\
\enddata
\tablecomments{$n_{im}$ is the number of images and $t_{exp}$ is the exposure time of each image.}
\tablenotetext{1}{Data from \cite{Hickel2002}}
%
\end{deluxetable}

\begin{deluxetable}{lcccccc}
\tablecolumns{7}
\tablewidth{0pc}
\tablecaption{Interstellar polarization and extinction in the observed fields
\label{tab:reducao}}
\tablehead{
\colhead{Field}    &   \colhead{$N_f$} &
\colhead{$\theta_P$} & \colhead{$\theta_B$} & \colhead{$\sigma_B$} & \colhead{$P_{ISM}$} & \colhead{A(V)} \\
 & & \colhead{(deg)} & \colhead{(deg)} & \colhead{(deg)} & \colhead{(\%)} & \colhead{(mag)} \\
}
\startdata
1 &  53 & 173 & 173 & 8.0 & 2.21 & 2.0\\
2\tablenotemark{1} &  166 & 136 & 136 & 7.0 & 1.74 & 3.3 \\
3 &  95& 108& 108& 14.3& 1.57& 3.8\\
4 &  69 & 12 & 12 & 17.1 & 0.73 & 1.7\\
5 &  336 & 51 & 51 & 11.5 & 1.87 & 5.9 \\
6 &  27 & 131 & 131 & 7.6 & 1.46 & 3.7\\
7  & 175 & 116 & 116 & 21.0 & 0.42 & 6.1\\
8 &  19 & 135 & 135 & 5.3 & 0.24 & 13.3\\
9 &  176 & 177 & 177 & 12.4& 1.04& 7.3\\
10 & 15&20&20&23.1&0.08& 2.6\\
11 &  18& 53& 53& 9.6& 0.32& 0.7\\
12A & 14 & 50 & 50 & 11.3 & 0.55 & 1.0\\
12B\tablenotemark{2} & 44 & 160 & 160 & 7.7 & 0.47 & 0.9\\
13 & 234 & 47/101 & 101 & 7.5 & 0.39 & 4.9\\
14 &  289& 24& 24& 8.3& 2.24& 6.6\\
15 &502&60/103&41&21.7&0.52& 17.9\\
16 & 509 & 86/170 & 171 & 11.4 & 0.58 & 13.5\\
17 & 559 & 59 & 59 & 7.0 & 2.56 & 4.7\\
18 &139&68&68&9.3&2.08& 4.6\\
19 & 72&57/150&150&16.5&0.29& 7.2\\
20 &  105& 142& 142& 7.1& 0.99& 7.6\\
21 & 129 &109 &109 &7.8 &2.92 & 4.7\\
22 &  279& 108& 108& 12.4& 1.03& 9.2\\
23 &  17& 39& 39& 18.2& 0.31& 1.2 \\
24A\tablenotemark{2} & 22&163&163&25.8& 0.41 & 1.4\\
24B\tablenotemark{2} & 38&151&151&12.4& 0.15 & 1.4\\
24AB & 60&152&152&16.9& 0.19& 1.4\\
25 & 44 & 161 & 161 & 29.9 & 0.32 & 0.6\\
26 &  497 & 102&102&9.3&1.31& 5.4\\
27 &  179 & 175 & 175 & 14.7 & 0.43 & 61.7\tablenotemark{3}\\
28 & 51& 83/147&151&7.9&0.90& 0.8\\
\enddata
\tablenotetext{1}{Data from \cite{Hickel2002}}
\tablenotetext{2}{Not used in the analysis. Kept to comparison.}
\tablenotetext{3}{Value not reliable (low Galactic latitude). This value is not included in the graphs.}
\end{deluxetable}

\begin{deluxetable}{cccccc}
\tablecolumns{6}
\tablewidth{0pc}
\tablecaption{Polarimetric catalog\label{tab:polarimetriccatalog}}
\tablehead{
\colhead{Field} & \colhead{ID} & \colhead{RA} & \colhead{Dec} & \colhead{P} & \colhead{PA} \\
 &  & (1950.0) & (1950.0) & (\%) & (deg) 
}
\startdata
\cutinhead{Field 01} 
1 &1 &5 42 14.94&+00 02 07.90 &1.83$\pm$0.12&165.0 \\
1 &2 &5 42 15.04&+00 01 26.40 &2.21$\pm$0.10&169.5 \\
1 &3 &5 42 15.24&+00 02 56.09 &1.96$\pm$0.04&166.8 \\
1 &4 &5 42 15.82&-00 06 09.76 &3.76$\pm$0.79&173.8 \\
1 &5 &5 42 17.40&-00 01 06.54 &2.82$\pm$0.51&162.9 \\
\enddata
\tablecomments{Table 4 is published in its entirety in the electronic 
edition of the {\it Astrophysical Journal}.  A portion is shown here 
for guidance regarding its form and content.}
\end{deluxetable}

\begin{deluxetable}{ccccccccc}
\tabletypesize\small
\tablecolumns{9}
\tablewidth{0pc}
\tablecaption{Jet Information \label{tab:jets}}
\tablehead{
\colhead{Jet}    &   \colhead{Distance} & 
\colhead{Reference} & \colhead{Extension\tablenotemark{a}} & \colhead{Reference\tablenotemark{b}} & \colhead{PA} & \colhead{Reference\tablenotemark{c}}\\
& \colhead{(pc)} & & \colhead{(pc)} &  & \colhead{(\degr)} \\
}
\startdata
24J,19/24K,27 & 450 & 21 & 0.95/0.68 & 21 & 132 & Visual (21,52) \\
24C,20/24E,24M & 450 & 21 & 1.09/0.11 & 21 & 153 & 40 \\
24G & 500 & 40 & 0.24 & 40 & 38 & 40 \\
22 & 450 & 21 & 0.40 & 21 & 76 & Visual (21) \\
23 & 450 & 21 & 0.37 & 21 & 8 & Knots (21) \\
24A/25B & 400 & 24 & 0.13 & This work (24,35) & 40 & 24,35 \\
25A/25D & 400 & 24 & 0.07 & This work (24,35) & 164 & 24,35 \\
26A/26B & 400 & 24 & 0.07 & This work (24,35) & 50 & 24,35 \\
46/47 & 460,450 & 18,61 & 1.3 & 30 & 56 & 64,71 \\
52,53,54 & 130 & 9 & 0.39 & This work (54) & 60 & Visual (54) \\
55 & 150-250 & 27 & 0.04 & 27 & 160 & 27 \\
56 & 700,900: & 55,9 & 0.38 & 65 & 36  & Knots  (65) \\
57 & 700,900: & 55,9 & 0.02 & This work (65) & 19 & Knots (65)\\
59 & 460 & 15 & 0.05 & 63 & 0 & 15 \\
60 & 460 & 15 & 0.47 & 15 & 124 & 15\\
68 & 460 & 68 & 0.15 & This work (4) & 156 & 15\\
69 & 460 & 15 & 0.26 & This work (63) & 158: & 15\\
72 & 1500 & 24 & 0.44 & This work (63) & 79 & 24\\
73 & 450 & 52 & 0.15 & 52 & 146 & Visual (63)\\
74 & 450 & 15 & 0.13 & This work (37,75) & 93 & Knots (37,75) \\
75 & 450,870 & 15,63 &  1\tablenotemark{d}  & This work (63) & 152 & 15\\
76 & 700,500-1000 & 6,63 & 0.04\tablenotemark{e} & 63 & 148 & 15 \\
77 & 700,750 & 6,15 & \nodata & \nodata & 119 & 15 \\
82 & 129,170 & 63,74 & 0.07 & This work (63) & 100 & 63 \\
83/83 & 470,480 & 5,39 & 0.43 & 40 & 125 & 40 \\
90-93,597,598 & 415 & 7 & 2.05 & 7 & 131 & 7\\
96-98,100,101 & 129,170 & 63,74 & $>$0.09 & This work (74) & 32 & Visual (74)\\
99,104C-D,730,860 & 129,170 & 63,74 & 0.26 & 74 & 58 & Visual (74)\\
120 & 450 & 13 & 0.01 & 70 & 110 & Visual (28) \\
133 & 870 & 46 & 0.32 & This work (46) & 105 & 46 \\
135/136 & 2700-2900 & 50 & 0.55 & 50 & 39 & 67  \\
137 & 2200 & 47 & 0.84 & This work (47) & 103 & 47 \\
138 & 2200 & 47 & 0.23 & This work (47) & 106 & Knots (47)\\
139 & 700,1500 & 6,29 & 0.005 or 0.01 & This work (6,29) & 100 & 6 \\
140 & 2900,700 & 56,6 & 1.8 or 0.43 & This work (56,6) & 135 & Visual (56)\\
141 & 2900,700 & 56,6 & 0.42 or 0.1 & This work (56,6) & 95 & Visual (56) \\
142,143 & 2900,700 & 56,6 & 1.31 or 0.32 & This work (56,6) & 135 & 56 \\
160 & 1150 & 59 & 1.8 & 53 & 60 & 73\\
171 & \nodata & \nodata & \nodata & \nodata & 60 & 48 \\
188 & 450 & 25 & 1.2 & 25 & 149 & 25\\
217 & 4300 & 41 &  0.40 & 41 & 60 & Visual (36,41,58) \\
240/241 & 460,500 & 45,18 & 0.40 & 18 & 103 & Visual (18) \\
246 & 450 & 26 & 0.02 & 26 & 115 & 26  \\
271-272 & 830 & 14 & 0.72\tablenotemark{f} & 14 & 165\tablenotemark{g} & Visual (8,14) \\
273 & 830 & 14 & \nodata & \nodata & \nodata & \nodata \\
289 & 470 & 38 & 0.62 & 38 & 65 & Visual (38) \\
320 & 200 & 11 & 0.06 & This work (24) & 144 & 24 \\
321 & 200 & 11 & 0.08 & This work (24) & 0 & 24  \\
399 & 1680,1670-2670 & 66,76 &  0.14 & 76 & 20 & Visual (76) \\
444 & 360-470 & 60 & 0.35 & 60 & 66 & Knots (38,60) \\
445 & 360-470 & 60 & 0.28 & 60 & 103 & Visual (60) \\
445X & 360-470 & 60 & 0.03 & This work (60) & 78 & Visual (60) \\
446 & 360-470 & 60 &  0.03 & 60 & 168 & Visual (60) \\
447 & 360-470 & 60 & 0.02 & 60 & 32 & Visual (60) \\
729 & 129,170 & 63,74 & 0.07 & This work (74) & 115 & 74 \\
731 & 129,170 & 63,74 & \nodata & \nodata & \nodata & \nodata \\
732 & 129,170 & 63,74 & $>$0.03 & This work (74) & 157 & Visual (74) \\
733 & 129,170 & 63,74 & 0.16 & This work (74) & 35 & Visual (74)\\
734 & 129,170 & 63,74 & \nodata & \nodata & 129 & Visual (74) \\
735,736 & 129,170 & 63,74 & \nodata & \nodata & \nodata & \nodata\\
\enddata
\tablecomments{The numbers in the first column correspond to the HHs that define a jet. HHs separated by ``/" correspond to jet and counter-jet. Colon is used to indicate uncertainty.}
\tablenotetext{a}{The extension represents the length from the YSO to the farther knot detected.}
\tablenotetext{b}{``This work'' denotes the cases in which we calculated the extension using data from previous references, which are also cited in this column.}
\tablenotetext{c}{``Visual'' stands for PA estimated by images. ``Knots'' stands for PA estimated by YSO and knots coordinates. The reference number stands for the source of the image or coordinates used.}
\tablenotetext{d}{If IRAS09094-4522 is the source and if it is at 450 pc.}
\tablenotetext{e}{Distance from Knot a to Knot b at 725 pc. Visual determination.} 
\tablenotetext{f}{It corresponds to the total extension of the jet, including the parts before and after the deflexion. }
\tablenotetext{g}{This angle corresponds to the direction before the deflection.}
\tablerefs{We use the same numeration for references in this table and in Table \ref{tab:ysos}. (1) \cite{Abraham2004a}; (2) \cite{Abraham2004}; (3) \cite{Andrews2004}; (4) \cite{Avila2001}; (5) \cite{Bally1994}; (6) \cite{Bally1999}; (7) \cite{Bally2002}; (8) \cite{Beltran2001}; (9) \cite{Berrilli1989}; (10) \cite{Bohigas1993}; (11) \cite{Bourke2001}; (12) \cite{Brittain2007}; (13) \cite{Caratti2006}; (14) \cite{Carballo1992}; (15) \cite{Cohen1990}; (16) \cite{Connelley2007}; (17) \cite{Corporon1997}; (18) \cite{Davis1997}; (19) \cite{Dent1998}; (20) \cite{Dobashi1998}; (21) \cite{Eisloffel1997}; (22) \cite{Felli1998}; (23) \cite{Forbrich2007}; (24) \cite{Gianini2004}; (25) \cite{Girart2007}; (26) \cite{Graham1986}; (27) \cite{Graham1994}; (28) \cite{Gredel1994}; (29) \cite{Gyulbudaghian2005}; (30) \cite{Hartigan2005}; (31) \cite{Heyer1990}; (32) \cite{Huelamo2007}; (33) \cite{Jijina1999}; (34) \cite{Lefloch2002}; (35) \cite{Lis1999}; (36) \citet{Liseau1992}; (37) \citet{Lorenzetti2002}; (38) \cite{Mader1999}; (39) \cite{Miesch1994}; (40) \cite{Mundt1991}; (41) \cite{Neckel1995}; (42) \cite{Nielbock2005}; (43) \cite{Nisini1996}; (44) \cite{Noriega-Crespo2004}; (45) \cite{O'Connell2004}; (46) \cite{Ogura1990}; (47) \cite{Ogura1993}; (48) \cite{Ogura1994}; (49) \cite{Ogura1991}; (50) \cite{Ogura1992}; (51) \cite{Persi1994}; (52) \cite{Podio2006}; (53) \cite{Poetzel1989}; (54) \citet{Porras2007}; (55) \cite{Prusti1993}; (56) \cite{Ray1994}; (57) \cite{Reipurth1989}; (58) \cite{Reipurth1994}; (59) \cite{Reipurth2000}; (60) \cite{rei1998}; (61) \cite{Reipurth1995}; (62) \cite{Reipurth1993}; (63) \cite{Reipurth1988}; (64) \cite{Reipurth1991}; (65) \cite{Reipurth1997}; (66) \cite{Rho2006}; (67) \cite{rod2007}; (68) \cite{Rodriguez1994}; (69) \cite{Rolph1990}; (70) \cite{Schwartz2003}; (71) \cite{Stanke1999}; (72) \cite{Thi2006}; (73) \cite{Velazquez2001}; (74) \cite{Wang2004}; (75)\citet{Wu2002}; (76) \cite{Yusef-Zadeh2005} }
\end{deluxetable}

\begin{deluxetable}{ccccccccc}
\tabletypesize\scriptsize
\tablecolumns{9}
\tablewidth{0pc}
\tablecaption{Information about the young stellar objects associated with the jets of our sample\label{tab:ysos}}
\tablehead{
\colhead{Jet} &  \colhead{Name} & \colhead{Reference} & \colhead{$L_{bol}$} 
& \colhead{Reference}& \colhead{Mass} & \colhead{Reference} & 
\colhead{Class} & \colhead{Reference} \\
 &  &  & \colhead{($L\odot$)}
}
\startdata
24J,19/24K,27 & SSV 63W & 21 & $<$20\tablenotemark{a} & 35,62 & \nodata & \nodata & I & 35 \\
24C,20/24E,24M & SSV 63E & 21 & $<$20\tablenotemark{a} & 35,62 & \nodata & \nodata & I & 32  \\
24G & SSV 63NE & 40 & $<$ 20\tablenotemark{a} & 35,62 & low & 32 & \nodata & \nodata\\
22 & \nodata & \nodata & \nodata & \nodata & \nodata & \nodata & \nodata & \nodata \\
23 & IRAS~05436-0007, V1647 Ori & 21,12 & 6 &  2 & low & 2 & I/II\tablenotemark{b} & 12,2  \\
24A/25B & HH24MMS & 24 &  5 & 24 & \nodata & \nodata & 0 & 24  \\
25A/25D & HH25MMS,VLA2 & 24,35 & 6,24 & 24,35 & \nodata & \nodata & 0 & 24 \\
26A/26B & HH26IR & 24 & 29 & 24 & \nodata & \nodata & I & 24 \\
46/47 & IRAS~08242-5050 & 62  & 24,12,19 & 9,44,62 & low & 19 & I & 44 \\
52,53,54 & IRAS~12496-7650\tablenotemark{c} & 43 &\nodata & \nodata & int\tablenotemark{c} & 43 & \nodata & \nodata \\
55 & IRAS~15533-3742,HH55star & 27,31 & $<$0.3 & 31,27,9  & low & 27 & II/III & 27 \\
56 & Re 13 & 55 & 50 & 55 & \nodata & \nodata & I & 55 \\
57 & V346 Nor & 42 & 135\tablenotemark{b} & 55 & low & 1 & I\tablenotemark{b} & 42 \\
59 & \nodata & \nodata & \nodata & \nodata & \nodata & \nodata & \nodata & \nodata \\
60 & IRAS~05299-0627c & 19 & 1 & 15 & \nodata & \nodata & \nodata & \nodata \\
68 & IRAS~05391-0627c:,HH68b: & 15,4 & 10 & 15 & \nodata & \nodata & \nodata & \nodata \\
69 & IRAS~05393-0632: & 15 & 25 & 15 & \nodata & \nodata & \nodata & \nodata \\
72 & IRAS~07180-2356 & 24 & 170,316 & 13,15 & \nodata & \nodata & I & 13 \\
73 & unknown & 52 & \nodata & \nodata & \nodata & \nodata & \nodata & \nodata\\
74 & IRAS~09003-4438C & 15 & 6 & 15 & \nodata & \nodata & \nodata & \nodata \\
75 & IRAS~09094-4522: & 15  & 130 & 15 & int & 72 & I & 72 \\
76 & IRAS~14563-6250 & 62 & 21 & 15 & \nodata & \nodata & I & 6 \\
77 & IRAS~14564-6254 & 15 & 47 & 63 & \nodata & \nodata & \nodata & \nodata \\
82 & S CrA & 63 & 2 & 62 & low\tablenotemark{d} & 63 & II\tablenotemark{d} & 63 \\
83 & IRAS~05311-0631 & 52 & 9\tablenotemark{e} & 33 & low\tablenotemark{d} & 5 & II\tablenotemark{d} & 16\\
90-93,597,598 & IRAS~05399-0121 & 7 & 10 & 7 & low & 7 & I & 16 \\
96-98,100,101 & IRS1/HH100-IRS & 74 & 3 & 23 & \nodata & \nodata & I & 74 \\
99,104C-D,730,860 & IRS 6 & 74 & 1 & 23 & \nodata & \nodata & II & 23 \\
120 & IRAS~08076-3556 & 13 & 13-19 & 13 & low & 51 & I & 13 \\
133 & \nodata & \nodata & \nodata & \nodata & low: & 46 & \nodata & \nodata \\
135/136 & IRAS~11101-5829 & 75 & 14000 & 50 & high & 50 & 0/I & 67\\
137,138\tablenotemark{f} & unknown & 47 & \nodata & \nodata & \nodata & \nodata & \nodata & \nodata\\
139 & IRAS~14568-6304 & 6 & 133 & 20\tablenotemark{g} & low\tablenotemark{d} & 6 & I,II\tablenotemark{d} & 6,29 \\
140 & IRAS~14592-6311 & 75 & 2400 & 56 & int & 56 & \nodata & \nodata\\
141 & \nodata & \nodata & \nodata & \nodata & \nodata & \nodata & \nodata & \nodata \\
142,143: & \nodata & \nodata & \nodata & \nodata & \nodata & \nodata & \nodata & \nodata \\
160 & Z CMa,IRAS~07013-1128  & 19 & 3500,3000 & 53,73 & int & 53 & I/II\tablenotemark{b} & 53 \\
171 & IRAS~09469-5443: & 48 & \nodata & \nodata & \nodata & \nodata & \nodata & \nodata \\
188 & IRAS~08194-4925 & 25 & 30 & 25 & low & 25 &  0/I & 25 \\
217 & IRAS~08159-3543 & 58 & 2400 & 22 & int & 22 & I & 22 \\
240/241 & IRAS~05173-0555 & 18 & 17-27 & 13 & low & 10 & I & 13\\
246 & HD180617 & 26 & \nodata & \nodata & low & 26 & \nodata & \nodata \\
271,272\tablenotemark{h} & Bretz 4,IRAS~06103-0612  & 14,8  & 4 & 8 &  low\tablenotemark{d} & 14 & II\tablenotemark{d} & 8,14\\
273 & \nodata & \nodata & \nodata & \nodata & \nodata & \nodata & \nodata & \nodata \\
289 & IRAS~05355-0146 & 38 & 13 & 38 & \nodata & \nodata & \nodata & \nodata \\
320 & BHR71(IRS2) & 13 & 1-3 & 13 & low & 17 & I & 13\\
321 & BHR71-MM(IRS1) & 13 & 8-10 & 13 & low & 17 & 0 & 13\\
399 & TC2 & 66 & 600 & 66 & high & 34 & 0/I & 66 \\
444 & V510 Ori & 38 &  variable & 60 & low\tablenotemark{d} & 3 & II\tablenotemark{d} & 38 \\
445 & A0976-357 & 60 & \nodata & \nodata & low\tablenotemark{d} & 60 & II\tablenotemark{d} & 60 \\
445X & A0976-357 & 60 & \nodata & \nodata & low\tablenotemark{d} & 60 & II\tablenotemark{d} & 60 \\
446 & \nodata & \nodata & \nodata & \nodata & low\tablenotemark{d} & 3 & II\tablenotemark{d} & 60 \\
447 & Haro5-39 & 60 & \nodata & \nodata & low\tablenotemark{d} & 60 & II\tablenotemark{d} & 60 \\
729 & S CrA & 74 & 2 & 63 & low\tablenotemark{d} & 63 & II\tablenotemark{d} & 63 \\
731 & IRS1, 2 or 5 & 74 & \nodata & \nodata & \nodata & \nodata & I & 74 \\
732 & \nodata & \nodata & \nodata & \nodata & \nodata & \nodata & \nodata & \nodata \\
733 & T CrA & 74 & 3 & 23 & low & 23 & II & 23 \\
734 &  K-ex or WMB 55 & 74 & \nodata & \nodata & \nodata & \nodata & \nodata & \nodata \\
735,736: & IRS7 or MMS19: & 74 & \nodata & \nodata & \nodata & \nodata & \nodata & \nodata \\
\enddata
\tablecomments{The numbers in the first column correspond to the HHs that define a jet. HHs separated by "/" correspond to jet and counter-jet. The luminosities were rounded to integer values. Colon indicates uncertainty.}
\tablenotetext{a}{The luminosity of the source as a whole is: $L_{bol}$(SSV~63) = 21 (Ref. 35), 24 (Ref. 62).}
\tablenotetext{b}{FU Orionis object: see the text.}
\tablenotetext{c}{There are several low mass sources in this region, but there is also an intermediate mass source (IRAS~12496-7650), which could be the source of the 3 HHs, because they are almost aligned (Ref. 43). We considered IRAS~12496-7650 as the source of the 3 HHs. }
\tablenotetext{d}{T Tauri star: classified as low mass and Class II.}
\tablenotetext{e}{There are several determinations. We adopt the value nearest to the average. The other values are: 7.3 (Ref. 16); 8 (Ref. 5); 10.6 (Ref. 49); 9.5 (Ref. 69); 10.5 (Ref. 57); 7.9 (Ref. 62). }
\tablenotetext{f}{It is not clear if they are physically connected or not (Ref. 47). }
\tablenotetext{g}{This luminosity was calculated using a distance of 1260 pc.}
\tablenotetext{h}{It is thought that they belong to the same physical system, being HH~272 a deflection of HH~271 by the medium.}
\tablerefs{We use the same numeration for references in this table and in Table \ref{tab:jets}. (1) \cite{Abraham2004}; (2) \cite{Abraham2004a}; (3) \cite{Andrews2004}; (4) \cite{Avila2001}; (5) \cite{Bally1994}; (6) \cite{Bally1999}; (7) \cite{Bally2002}; (8) \cite{Beltran2001}; (9) \cite{Berrilli1989}; (10) \cite{Bohigas1993}; (11) \cite{Bourke2001}; (12) \cite{Brittain2007}; (13) \cite{Caratti2006}; (14) \cite{Carballo1992}; (15) \cite{Cohen1990}; (16) \cite{Connelley2007}; (17) \cite{Corporon1997}; (18) \cite{Davis1997}; (19) \cite{Dent1998}; (20) \cite{Dobashi1998}; (21) \cite{Eisloffel1997}; (22) \cite{Felli1998}; (23) \cite{Forbrich2007}; (24) \cite{Gianini2004}; (25) \cite{Girart2007}; (26) \cite{Graham1986}; (27) \cite{Graham1994}; (28) \cite{Gredel1994}; (29) \cite{Gyulbudaghian2005}; (30) \cite{Hartigan2005}; (31) \cite{Heyer1990}; (32) \cite{Huelamo2007}; (33) \cite{Jijina1999}; (34) \cite{Lefloch2002}; (35) \cite{Lis1999}; (36) \citet{Liseau1992}; (37) \citet{Lorenzetti2002}; (38) \cite{Mader1999}; (39) \cite{Miesch1994}; (40) \cite{Mundt1991}; (41) \cite{Neckel1995}; (42) \cite{Nielbock2005}; (43) \cite{Nisini1996}; (44) \cite{Noriega-Crespo2004}; (45) \cite{O'Connell2004}; (46) \cite{Ogura1990}; (47) \cite{Ogura1993}; (48) \cite{Ogura1994}; (49) \cite{Ogura1991}; (50) \cite{Ogura1992}; (51) \cite{Persi1994}; (52) \cite{Podio2006}; (53) \cite{Poetzel1989}; (54) \citet{Porras2007}; (55) \cite{Prusti1993}; (56) \cite{Ray1994}; (57) \cite{Reipurth1989}; (58) \cite{Reipurth1994}; (59) \cite{Reipurth2000}; (60) \cite{rei1998}; (61) \cite{Reipurth1995}; (62) \cite{Reipurth1993}; (63) \cite{Reipurth1988}; (64) \cite{Reipurth1991}; (65) \cite{Reipurth1997}; (66) \cite{Rho2006}; (67) \cite{rod2007}; (68) \cite{Rodriguez1994}; (69) \cite{Rolph1990}; (70) \cite{Schwartz2003}; (71) \cite{Stanke1999}; (72) \cite{Thi2006}; (73) \cite{Velazquez2001}; (74) \cite{Wang2004}; (75)\citet{Wu2002}; (76) \cite{Yusef-Zadeh2005} }
\end{deluxetable}

\begin{deluxetable}{cccc}
\tablecolumns{4}
\tablewidth{0pc}
\tablecaption{Proposed new coordinates for HH~140-143 \label{tab_hh140}}
\tablehead{
\colhead{\footnotesize HH} & \colhead{\footnotesize Knot} & \colhead{\footnotesize AR (1950.0)}  & \colhead{\footnotesize DEC (1950.0)} \\}
\startdata
140 & C & 14 59 12 & -63 10 44 \\
 & D & 14 59 26 & -63 12 43 \\
141 & A & 14 59 15 & -63 12 33 \\
 & D & 14 59 10 & -63 12 28 \\
142 & & 14 59 25 & -63 10 38 \\
143 &  & 14 59 35 & -63 11 52 \\
\enddata
\end{deluxetable}

\clearpage

\Huge

\center{From this page on, complementary graphs of Figure 1. They will appear as online-only material.}

\normalsize

\clearpage

\begin{figure}
\includegraphics[width=13cm,angle=0]{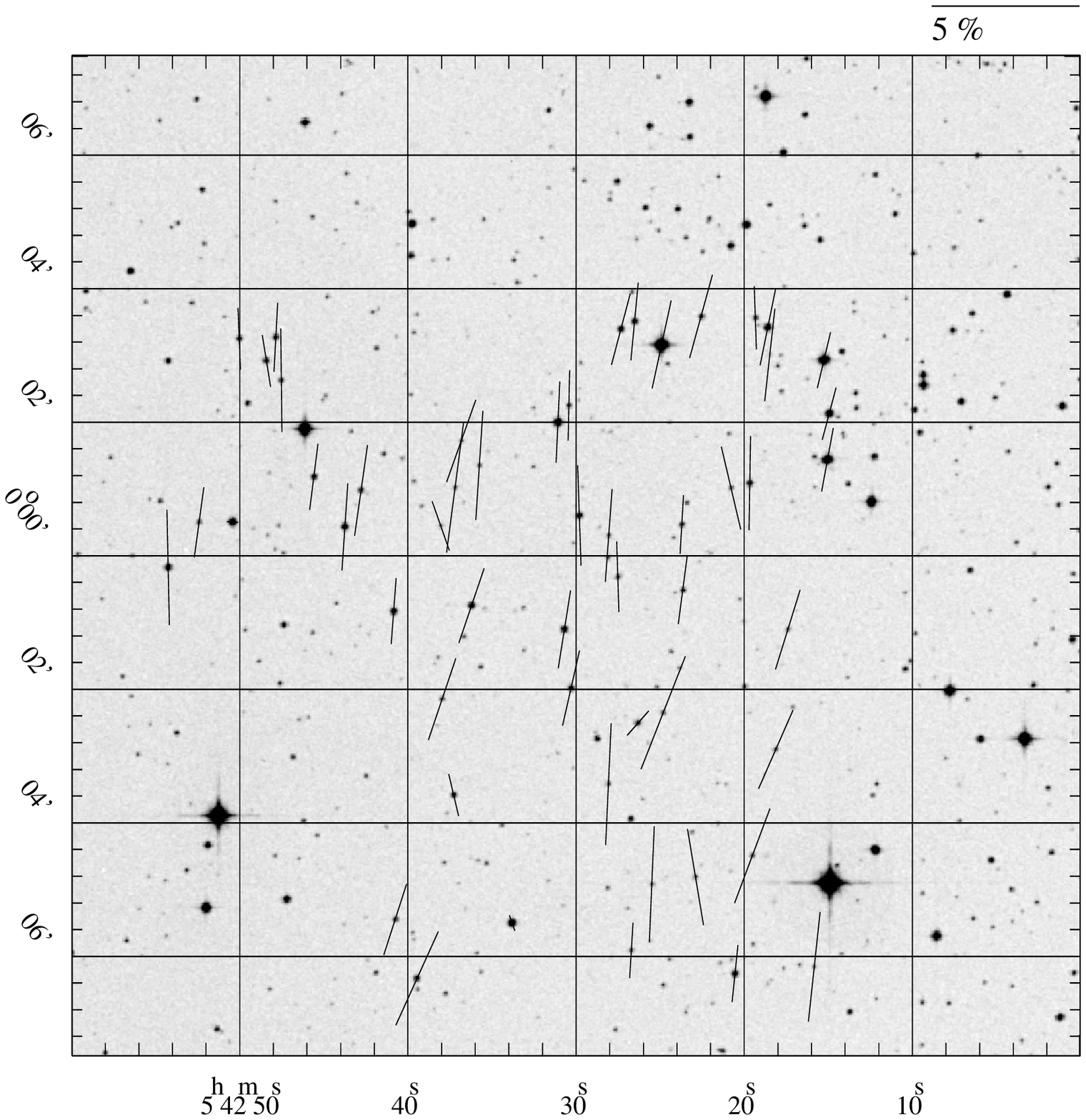}

\vspace{-5cm}
\includegraphics[width=6cm,angle=-90.]{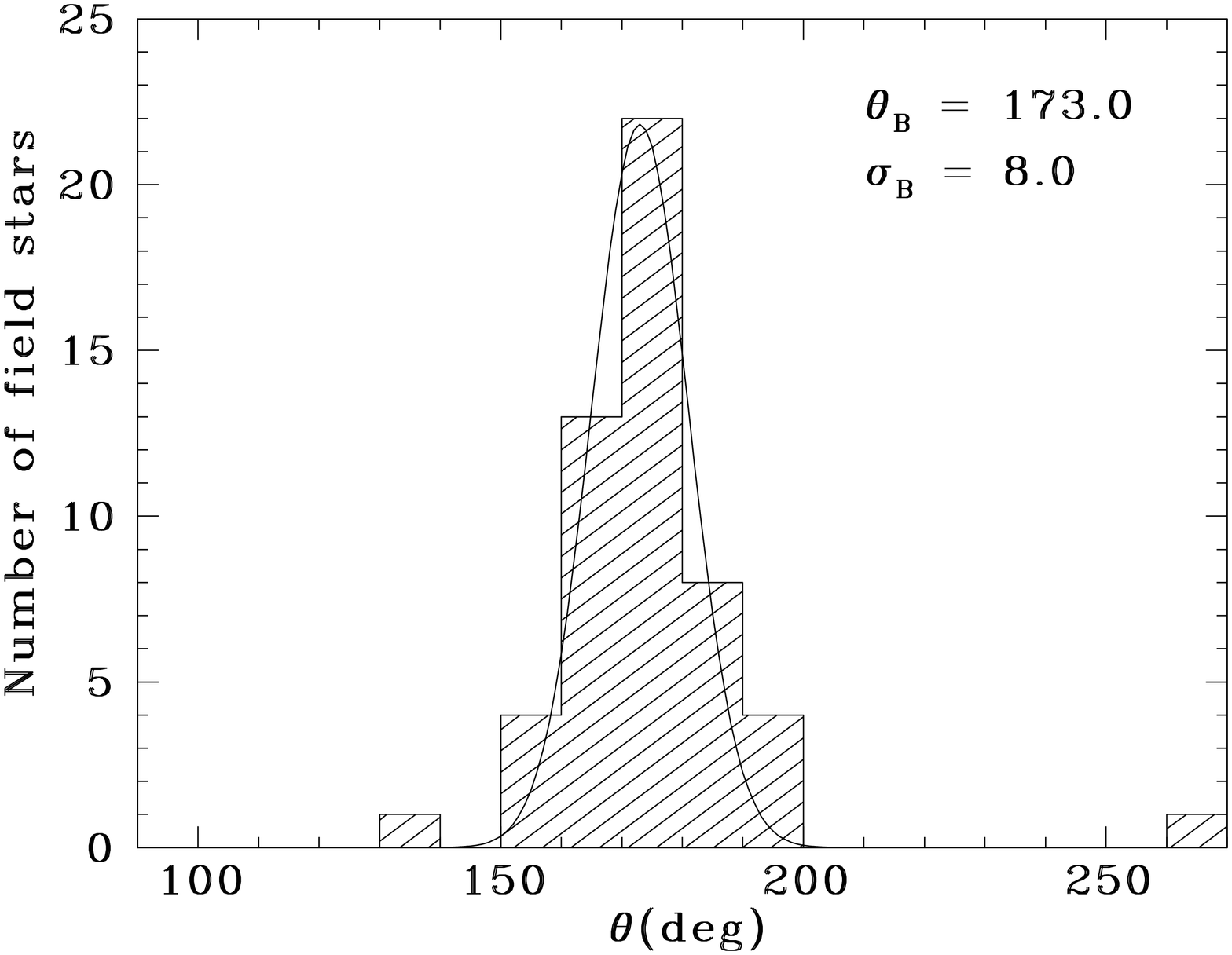}
\caption*{Polarimetry of Field~01. Upper panel: The observed polarization vectors overplotted on a DSS2 red image. The coordinates are B1950. Lower panel: The histogram of the corresponding position angles of the polarization, $\theta$. In the upper right corner, it is shown the average and the dispersion of the interstellar magnetic field used in the analysis. A Gaussian curve using these values is also depicted. }
\end{figure}

\clearpage

\begin{figure}
\includegraphics[width=13cm,angle=0]{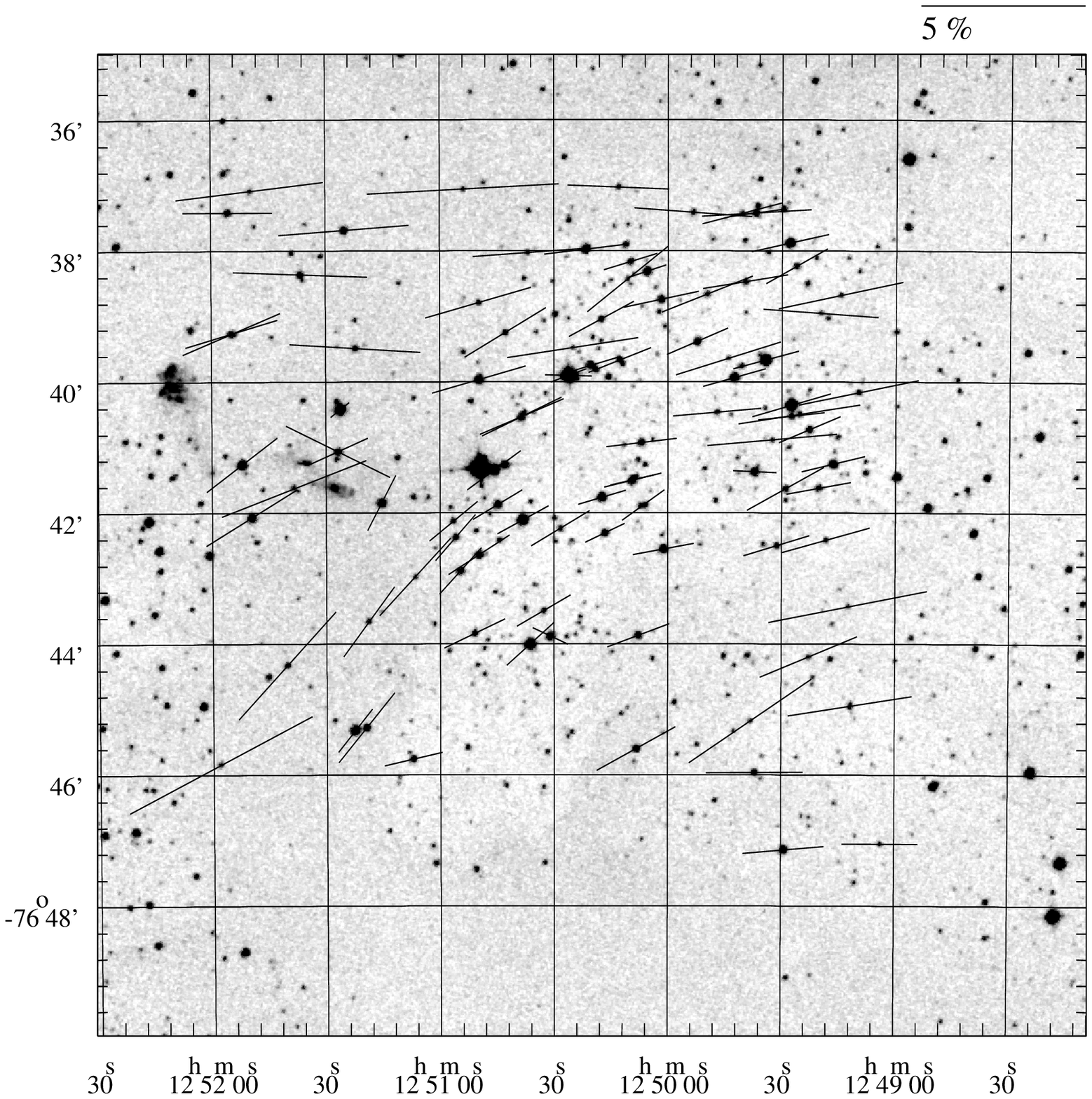}

\vspace{-5cm}
\includegraphics[width=6cm,angle=-90.]{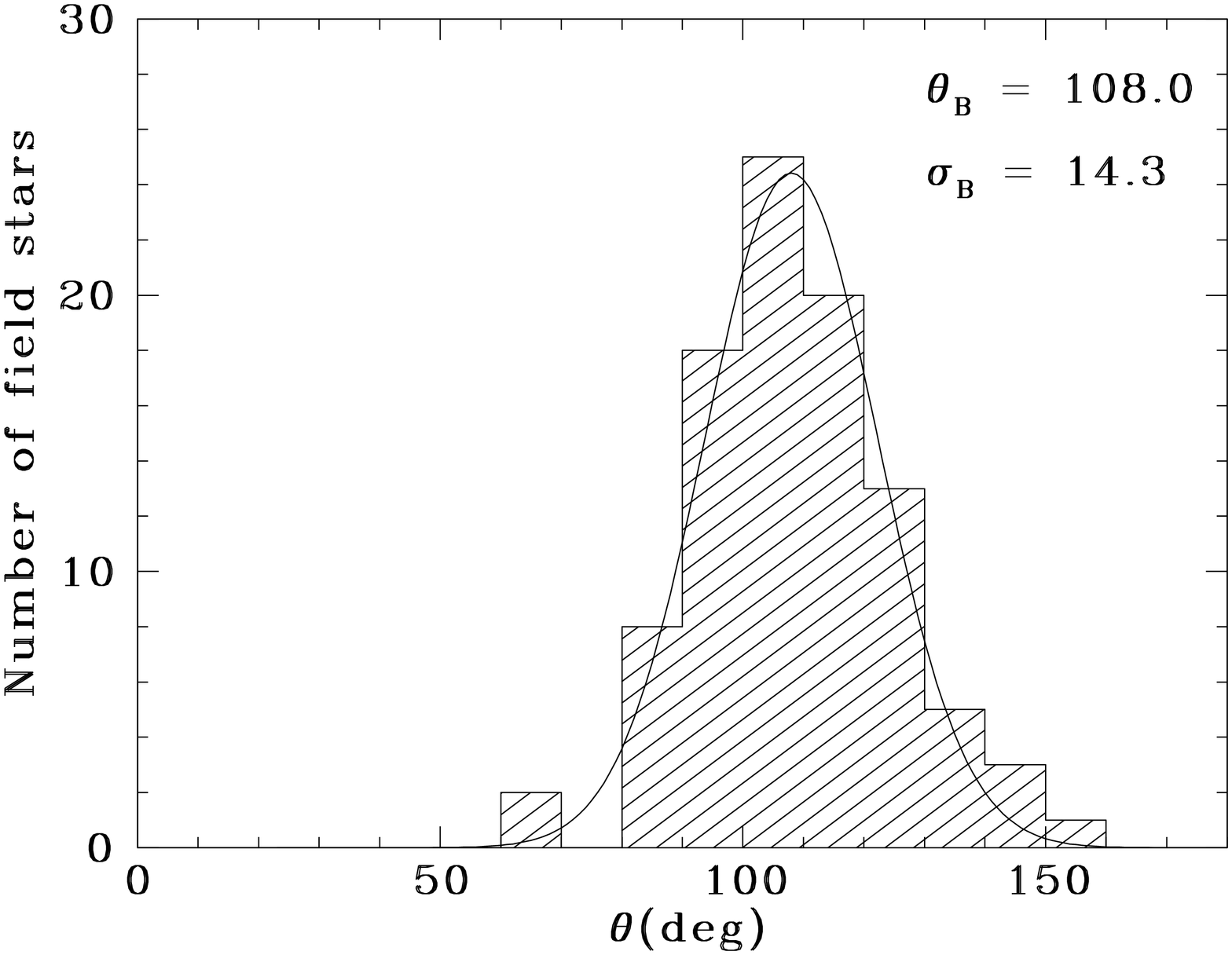}
\caption*{The same of Figure \ref{fig:hh139} for Field~03.}
\label{fig:field03}
\end{figure}

\clearpage

\begin{figure}
\includegraphics[width=13cm,angle=0]{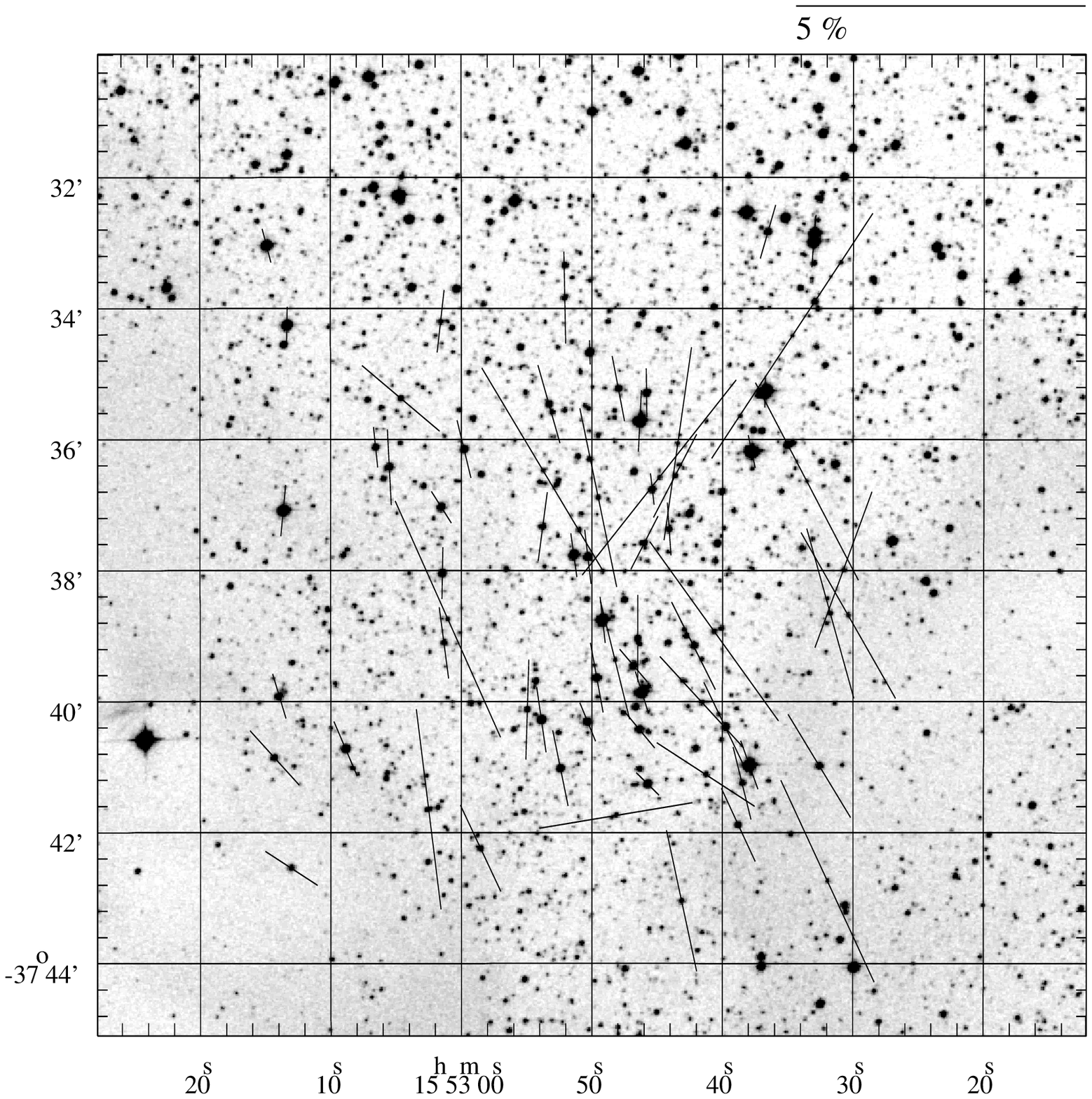}

\vspace{-5cm}
\includegraphics[width=6cm,angle=-90.]{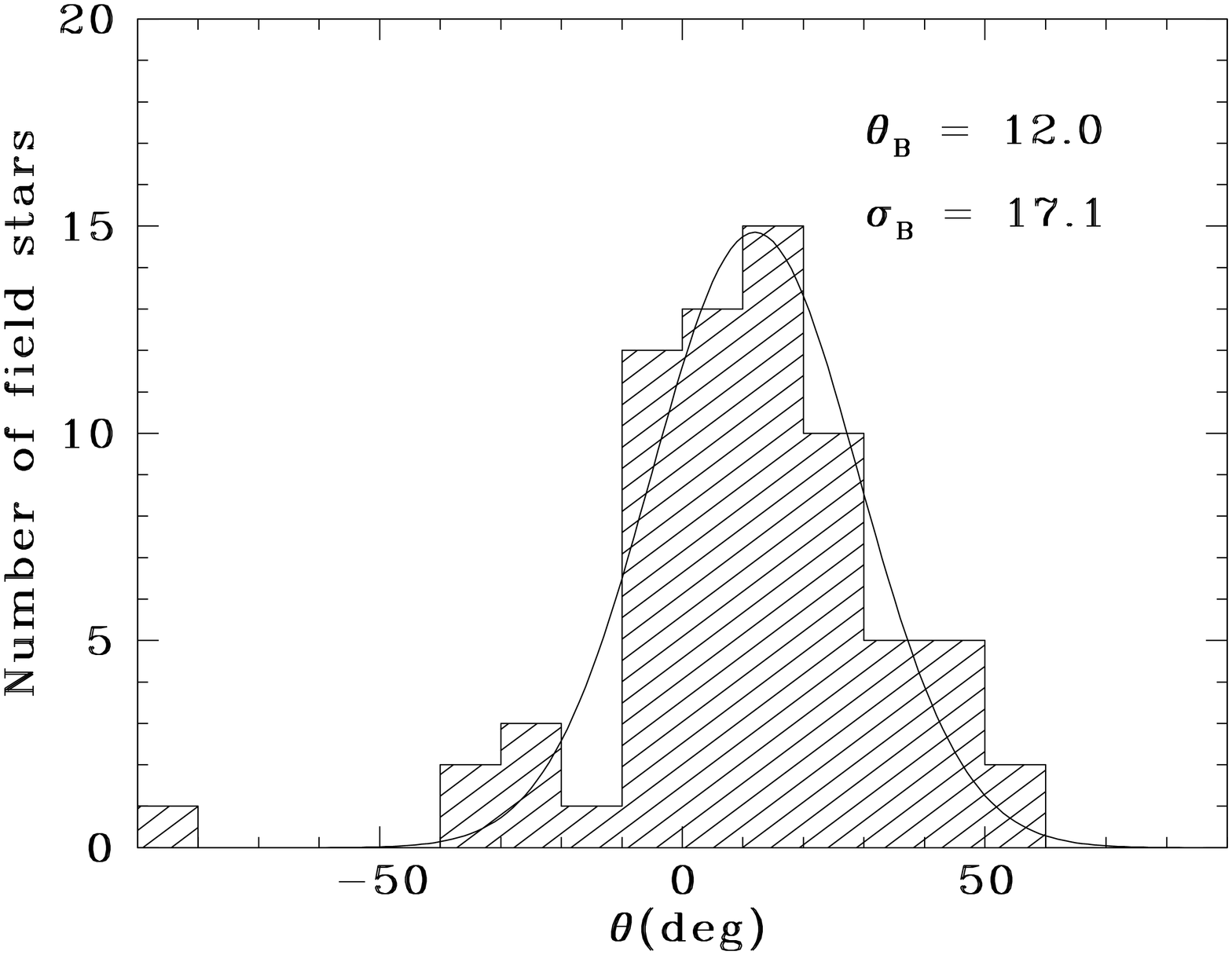}
\caption*{The same of Figure \ref{fig:hh139} for Field~04.}
\label{fig:field04}
\end{figure}

\clearpage

\begin{figure}
\includegraphics[width=13cm,angle=0]{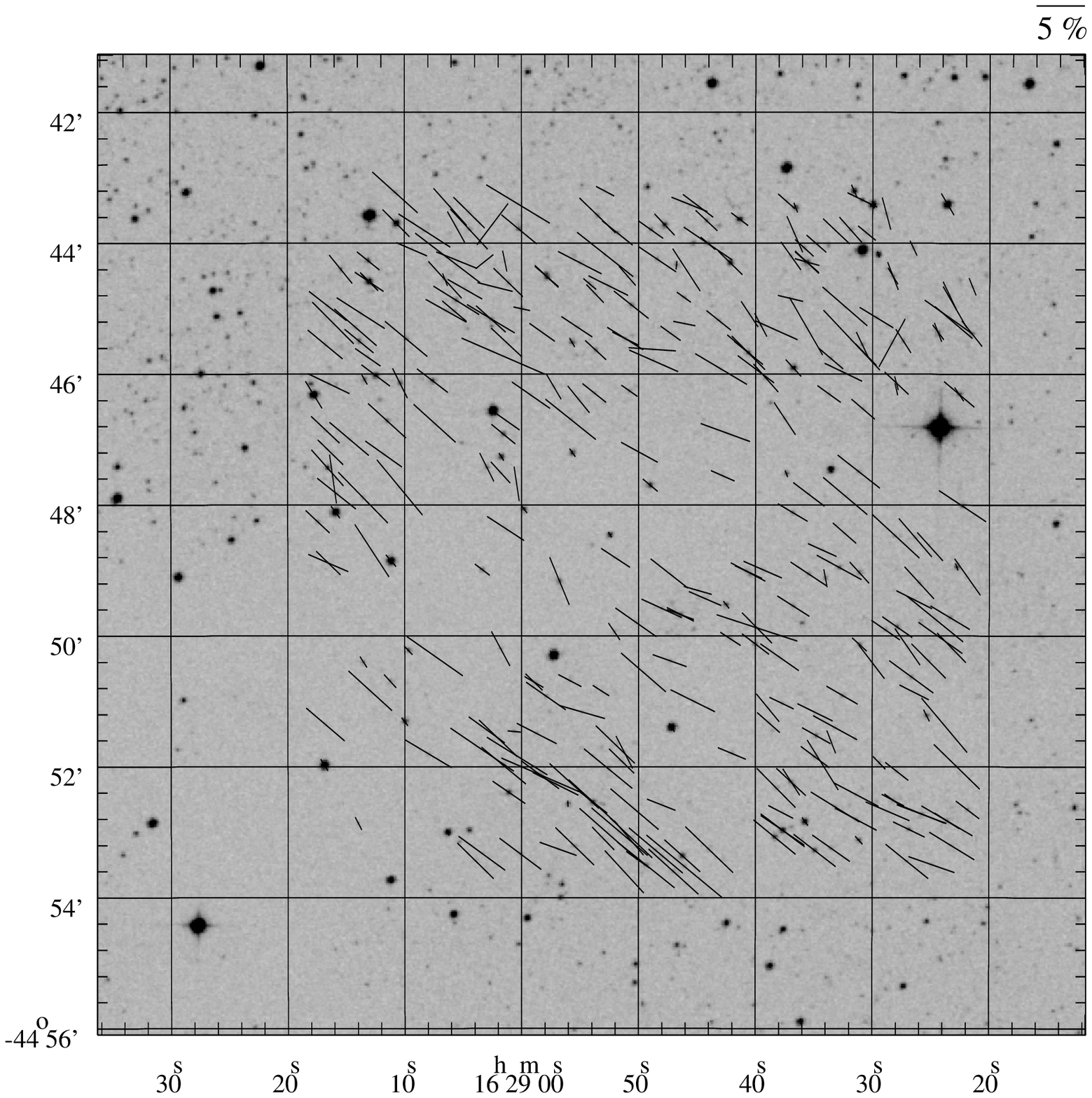}

\vspace{-5cm}
\includegraphics[width=6cm,angle=-90.]{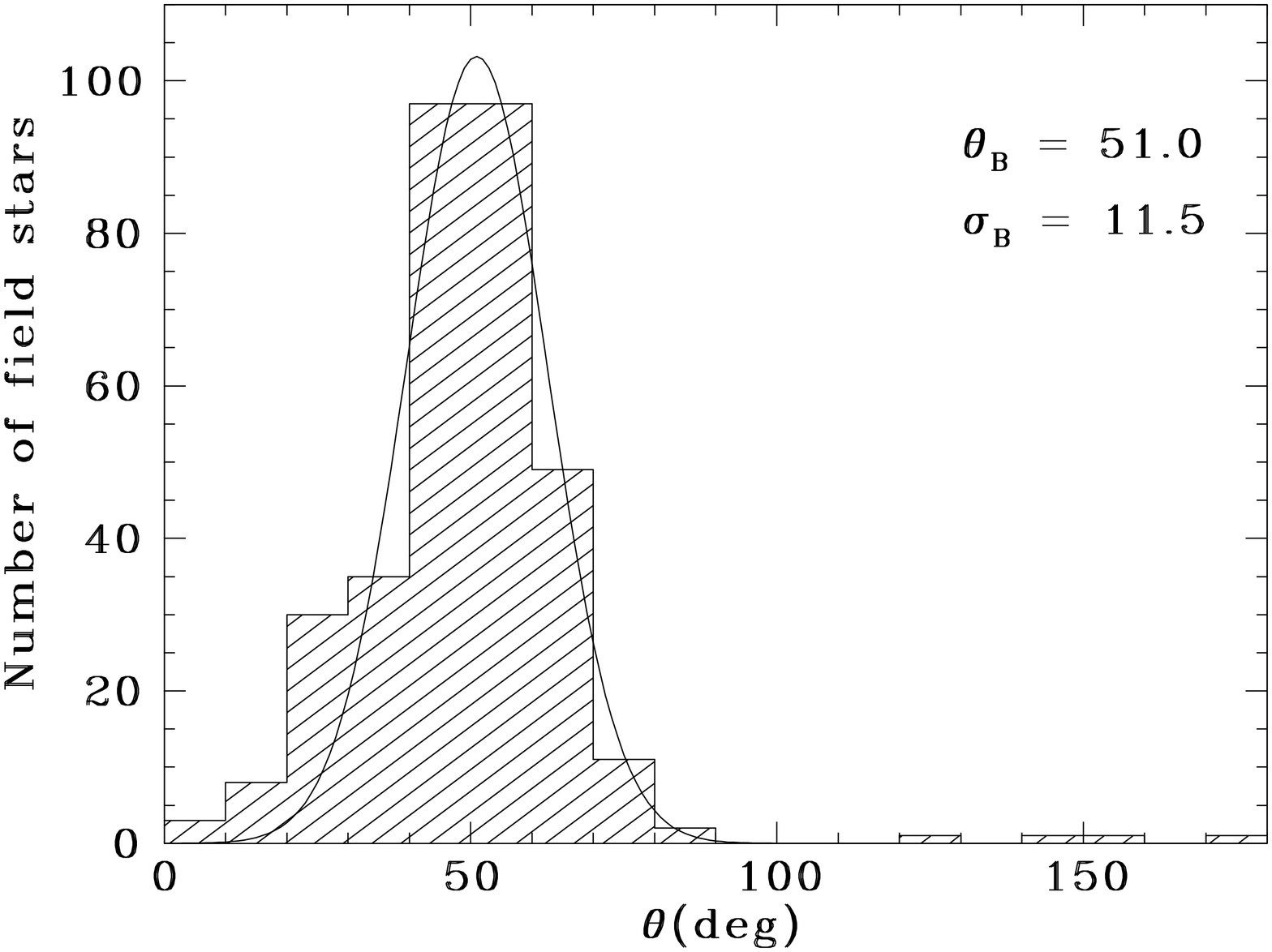}
\caption*{The same of Figure \ref{fig:hh139} for Field~05.}
\label{fig:field05}
\end{figure}

\clearpage

\begin{figure}
\includegraphics[width=13cm,angle=0]{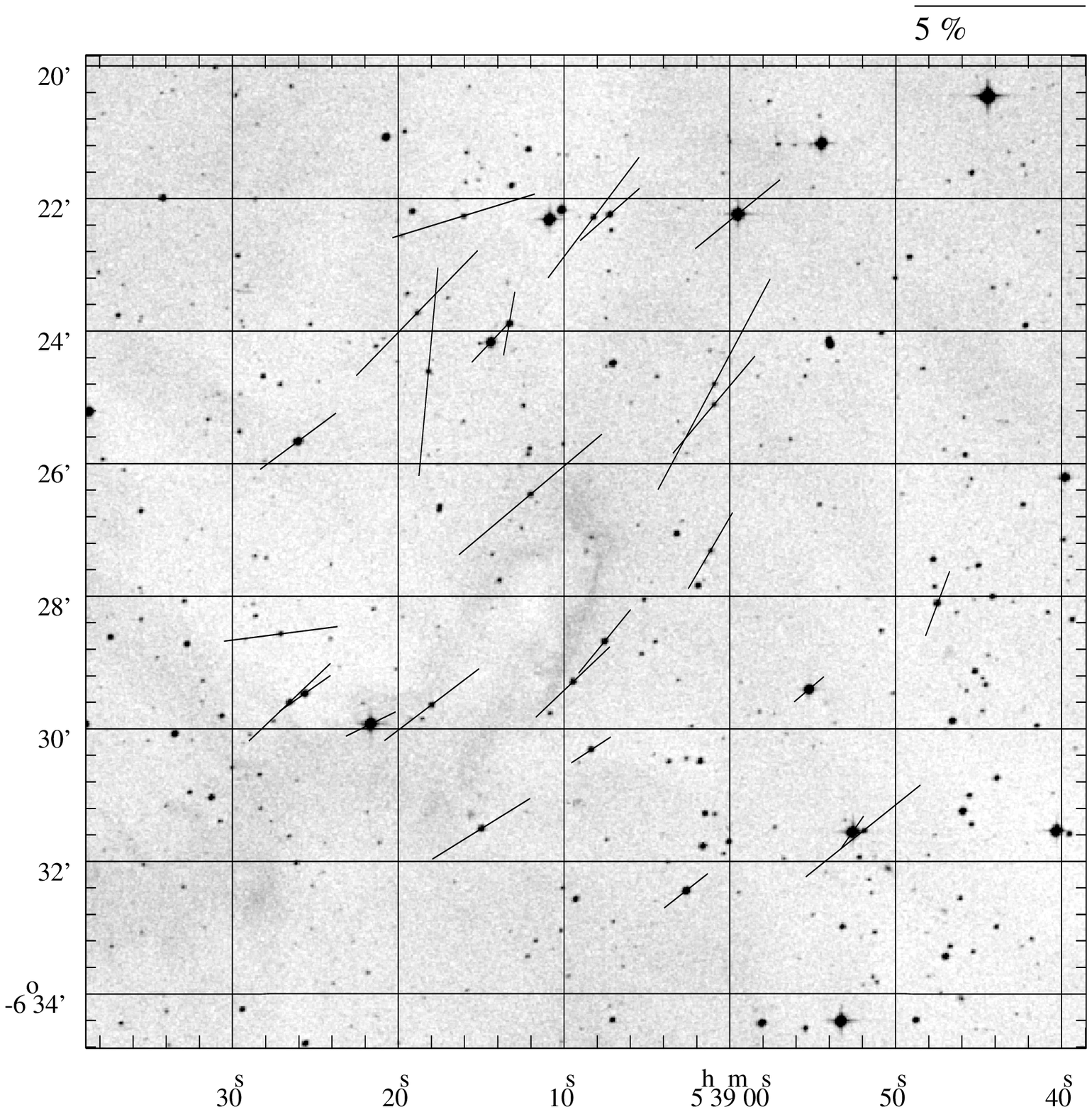}

\vspace{-5cm}
\includegraphics[width=6cm,angle=-90.]{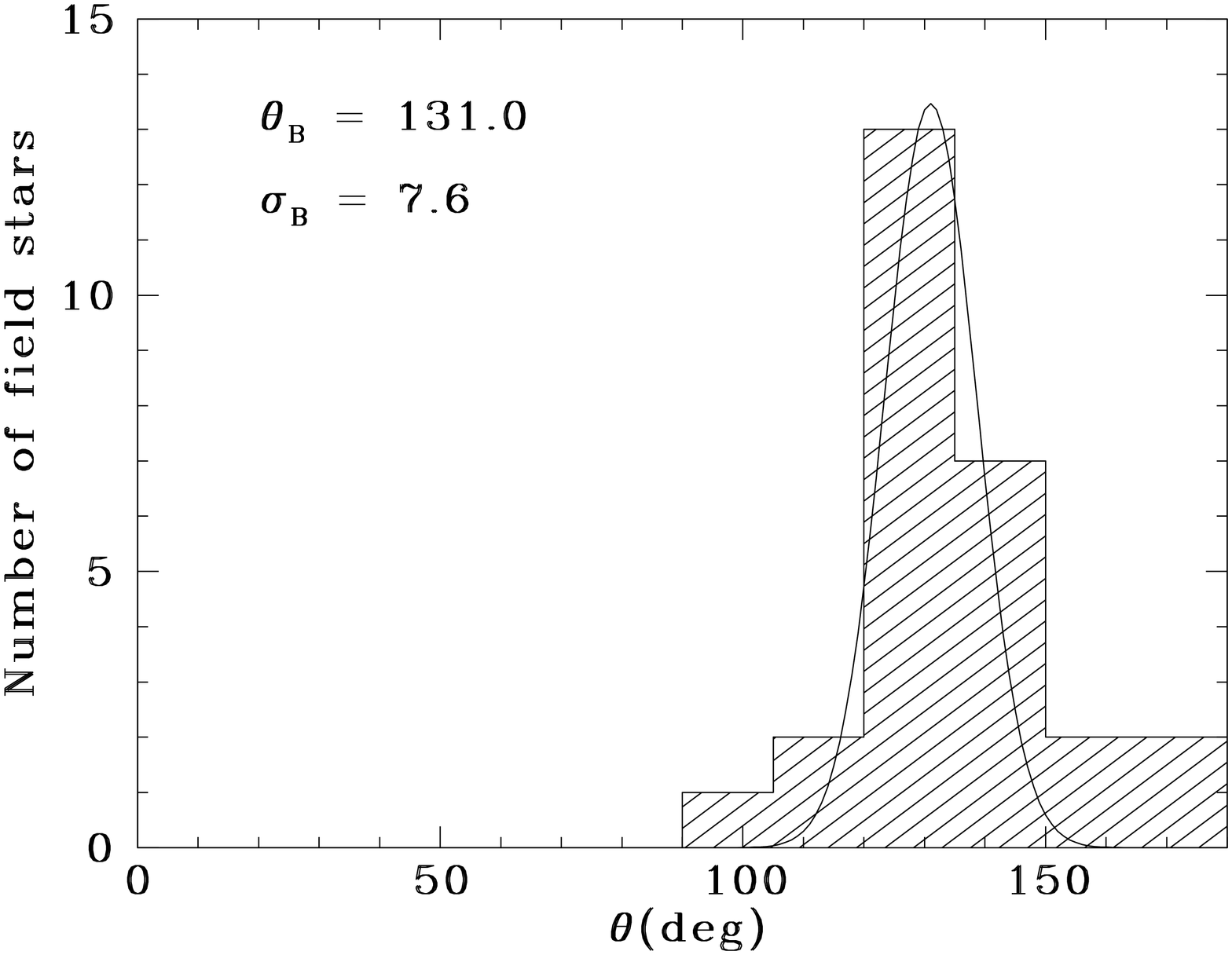}
\caption*{The same of Figure \ref{fig:hh139} for Field~06.}
\label{fig:field06}
\end{figure}

\clearpage

\begin{figure}
\includegraphics[width=13cm,angle=0]{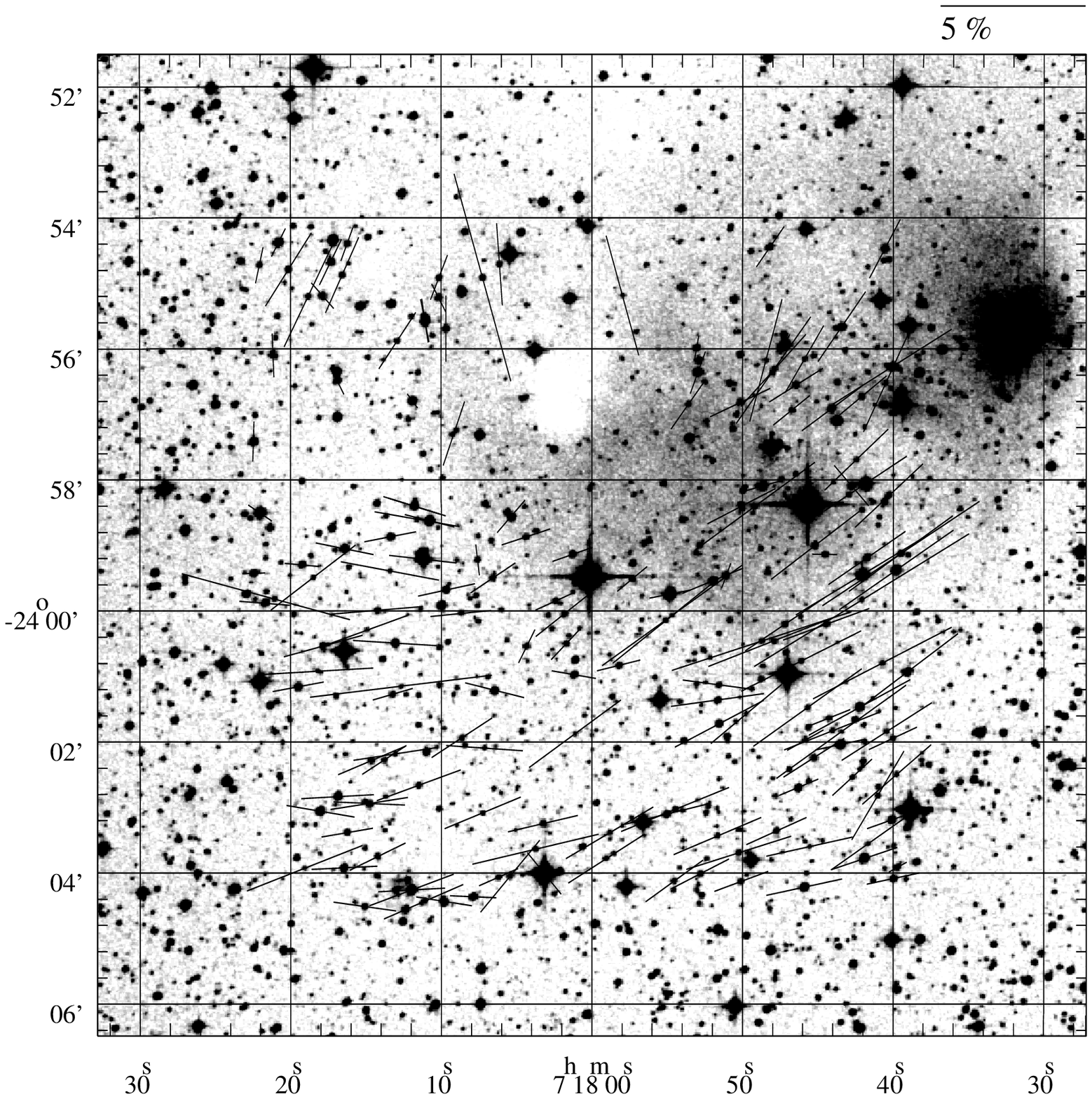}

\vspace{-5cm}
\includegraphics[width=6cm,angle=-90.]{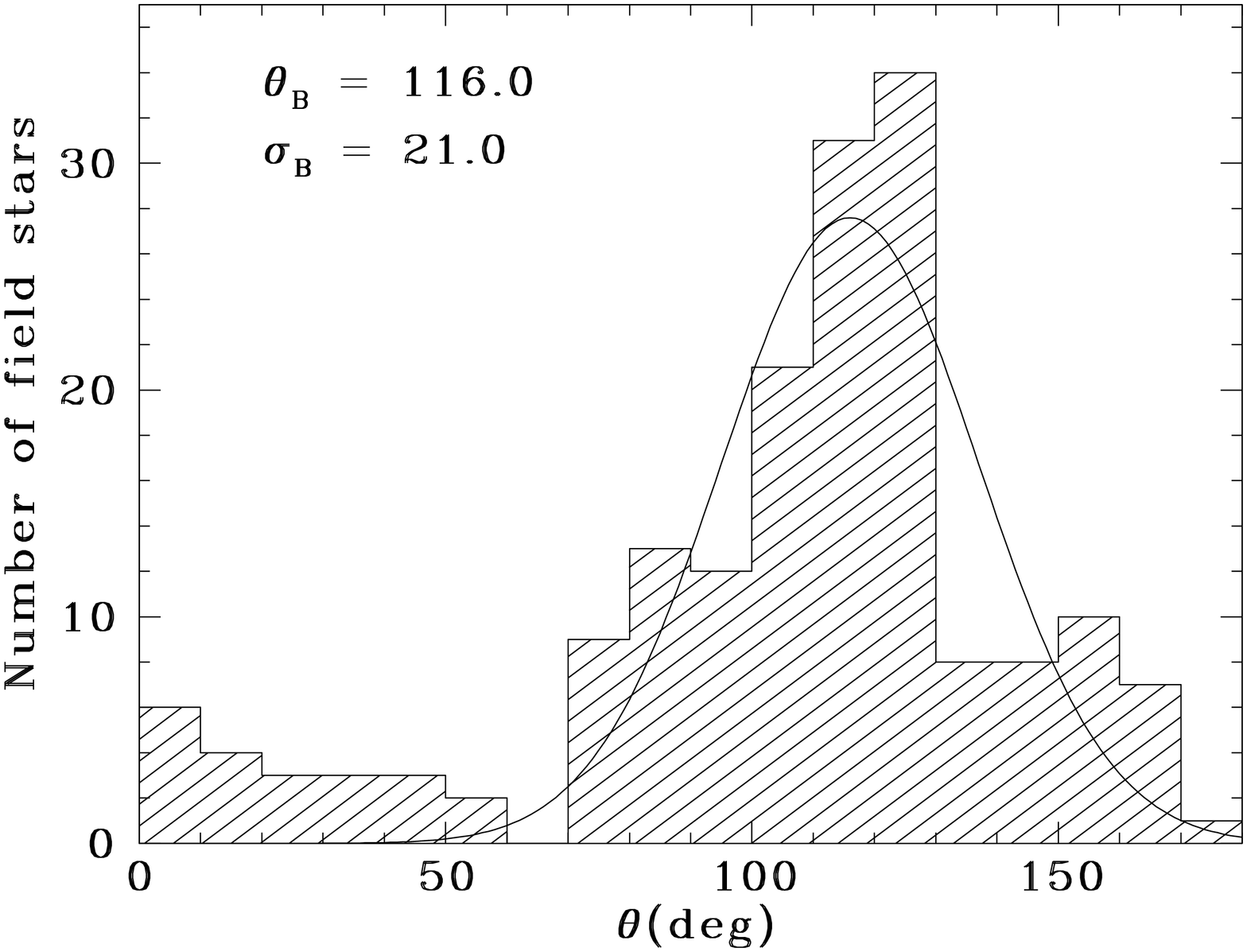}
\caption*{The same of Figure \ref{fig:hh139} for Field~07.}
\label{fig:field07}
\end{figure}

\clearpage

\begin{figure}
\includegraphics[width=13cm,angle=0]{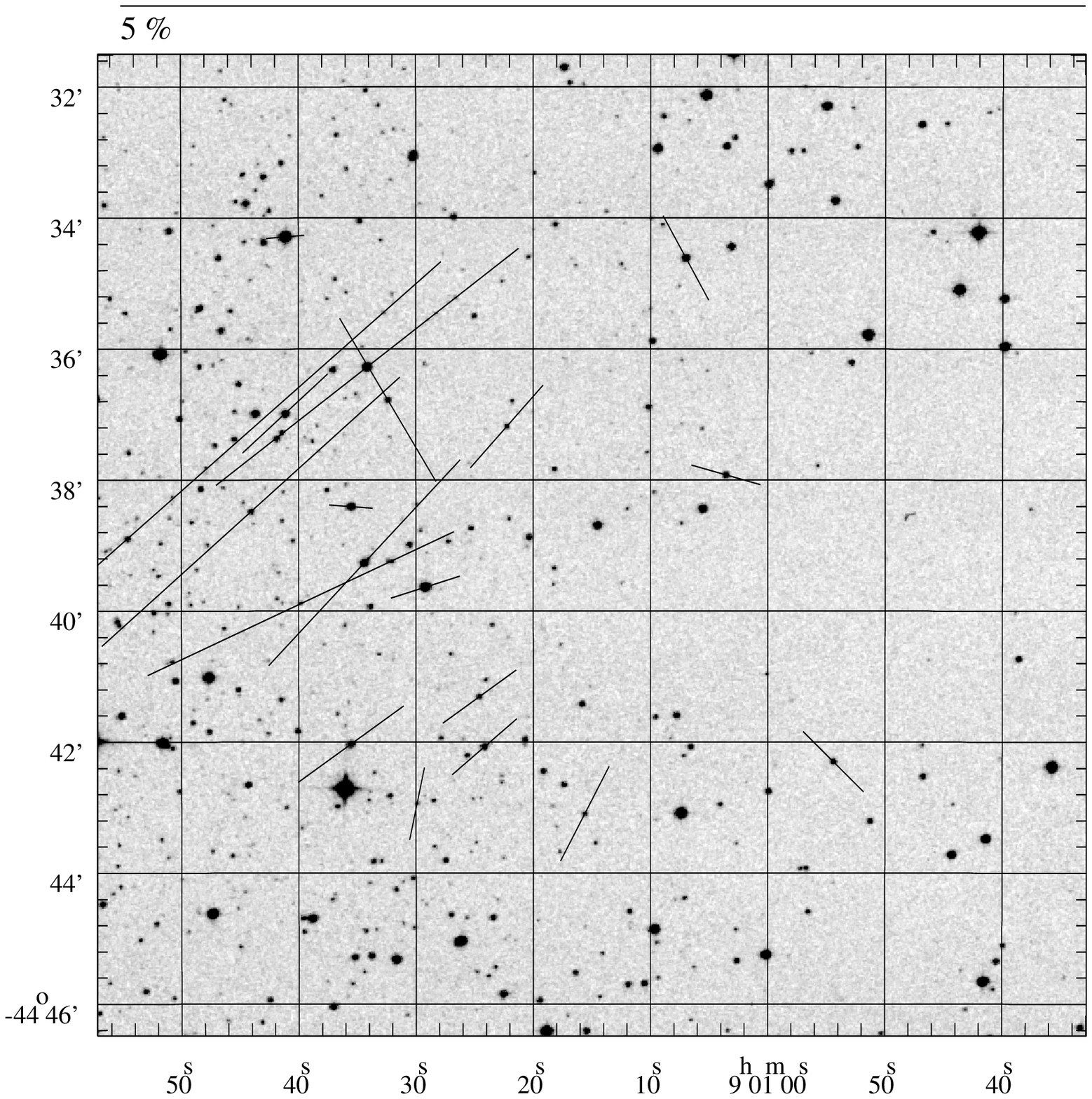}

\vspace{-5cm}
\includegraphics[width=6cm,angle=-90.]{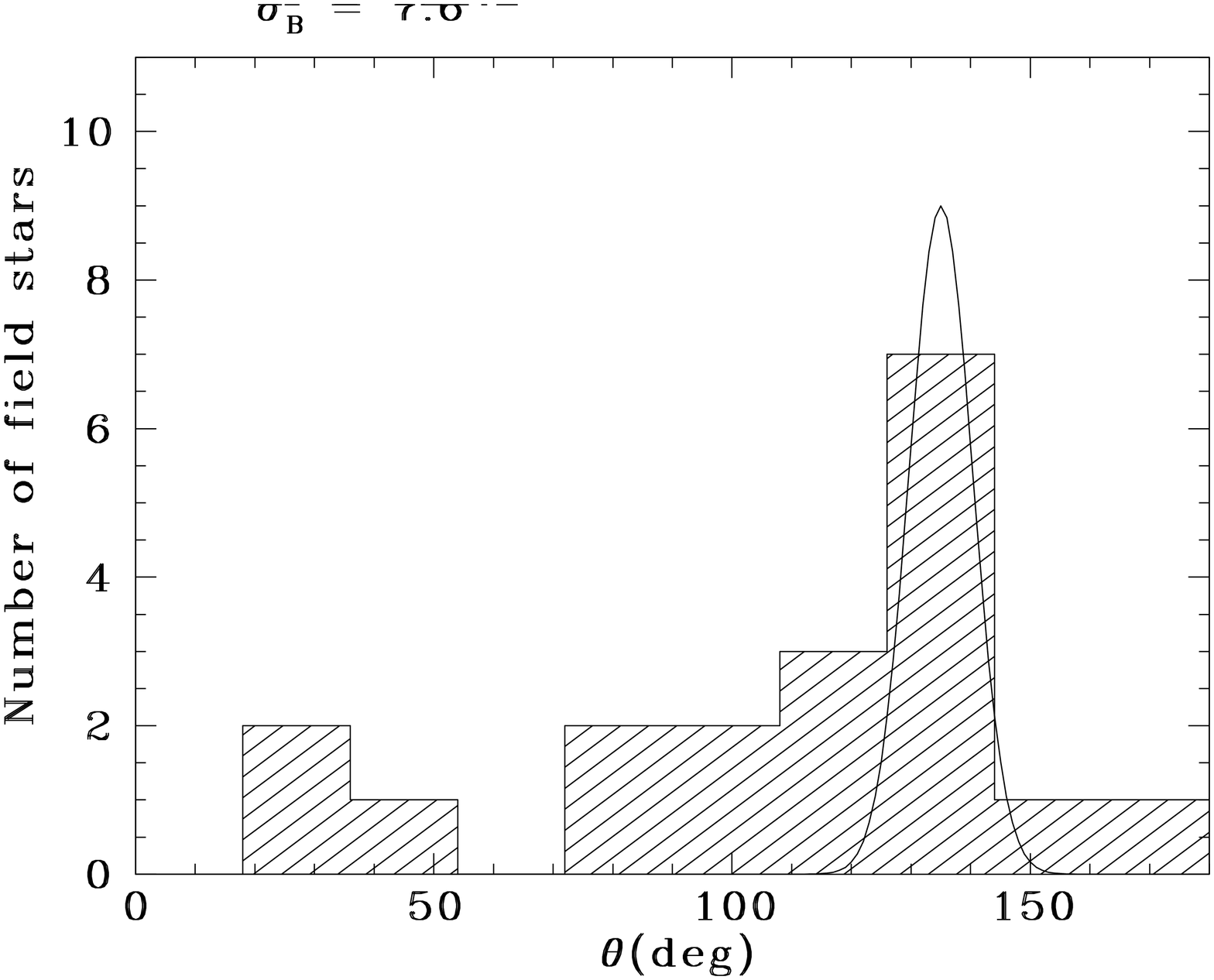}
\caption*{The same of Figure \ref{fig:hh139} for Field~08.}
\label{fig:field08}
\end{figure}

\clearpage

\begin{figure}
\includegraphics[width=13cm,angle=0]{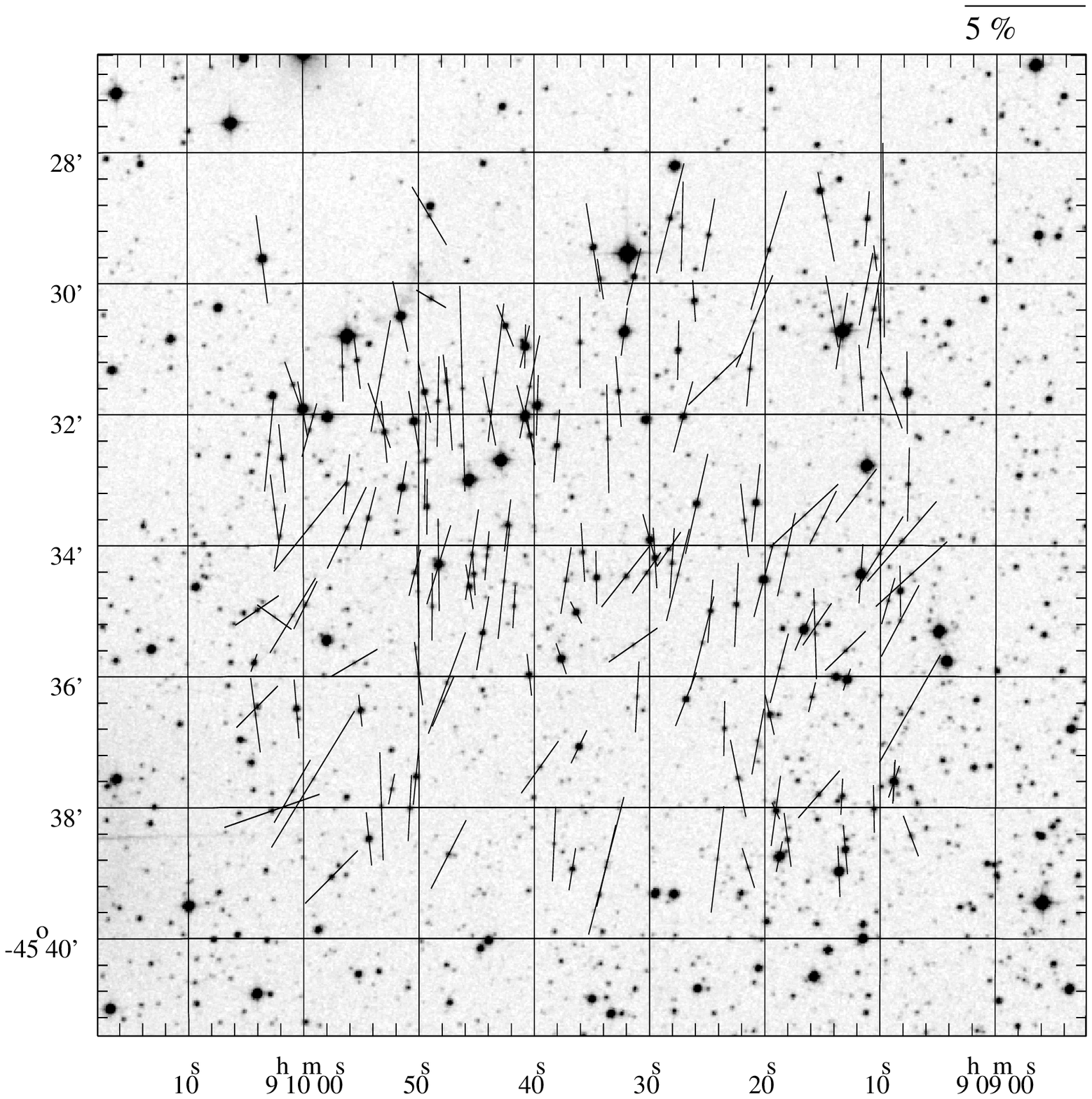}

\vspace{-5cm}
\includegraphics[width=6cm,angle=-90.]{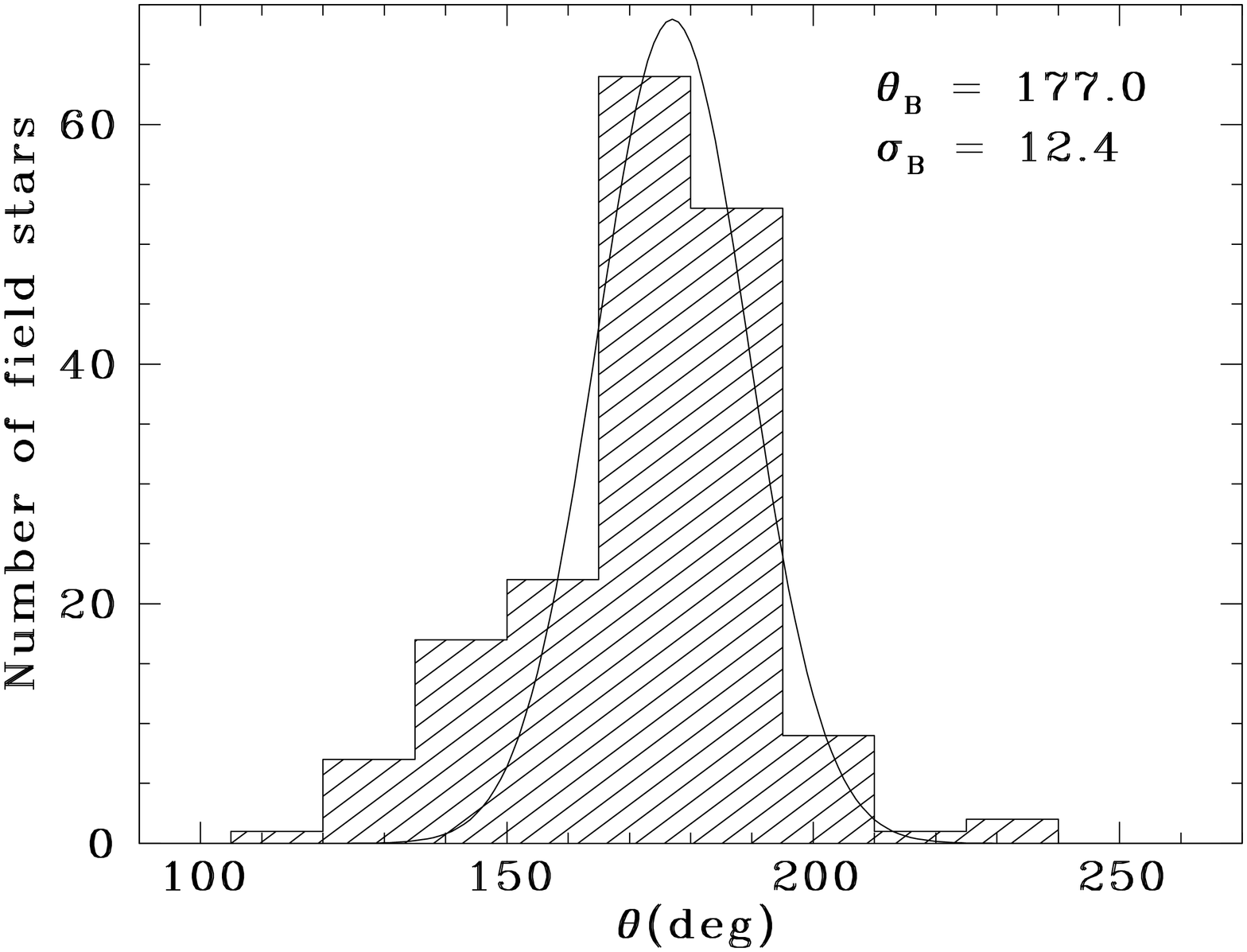}
\caption*{The same of Figure \ref{fig:hh139} for Field~09.}
\label{fig:field09}
\end{figure}

\clearpage

\begin{figure}
\includegraphics[width=13cm,angle=0]{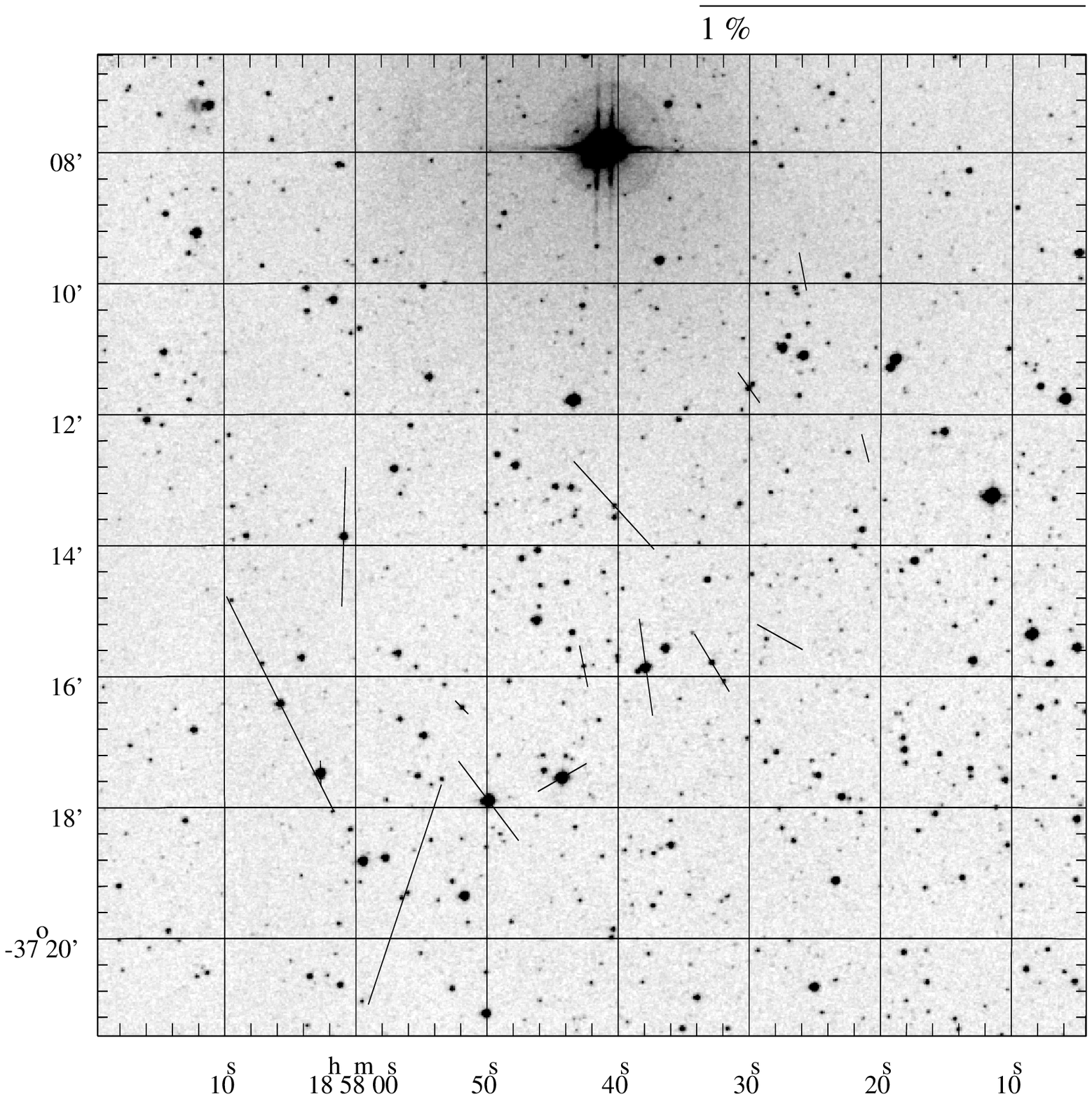}

\vspace{-5cm}
\includegraphics[width=6cm,angle=-90.]{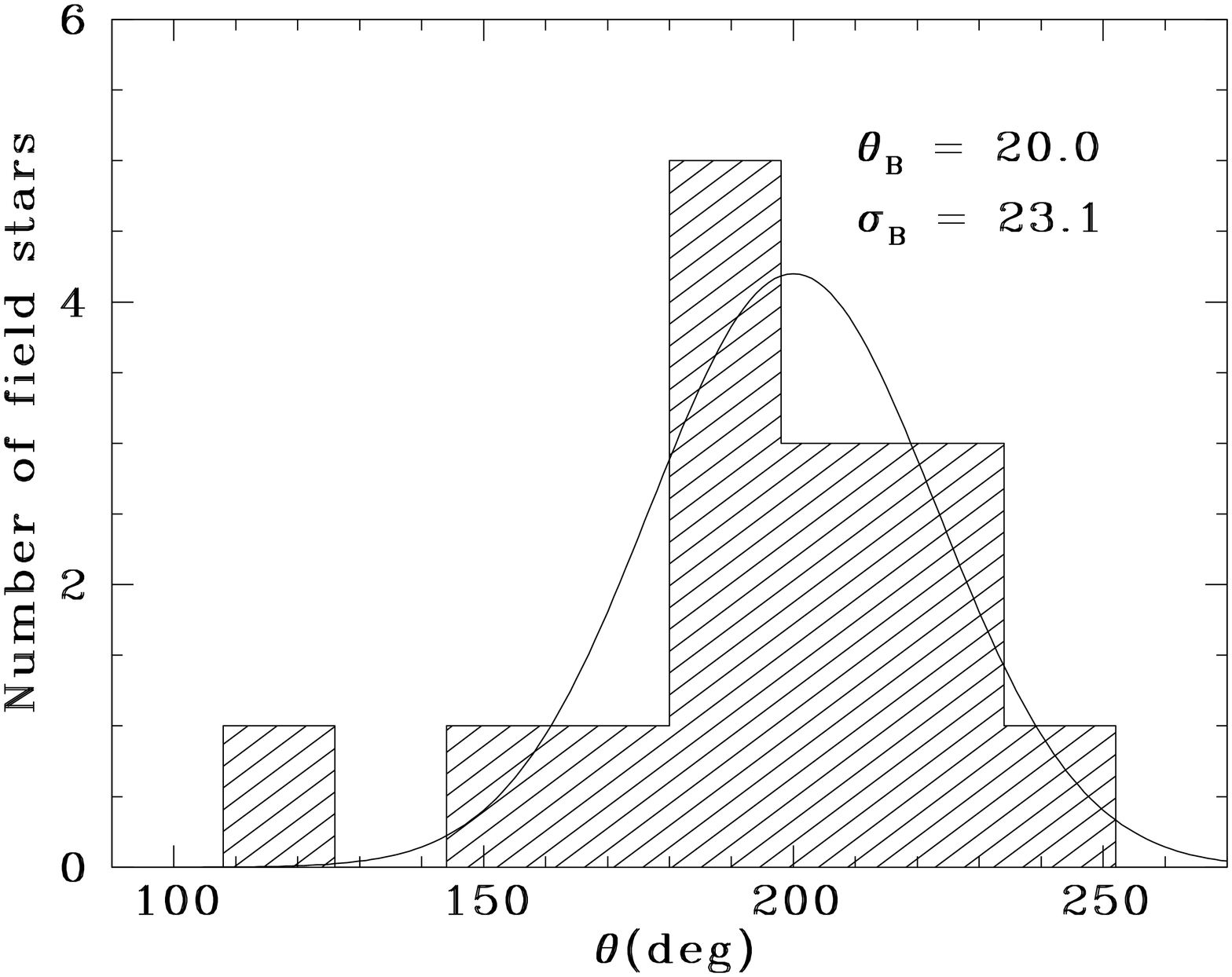}
\caption*{The same of Figure \ref{fig:hh139} for Field~10.}
\label{fig:field10}
\end{figure}

\clearpage

\begin{figure}
\includegraphics[width=13cm,angle=0]{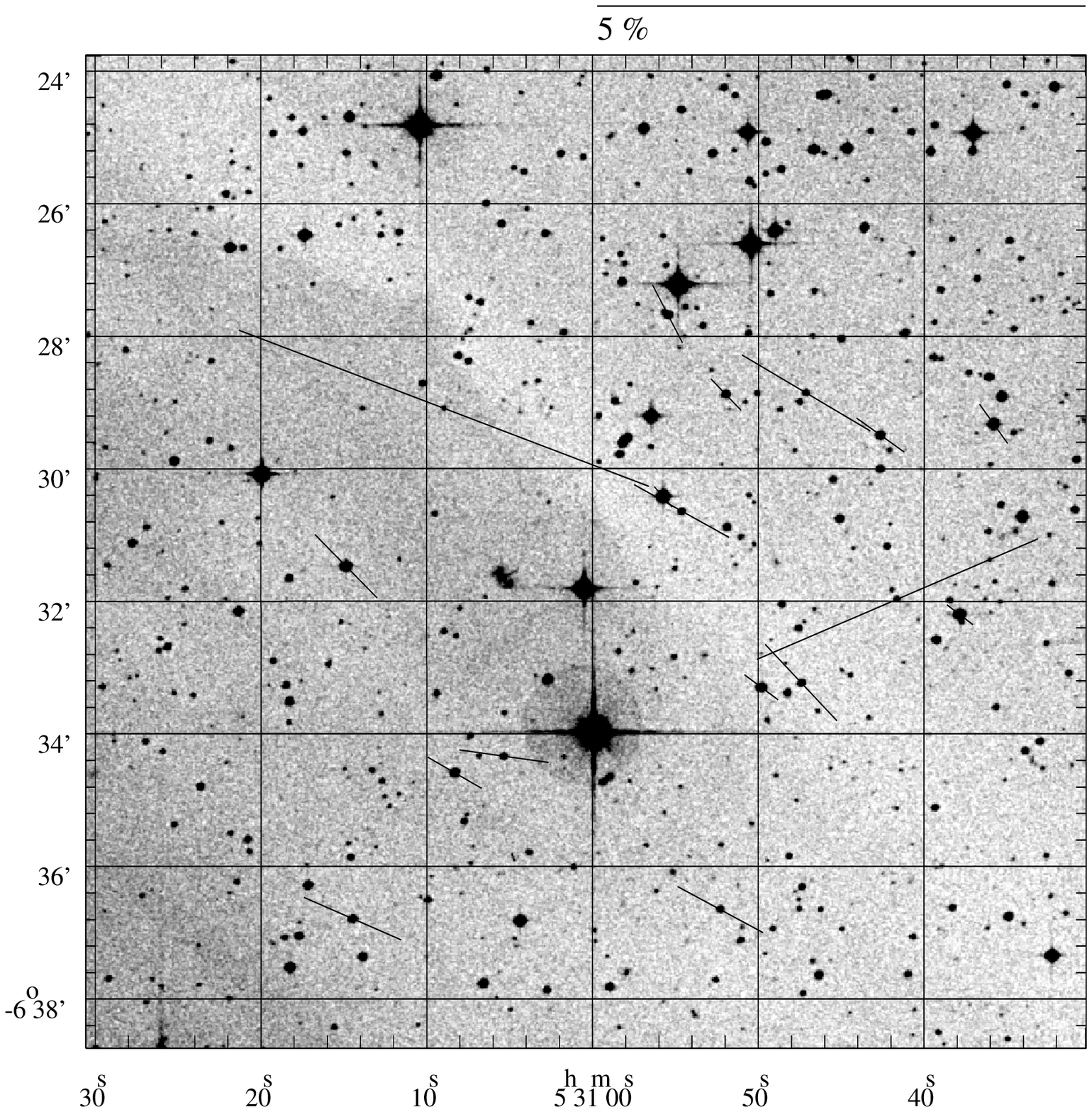}

\vspace{-5cm}
\includegraphics[width=6cm,angle=-90.]{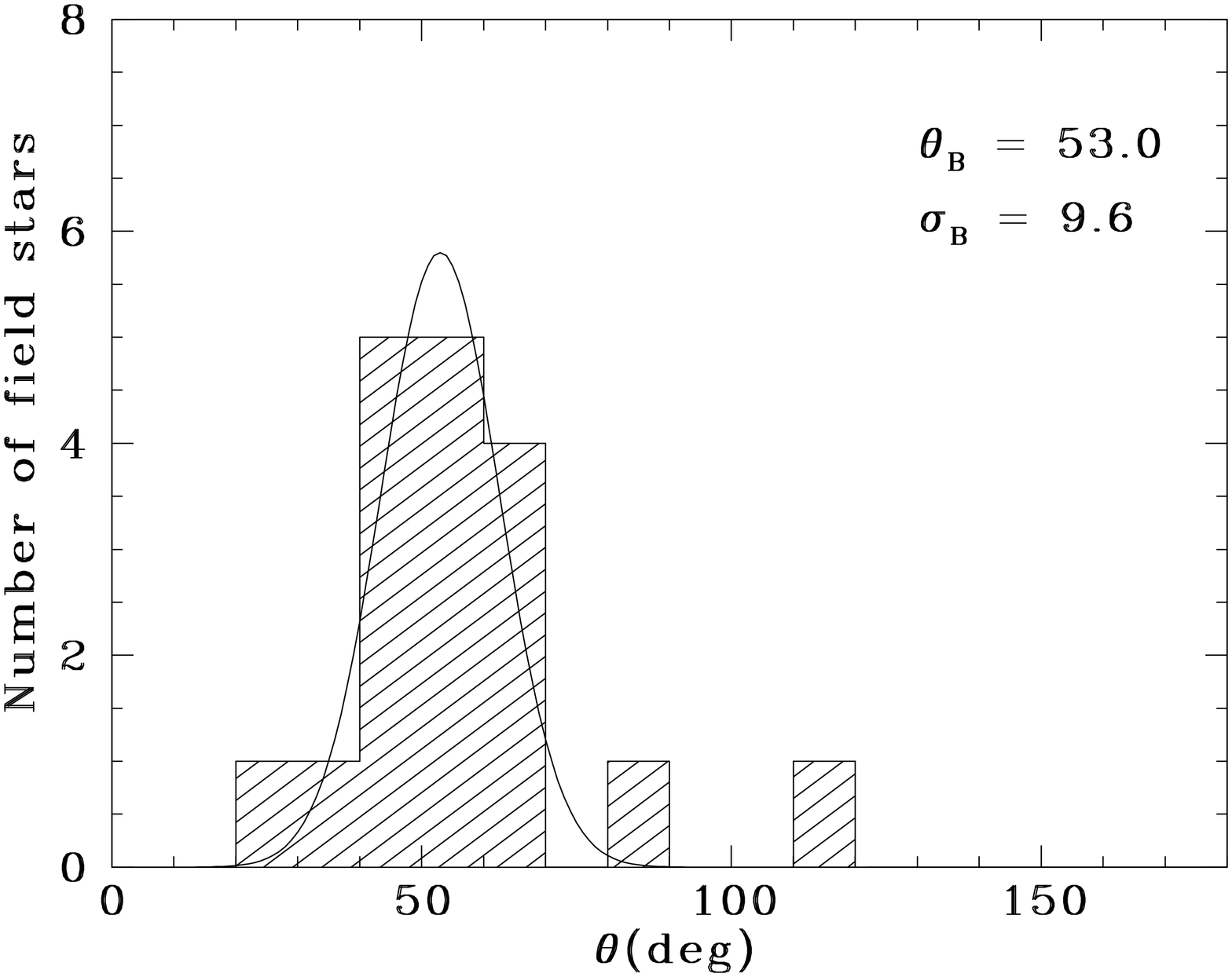}
\caption*{The same of Figure \ref{fig:hh139} for Field~11.}
\label{fig:field11}
\end{figure}

\clearpage

\begin{figure}
\includegraphics[width=13cm,angle=0]{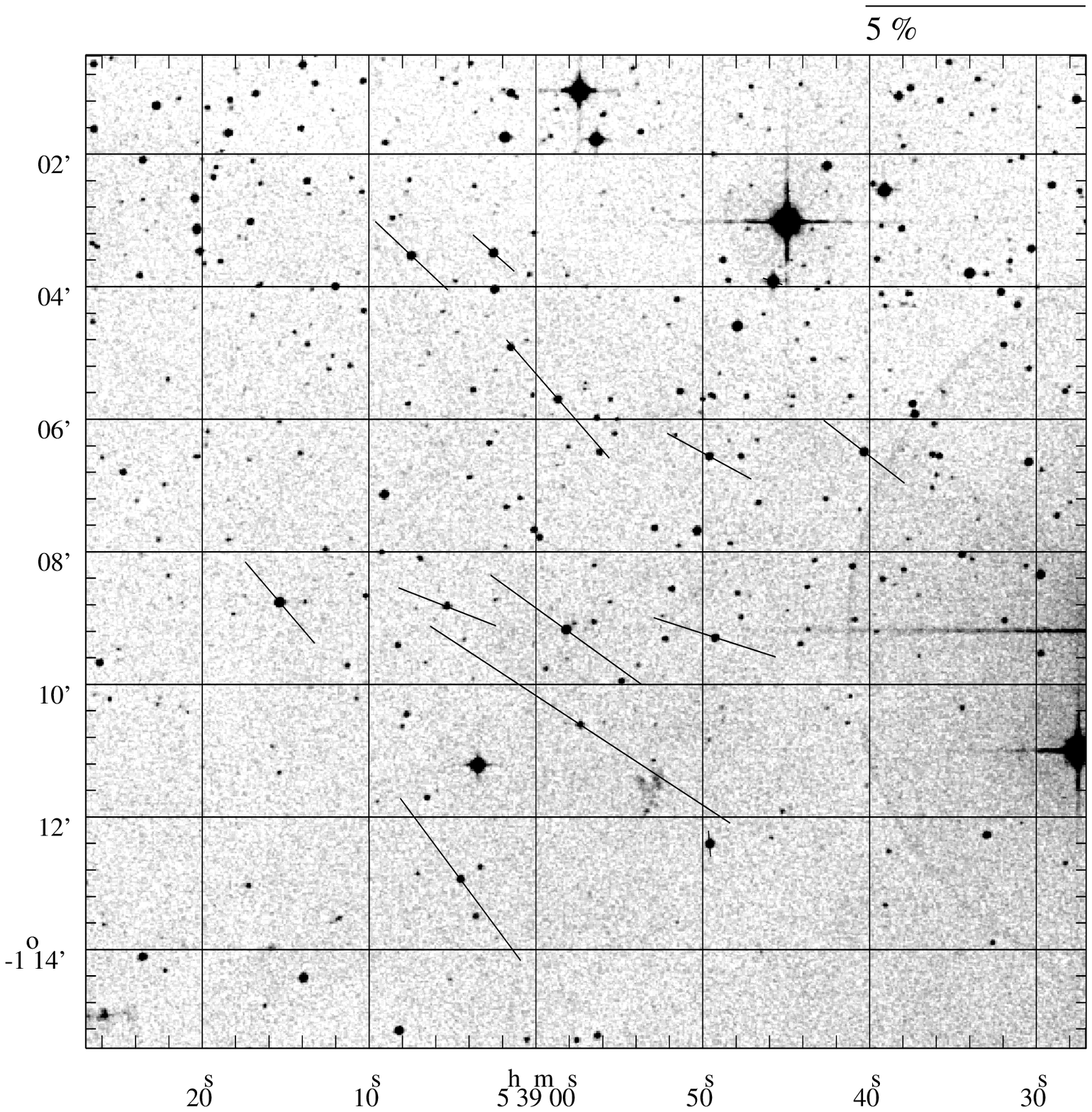}

\vspace{-5cm}
\includegraphics[width=6cm,angle=-90.]{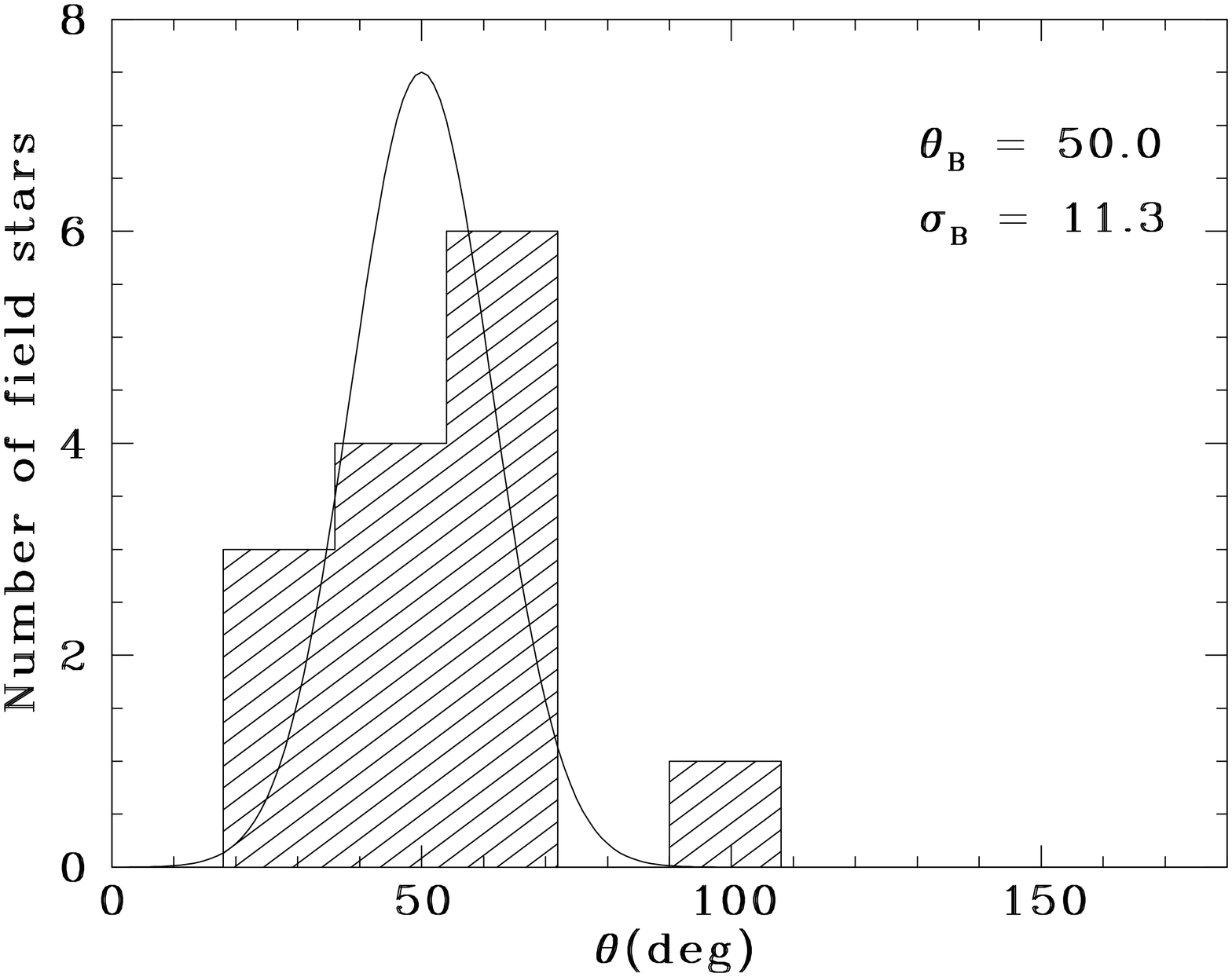}
\caption*{The same of Figure \ref{fig:hh139} for Field~12A.}
\label{fig:field12a}
\end{figure}

\clearpage

\begin{figure}
\includegraphics[width=13cm,angle=0]{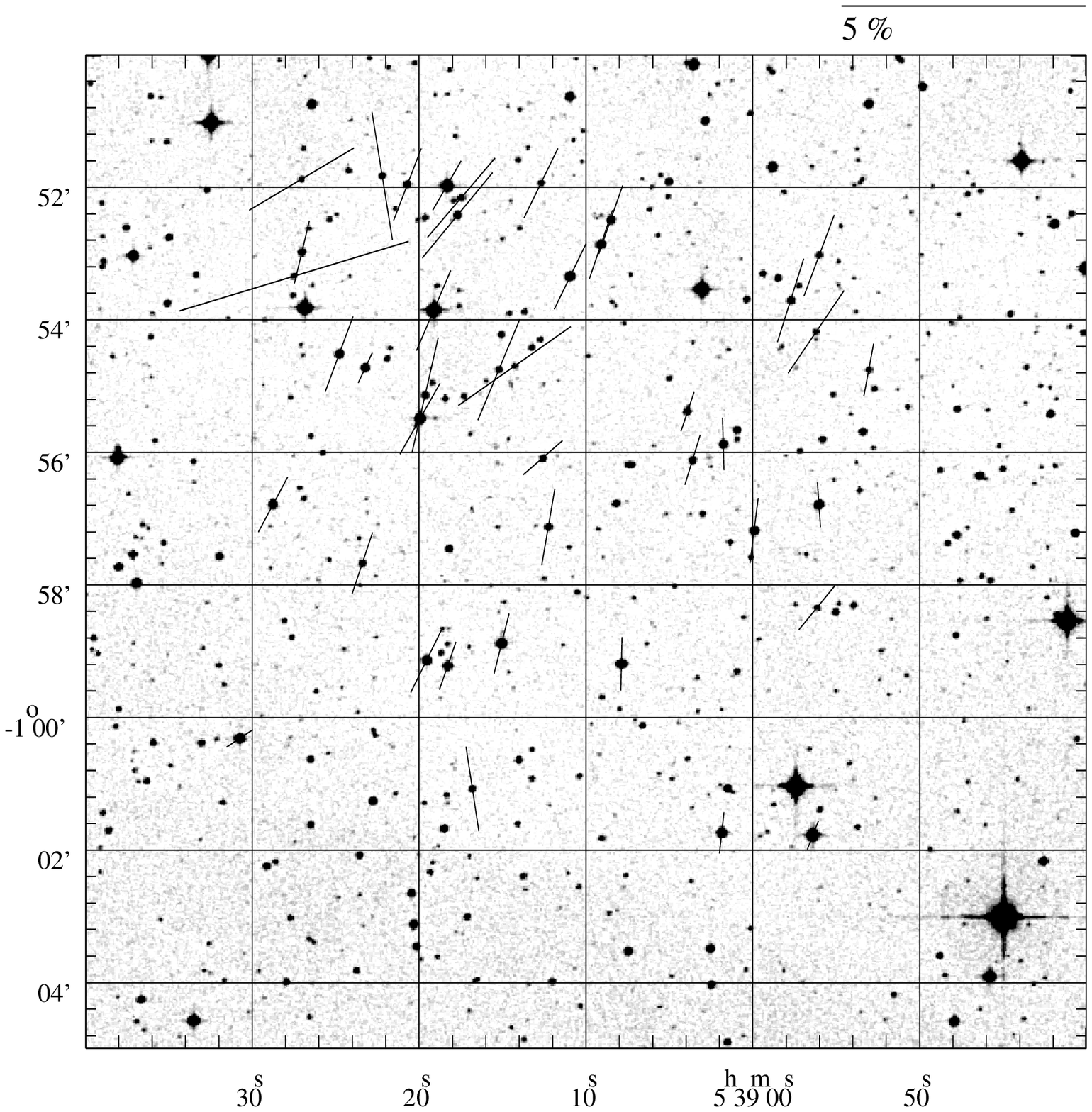}

\vspace{-5cm}
\includegraphics[width=6cm,angle=-90.]{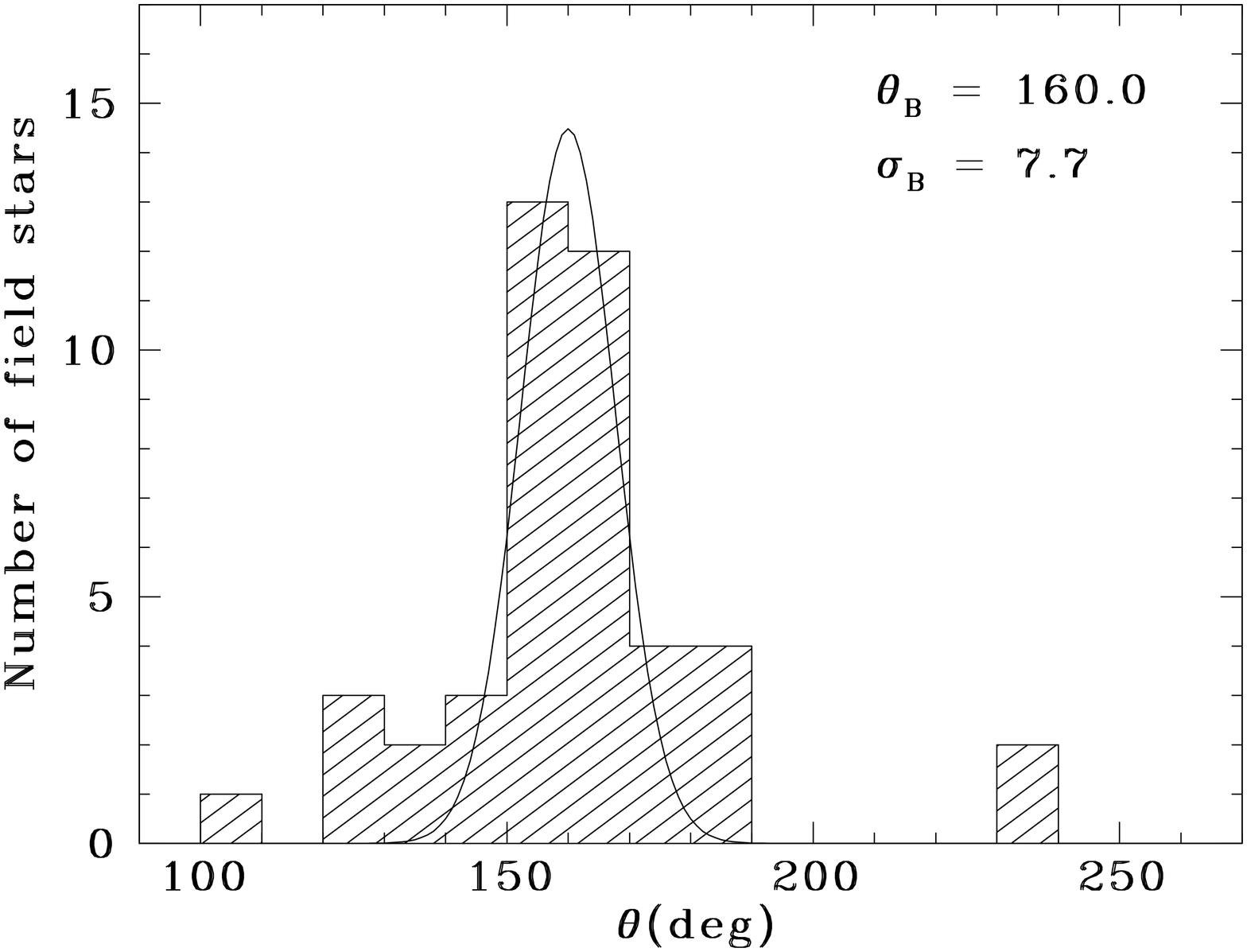}
\caption*{The same of Figure \ref{fig:hh139} for Field~12B.}
\label{fig:field12b}
\end{figure}

\clearpage

\begin{figure}
\includegraphics[width=13cm,angle=0]{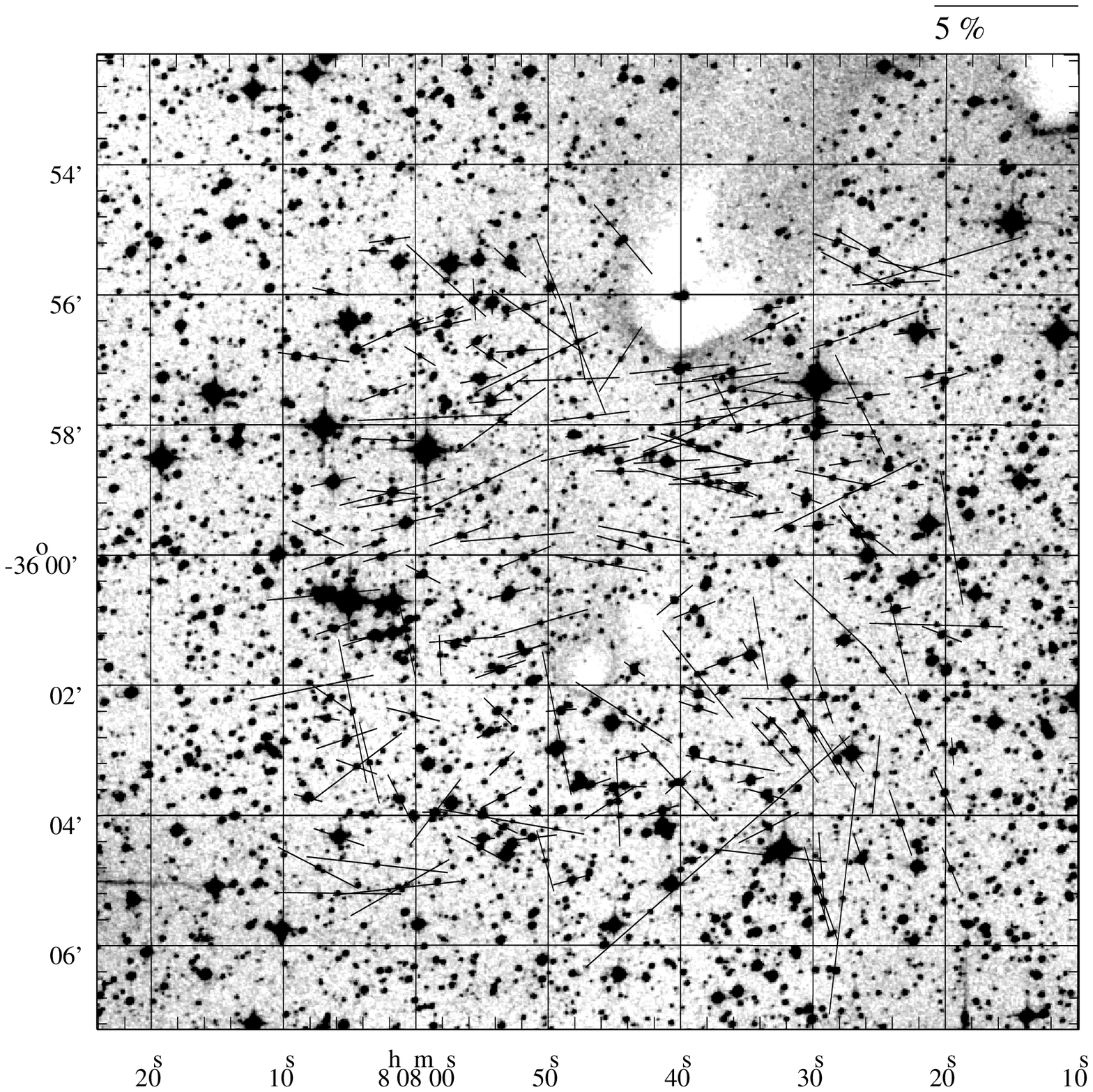}

\vspace{-5cm}
\includegraphics[width=6cm,angle=-90.]{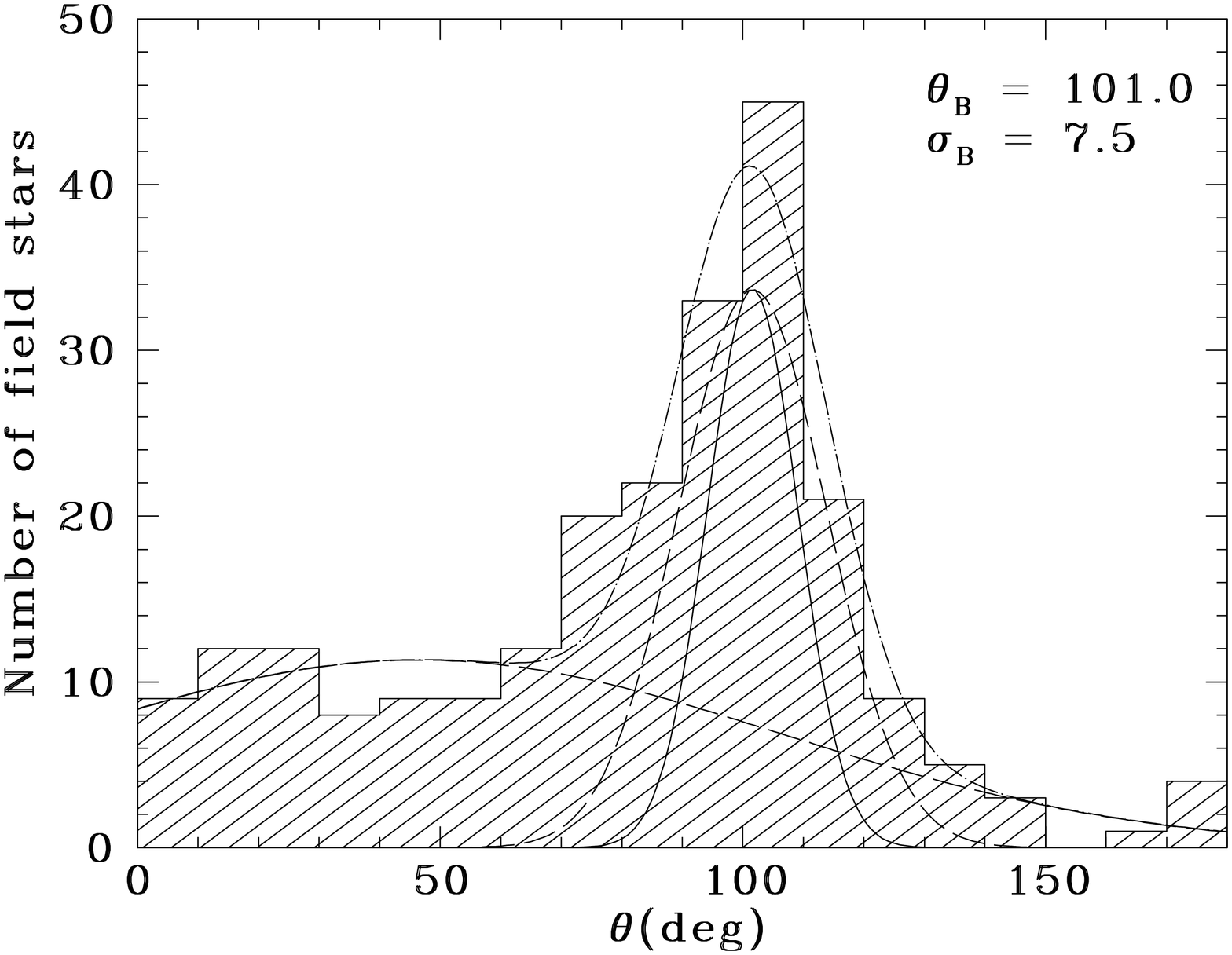}
\caption{Polarimetry of Field~13. Upper panel: The observed polarization vectors overplotted on a DSS2 red image. The coordinates are B1950. Lower panel: The histogram of the corresponding position angles of the polarization, $\theta$. In the upper right corner, it is shown the average and the unbiased dispersion of the interstellar magnetic field used in the analysis. The solid line is the respective Gaussian curve. The dashed lines represent the two Gaussian fits used to separated the polarimetric components and the dot-dash line is the sum of these two components.}
\label{fig:field13}
\end{figure}

\clearpage

\begin{figure}
\includegraphics[width=13cm,angle=0]{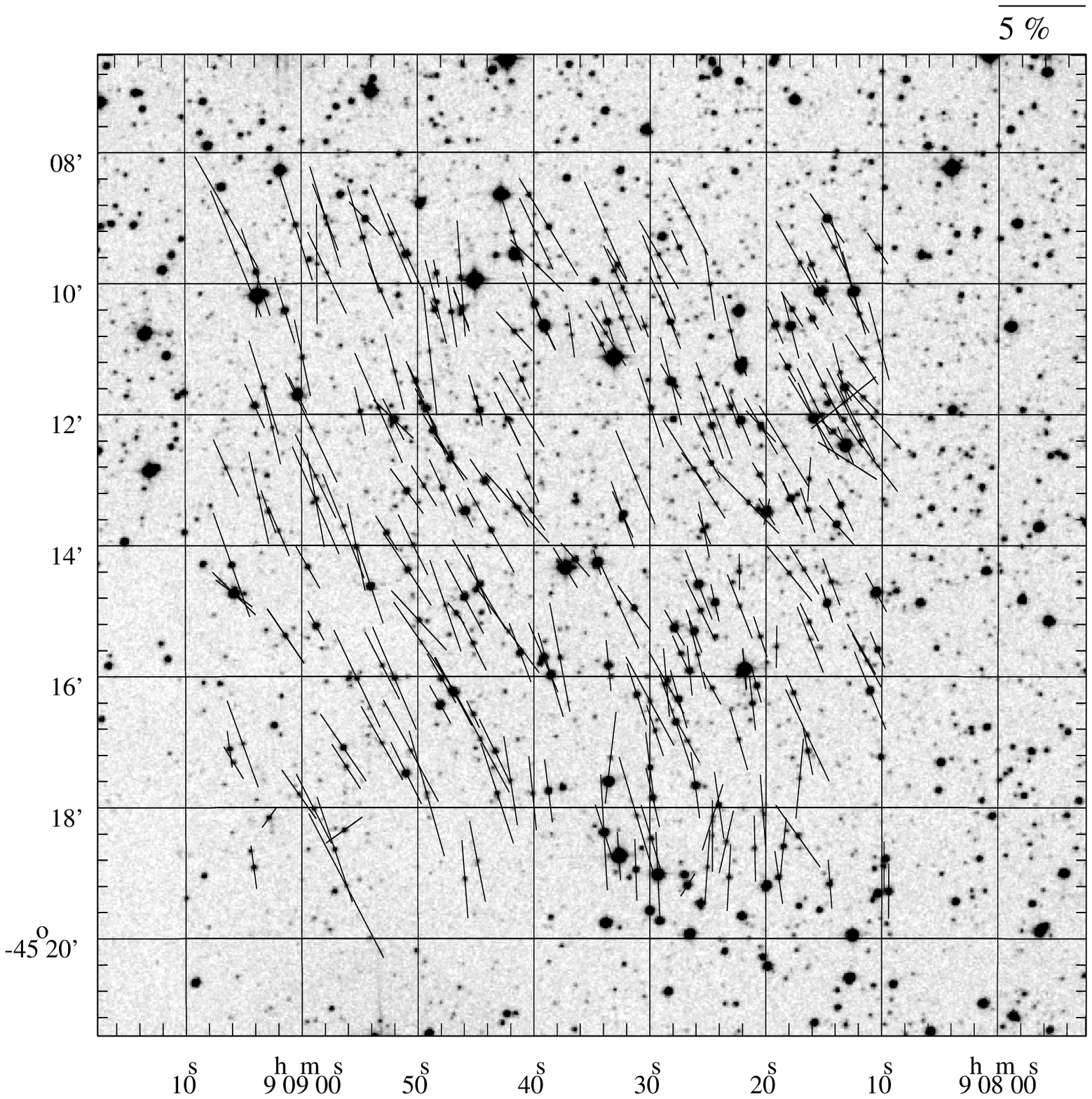}

\vspace{-5cm}
\includegraphics[width=6cm,angle=-90.]{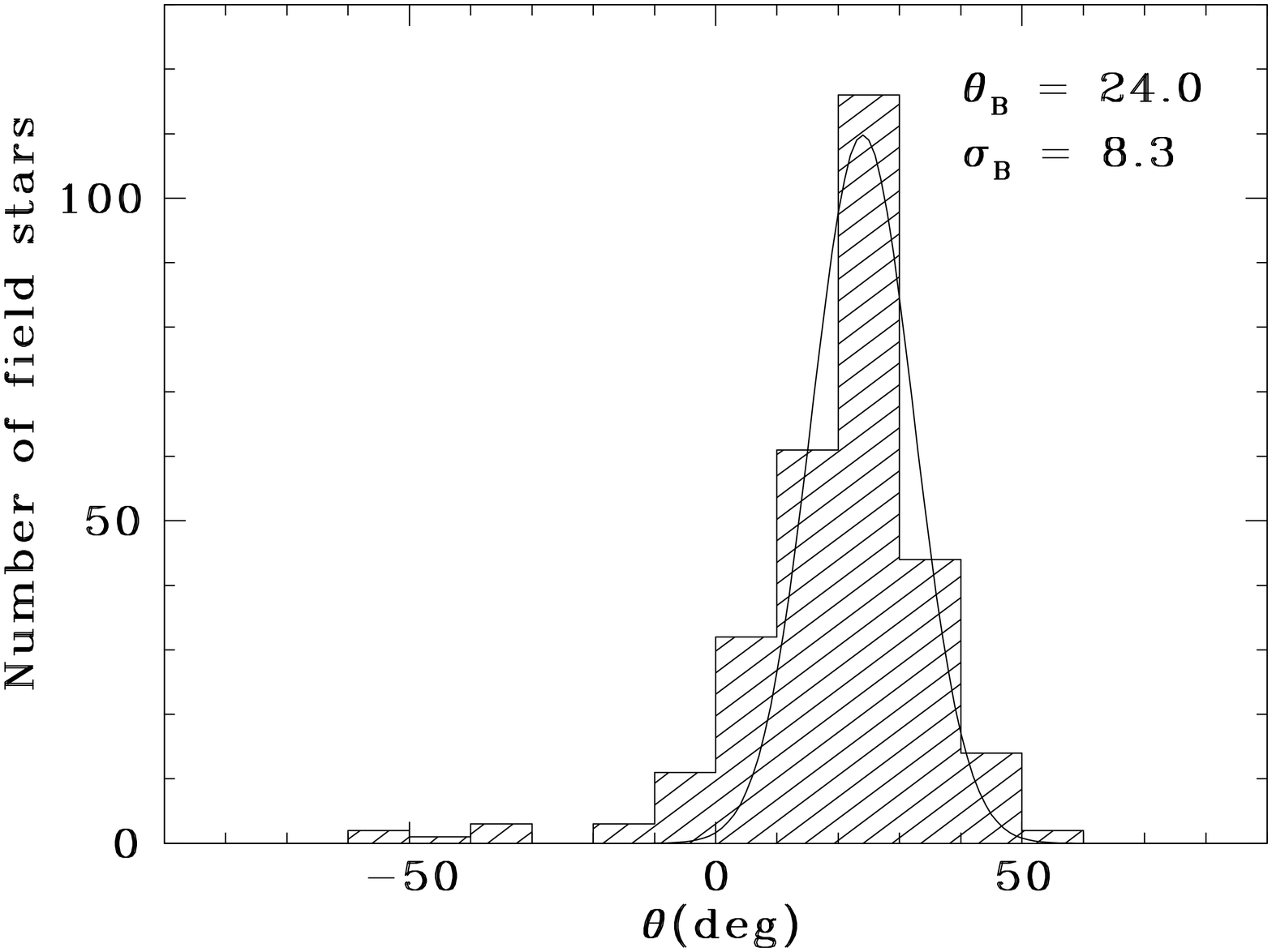}
\caption*{The same of Figure \ref{fig:hh139} for Field~14.}
\label{fig:field14}
\end{figure}

\clearpage

\begin{figure}
\includegraphics[width=13cm,angle=0]{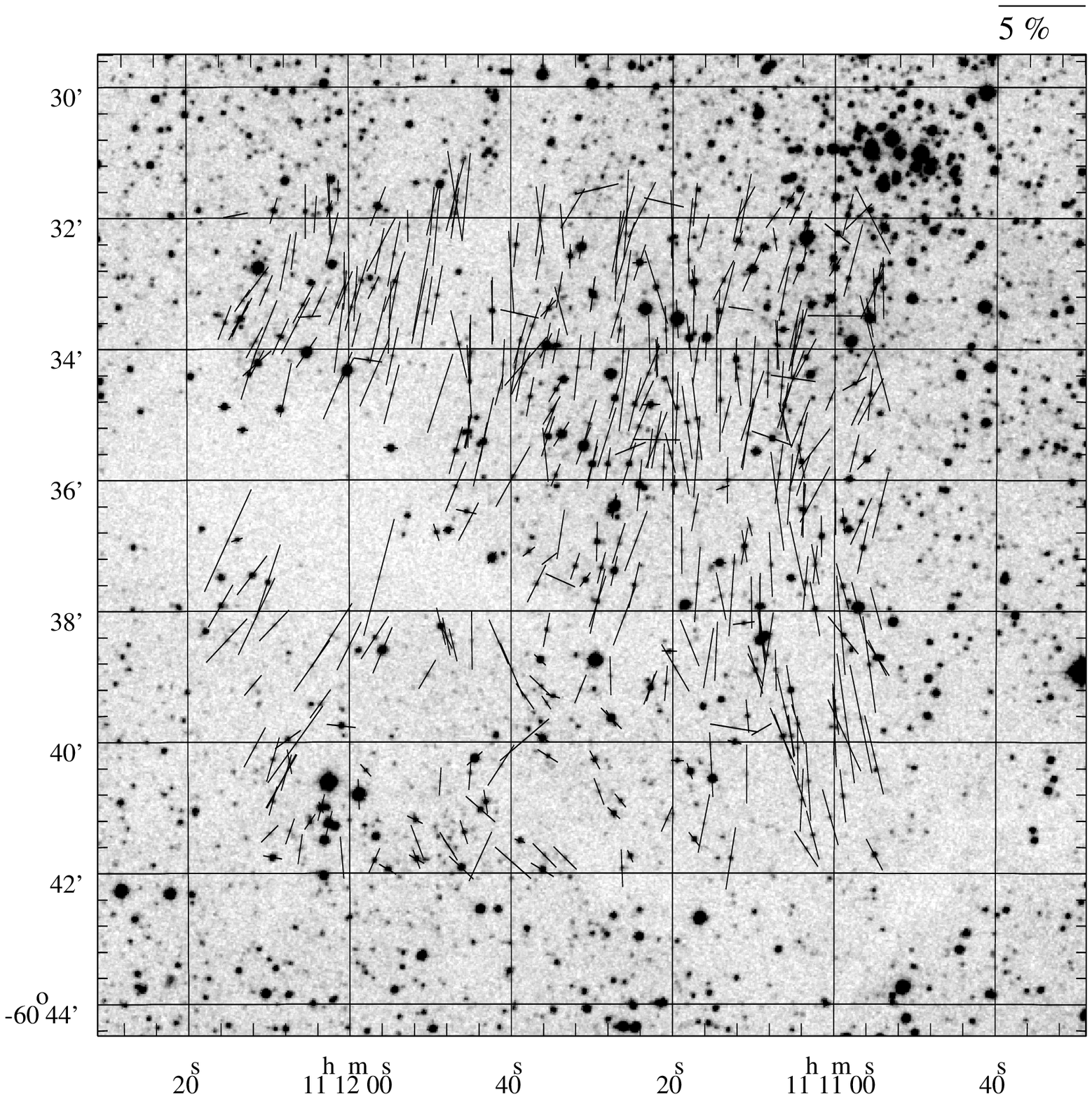}

\vspace{-5cm}
\includegraphics[width=6cm,angle=-90.]{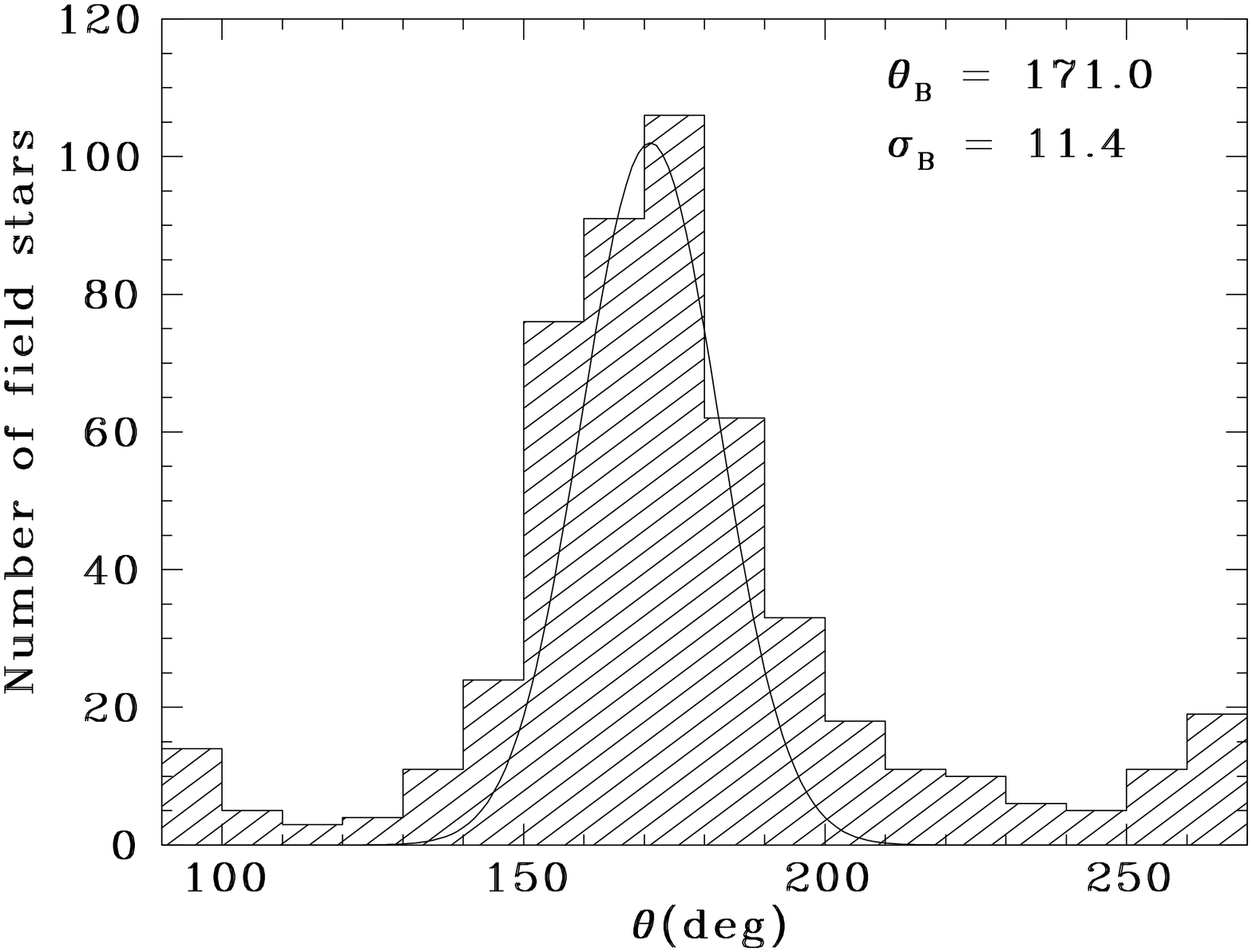}
\caption*{The same of Figure \ref{fig:hh139} for Field~16.}
\label{fig:field16}
\end{figure}

\clearpage

\begin{figure}
\includegraphics[width=13cm,angle=0]{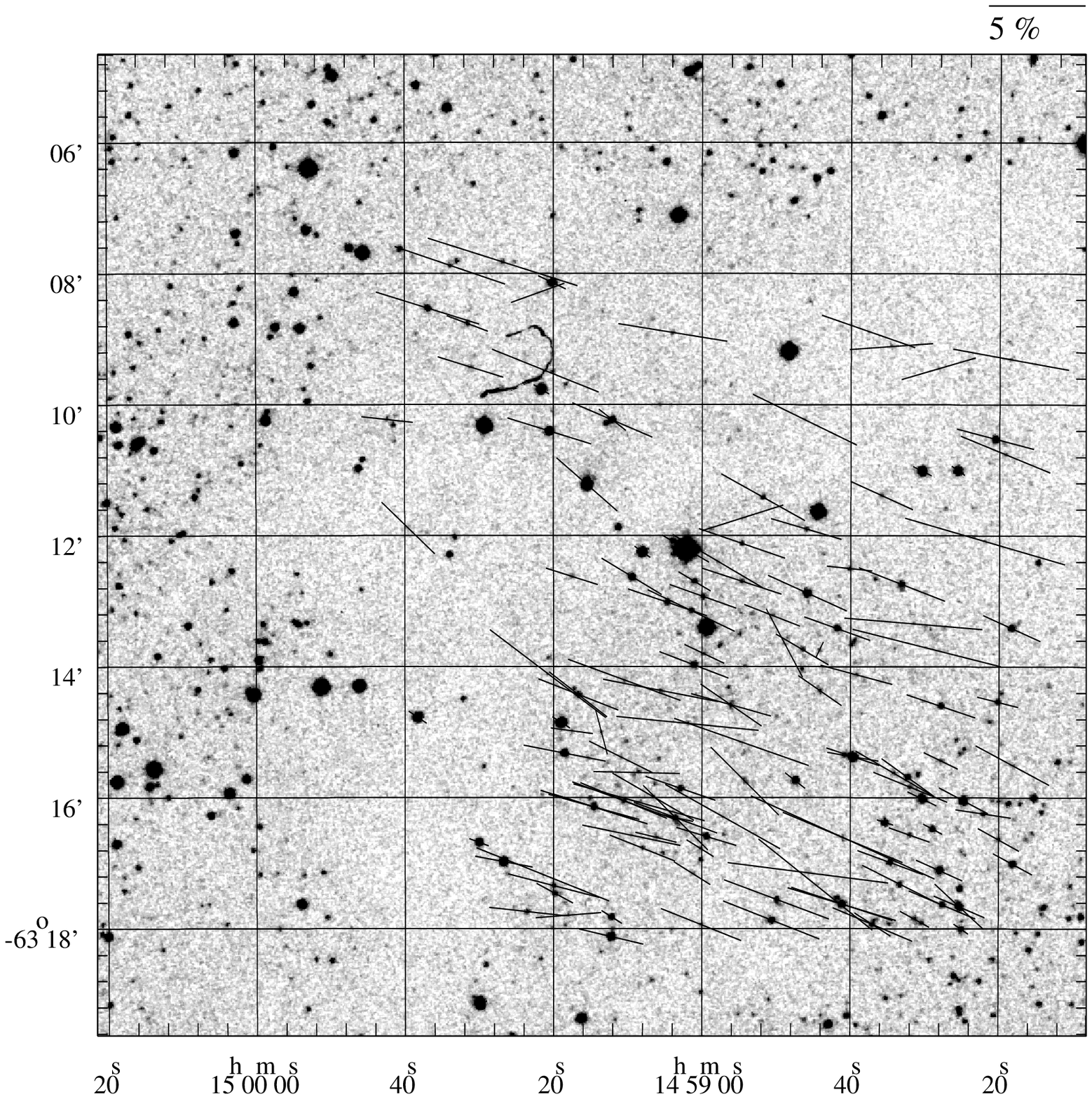}

\vspace{-5cm}
\includegraphics[width=6cm,angle=-90.]{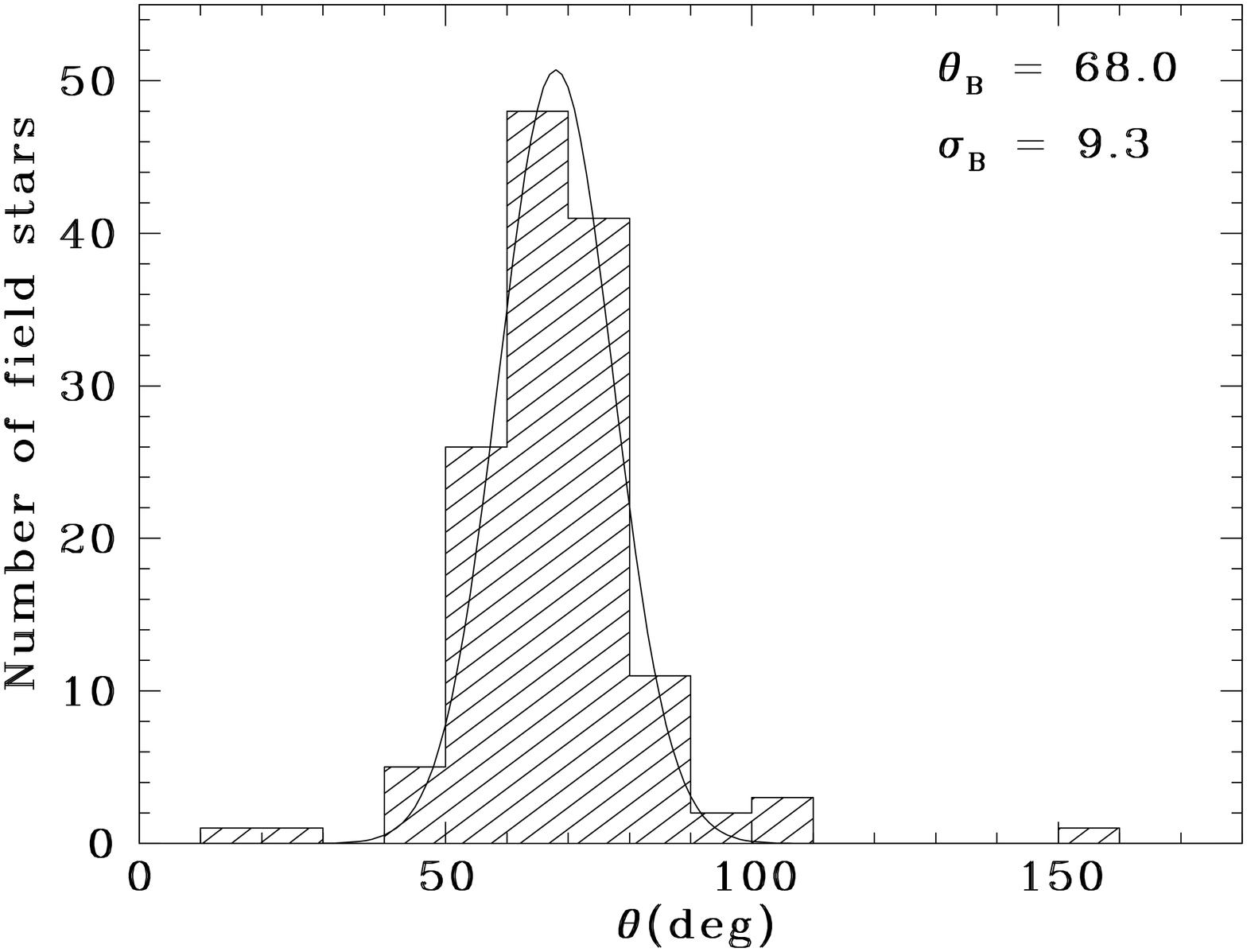}
\caption*{The same of Figure \ref{fig:hh139} for Field~18.}
\label{fig:field18}
\end{figure}

\clearpage

\begin{figure}
\includegraphics[width=13cm,angle=0]{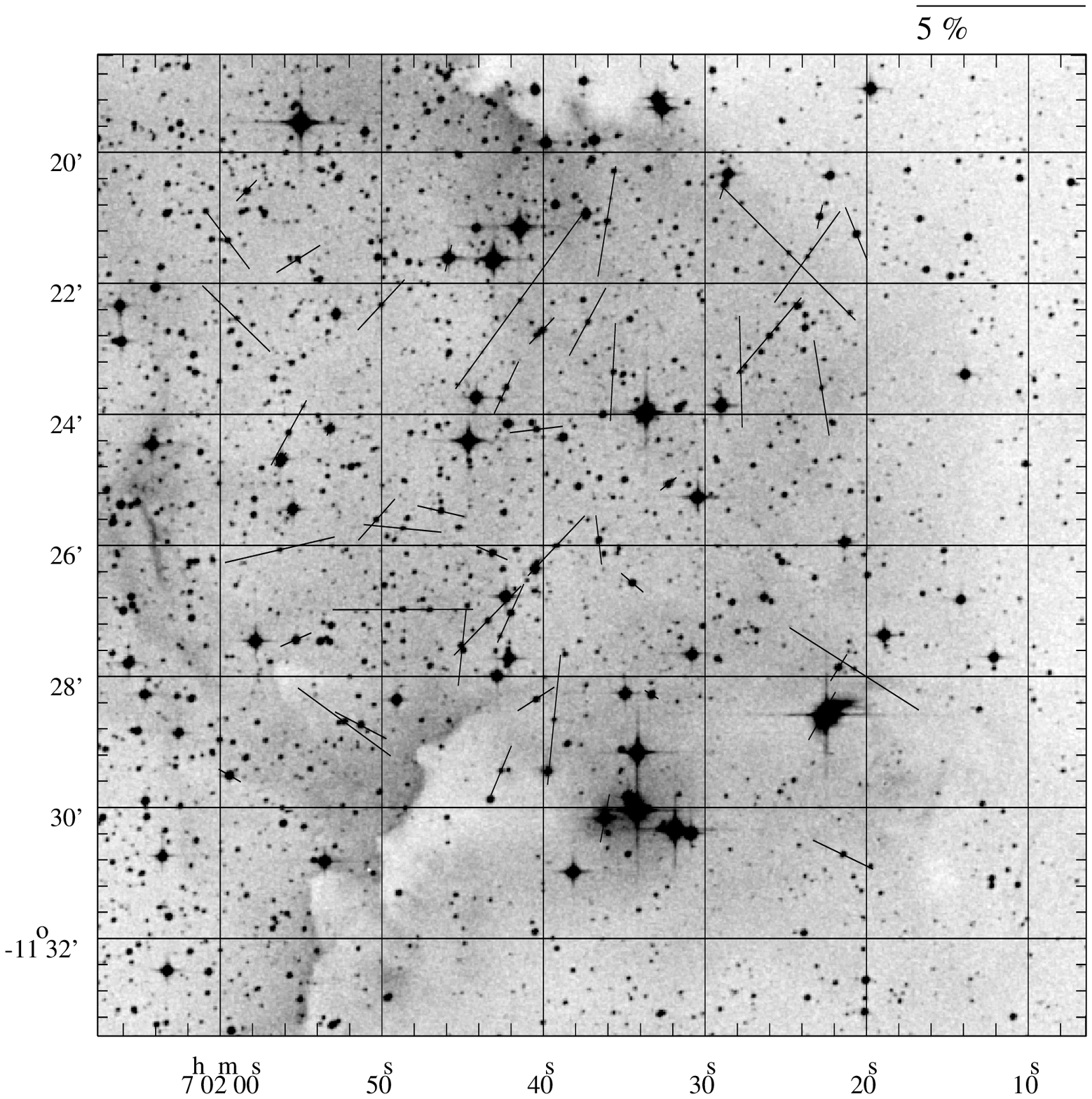}

\vspace{-5cm}
\includegraphics[width=6cm,angle=-90.]{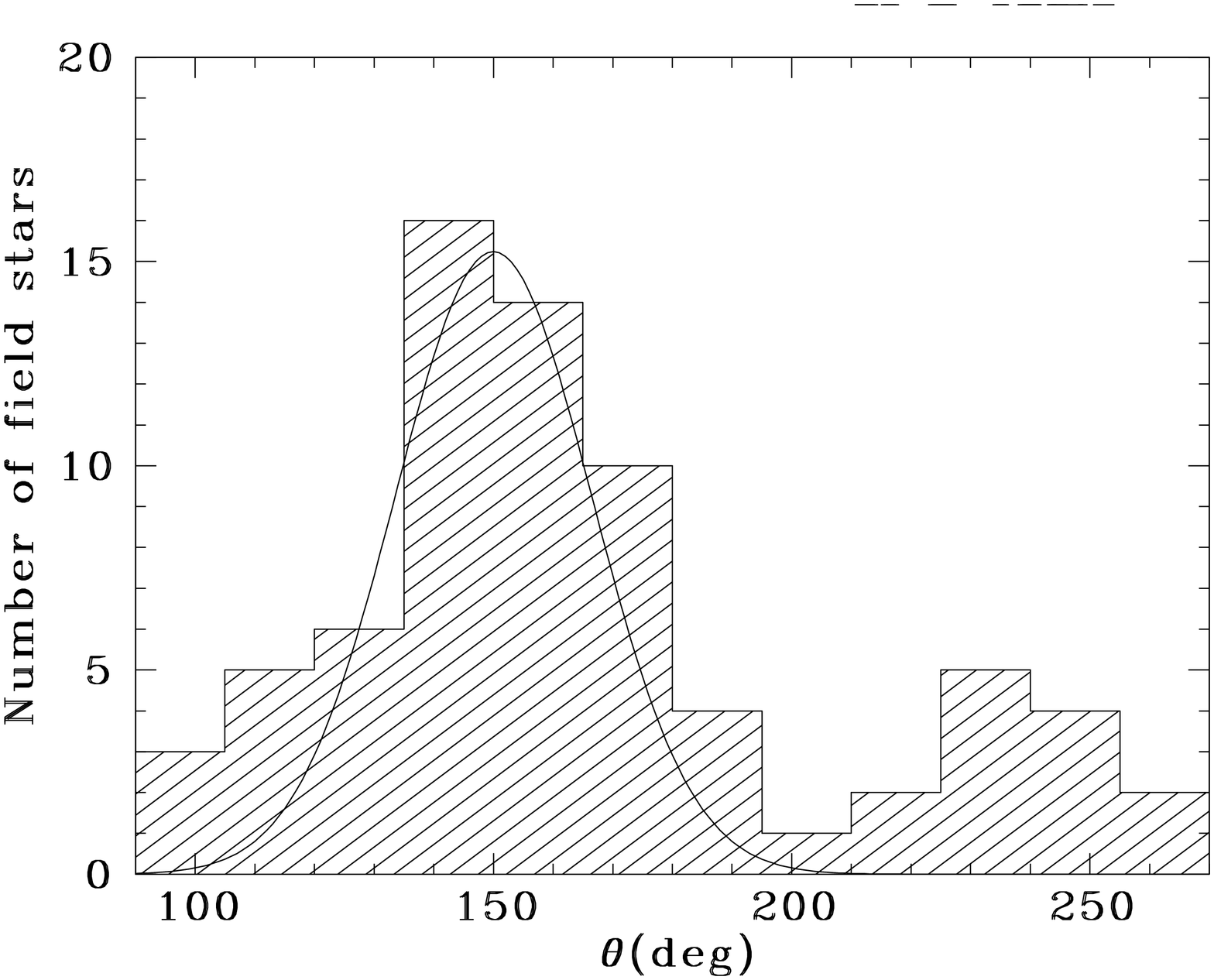}
\caption*{The same of Figure \ref{fig:hh139} for Field~19.}
\label{fig:field19}
\end{figure}

\clearpage

\begin{figure}
\includegraphics[width=13cm,angle=0]{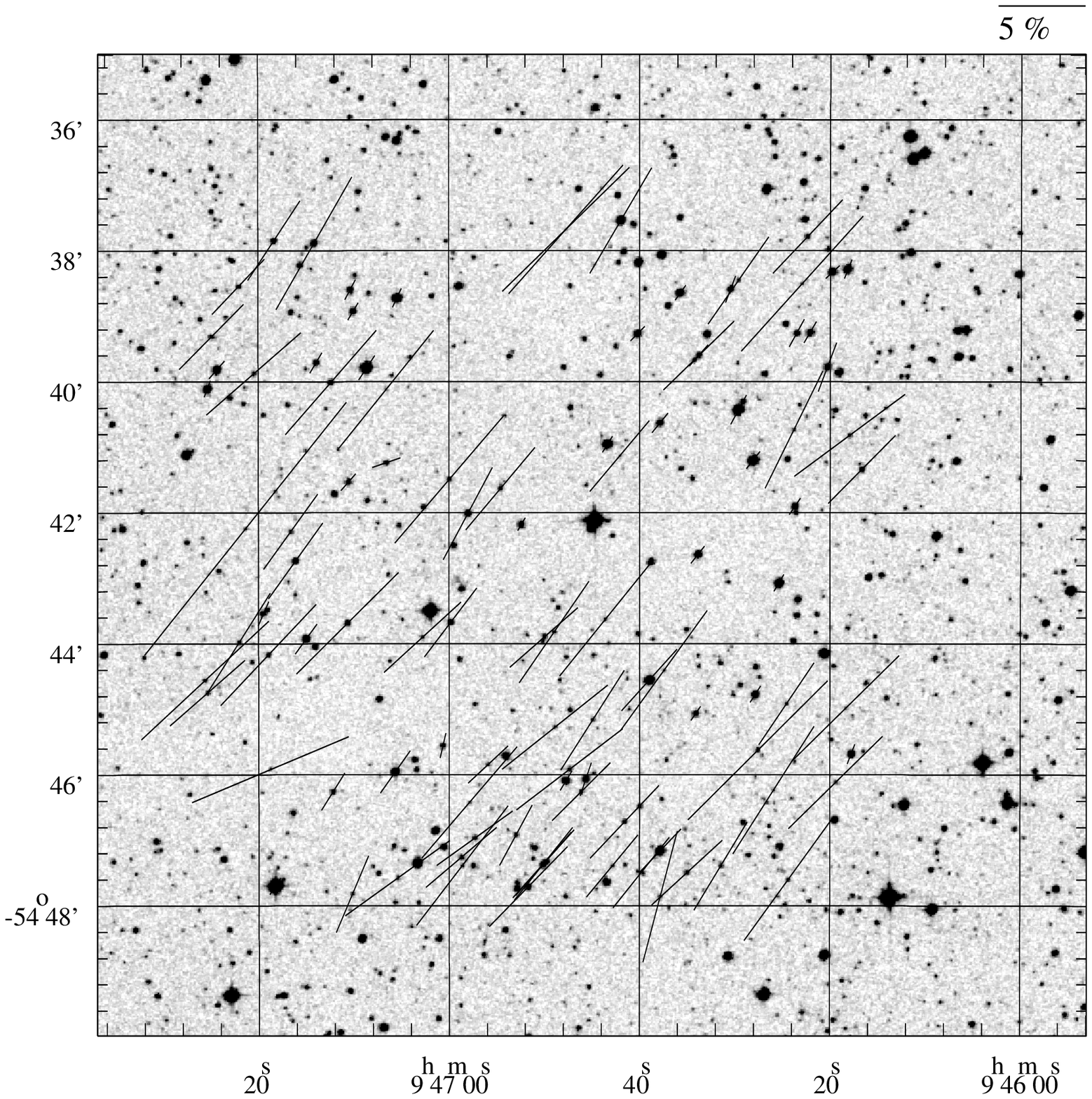}

\vspace{-5cm}
\includegraphics[width=6cm,angle=-90.]{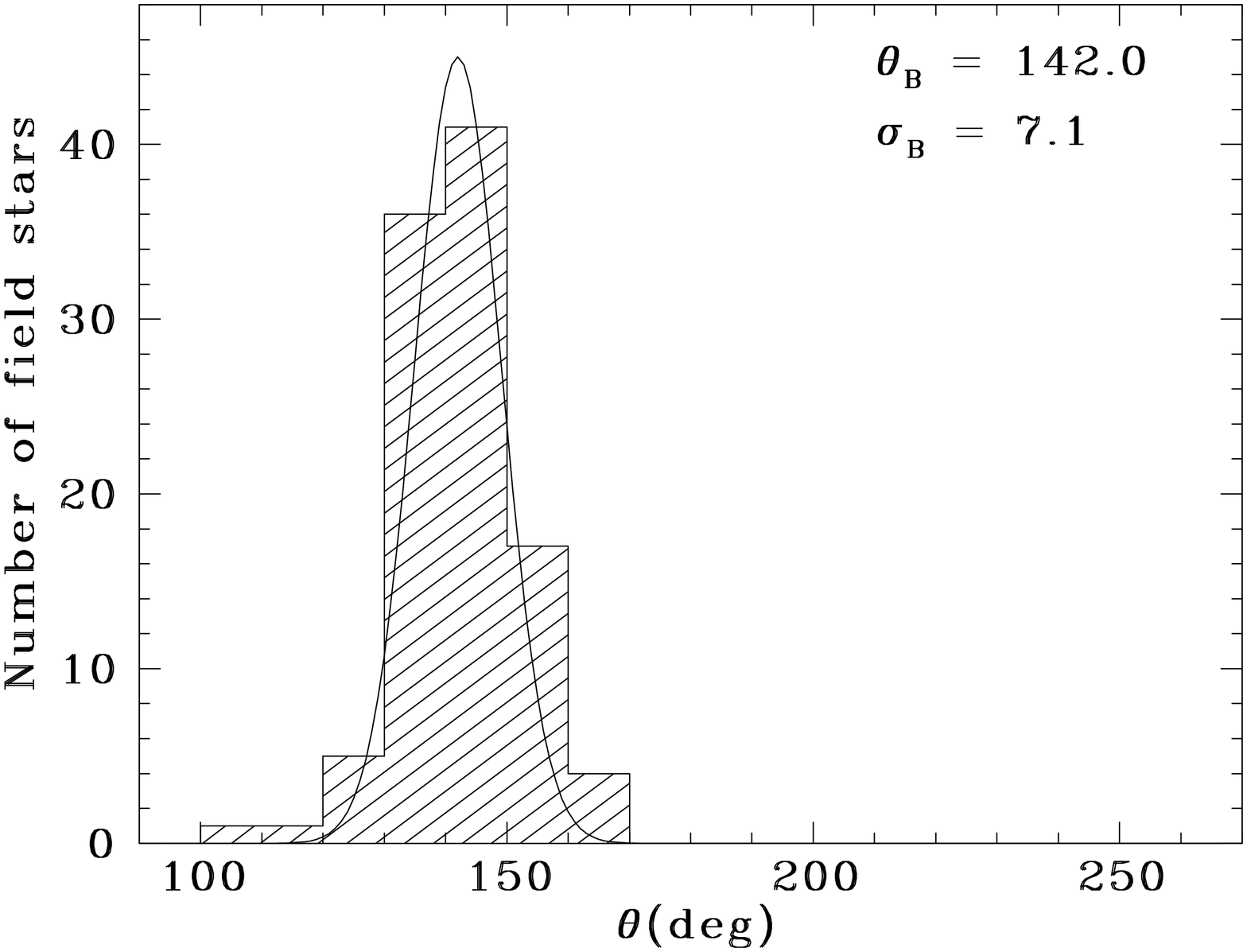}
\caption*{The same of Figure \ref{fig:hh139} for Field~20.}
\label{fig:field20}
\end{figure}

\clearpage

\begin{figure}
\includegraphics[width=13cm,angle=0]{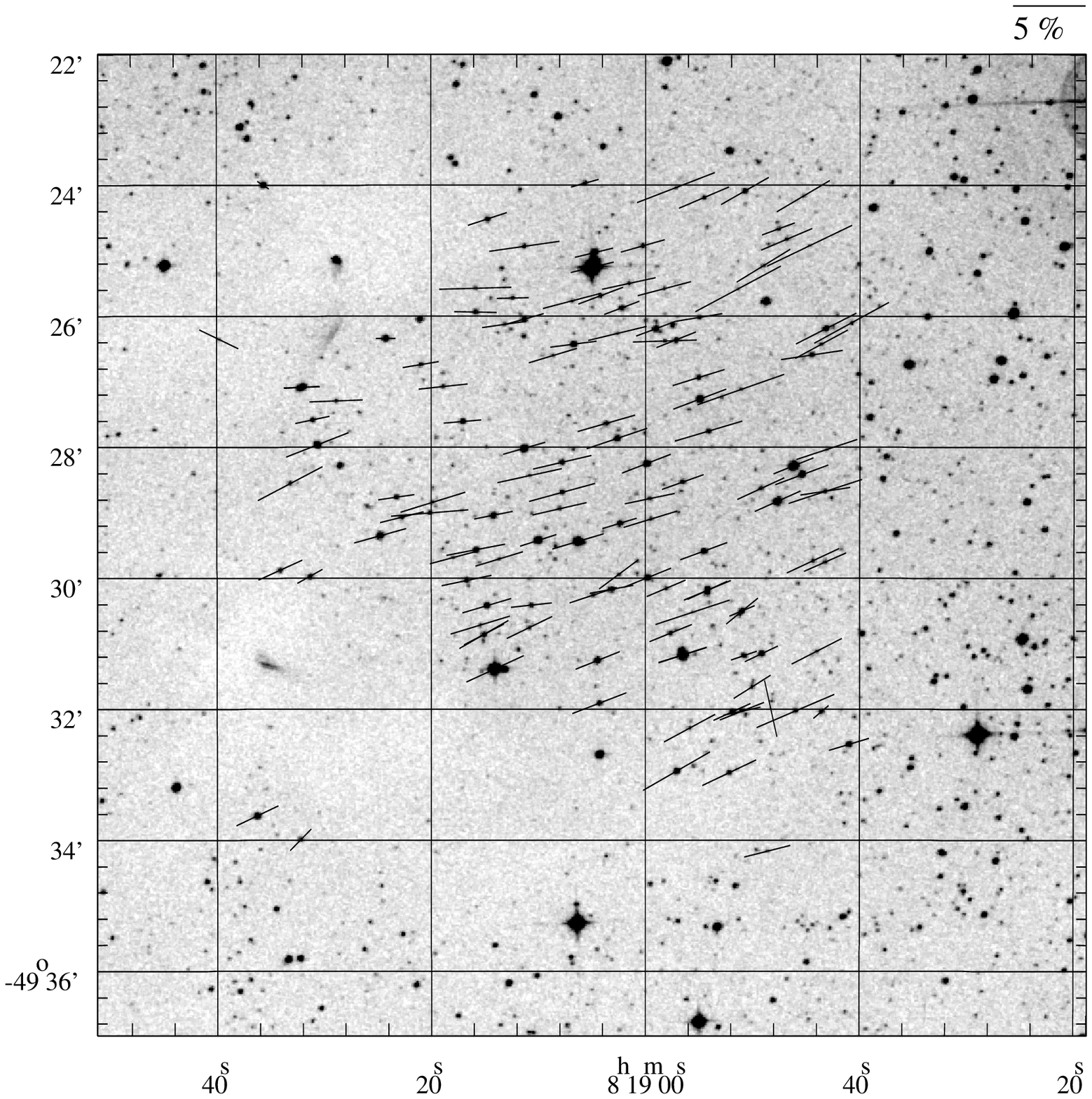}

\vspace{-5cm}
\includegraphics[width=6cm,angle=-90.]{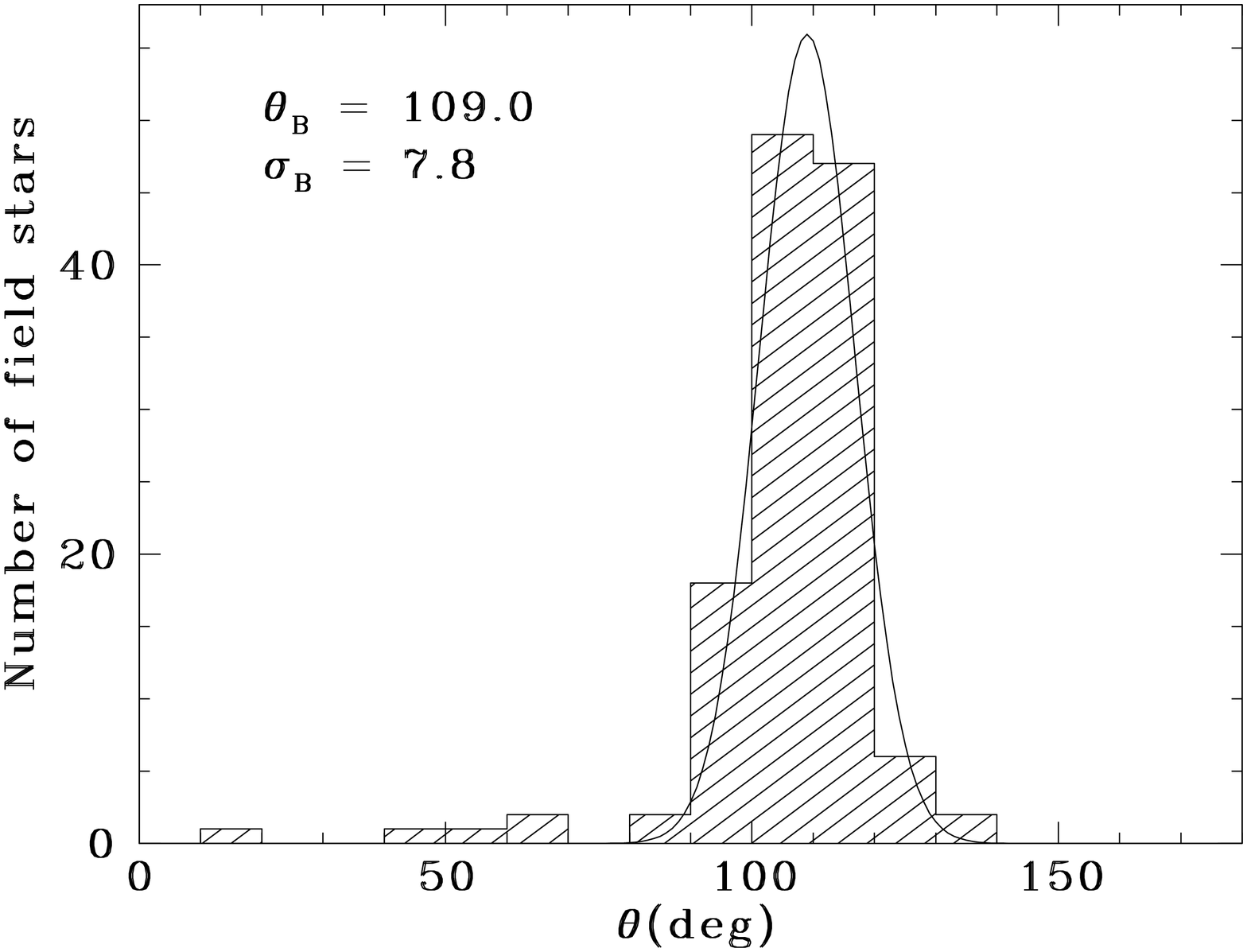}
\caption*{The same of Figure \ref{fig:hh139} for Field~21.}
\label{fig:field21}
\end{figure}

\clearpage

\begin{figure}
\includegraphics[width=13cm,angle=0]{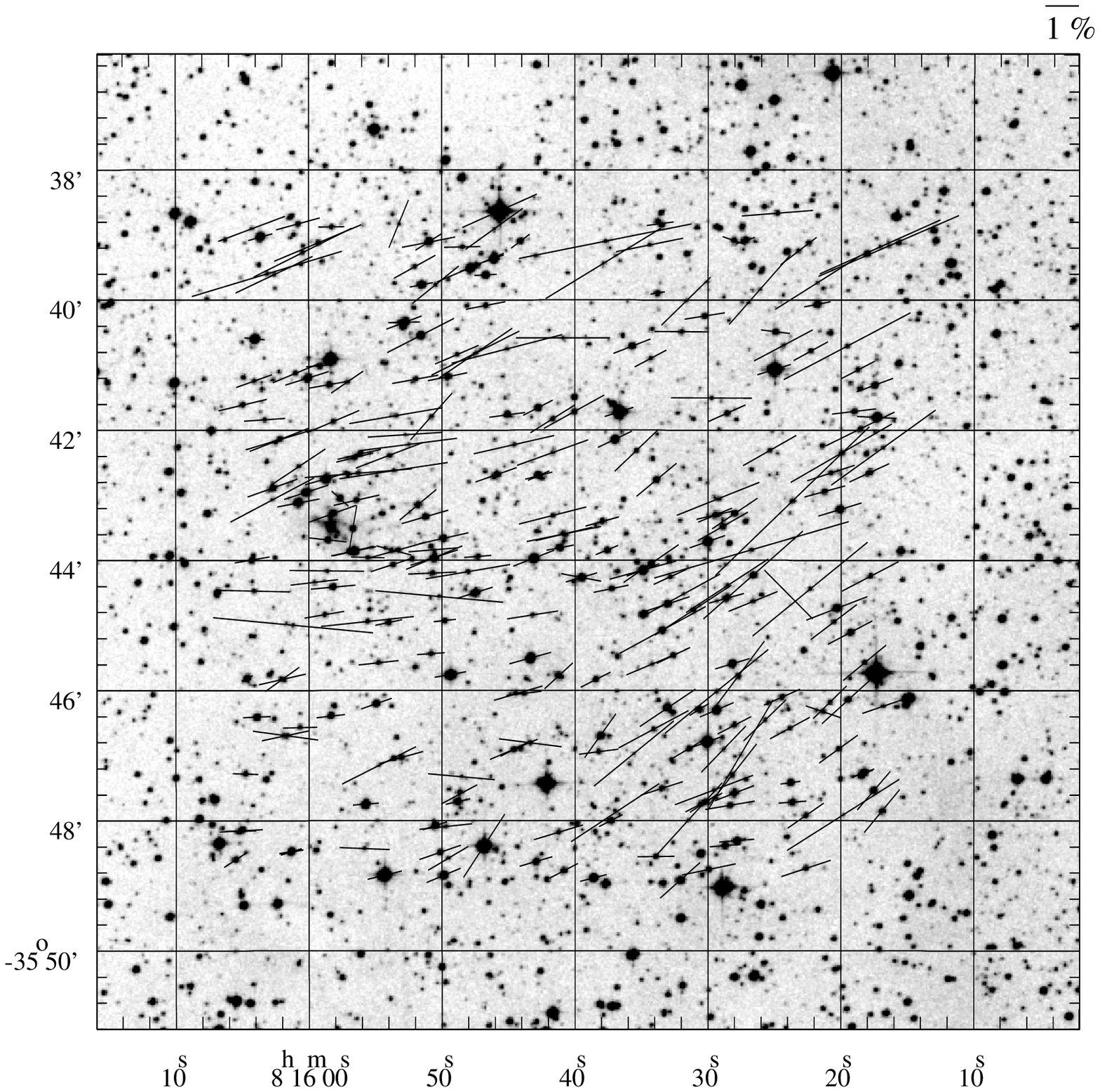}

\vspace{-5cm}
\includegraphics[width=6cm,angle=-90.]{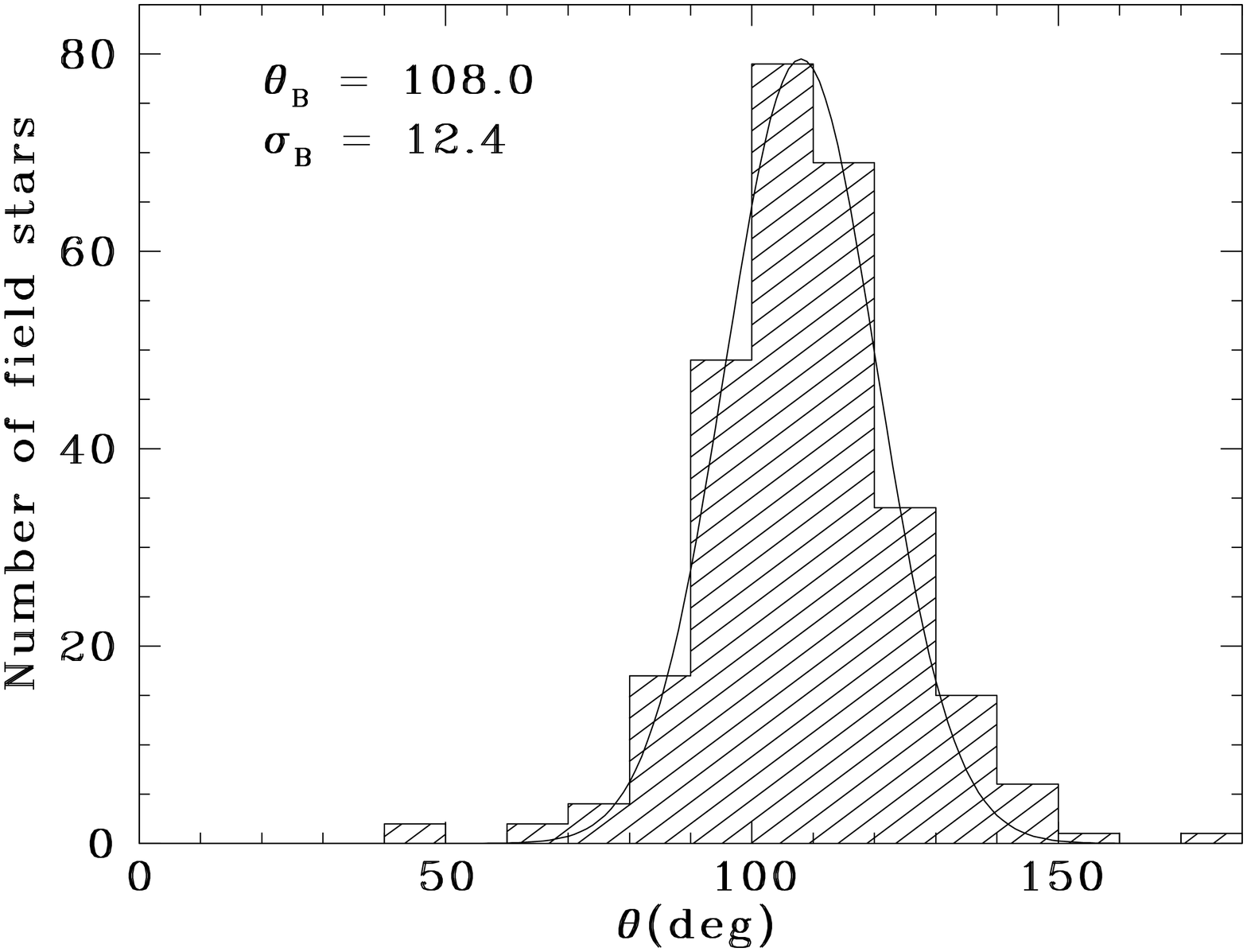}
\caption*{The same of Figure \ref{fig:hh139} for Field~22.}
\label{fig:field22}
\end{figure}

\clearpage

\begin{figure}
\includegraphics[width=13cm,angle=0]{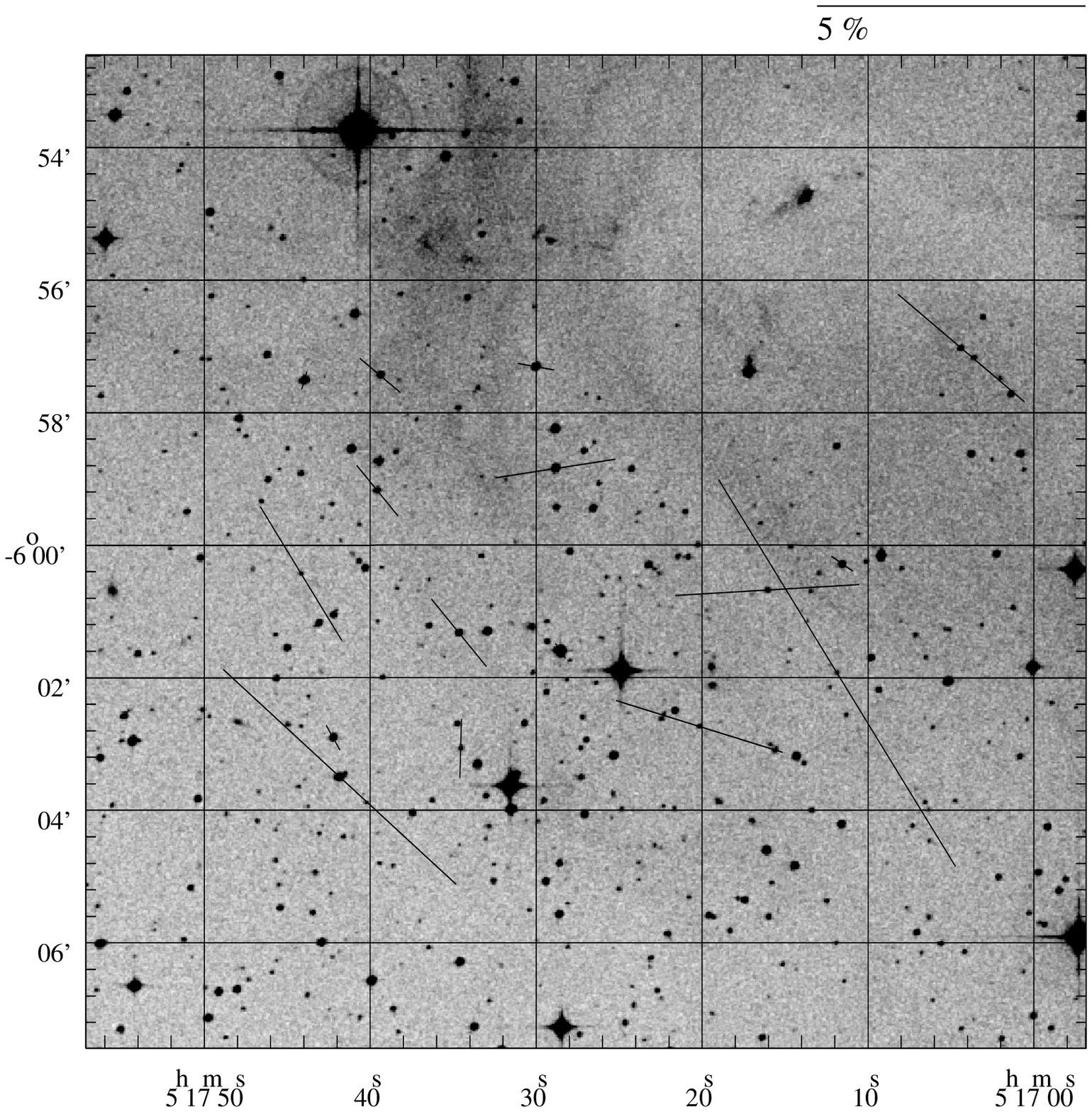}

\vspace{-5cm}
\includegraphics[width=6cm,angle=-90.]{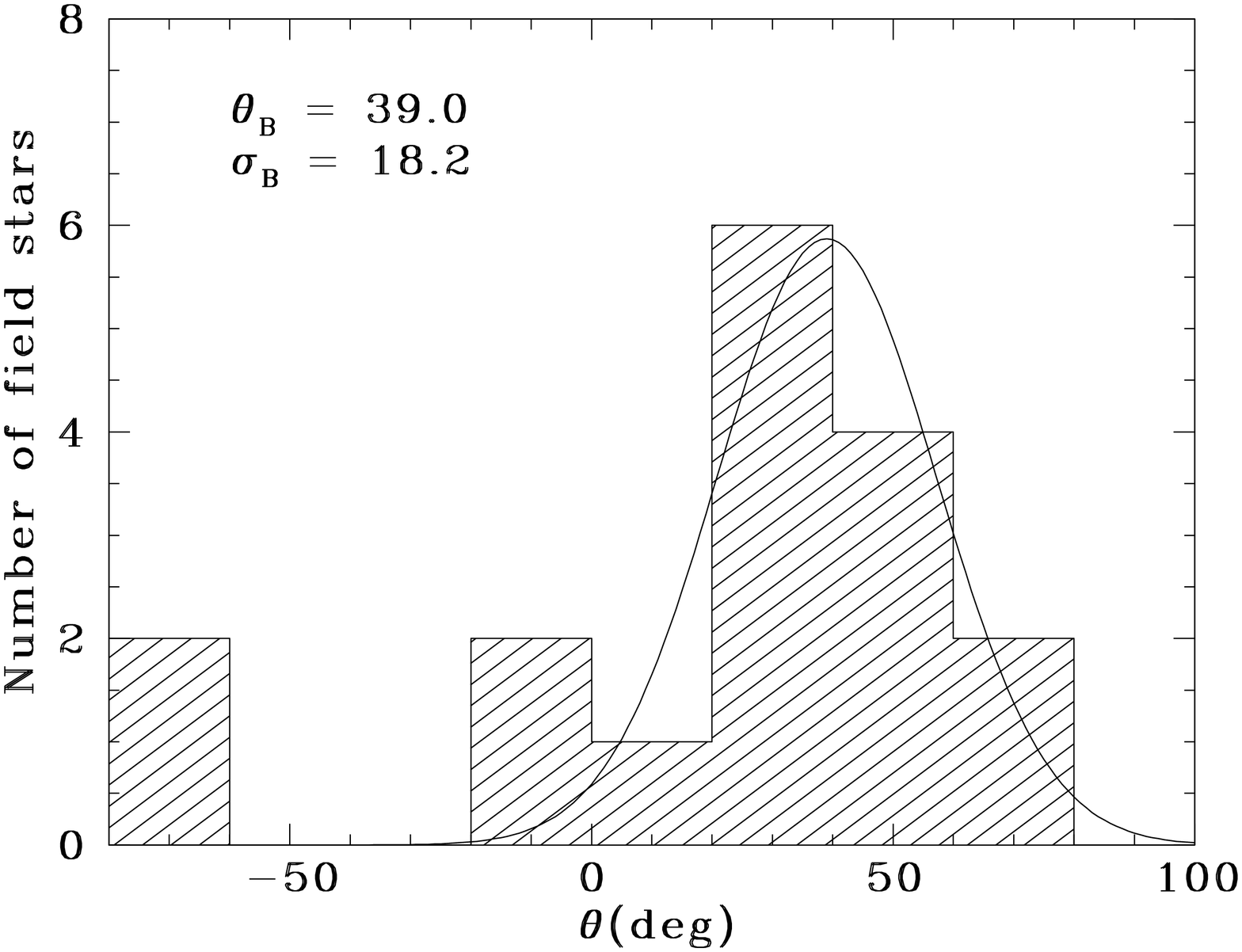}
\caption*{The same of Figure \ref{fig:hh139} for Field~23.}
\label{fig:field23}
\end{figure}

\clearpage

\begin{figure}
\includegraphics[width=9cm,angle=0]{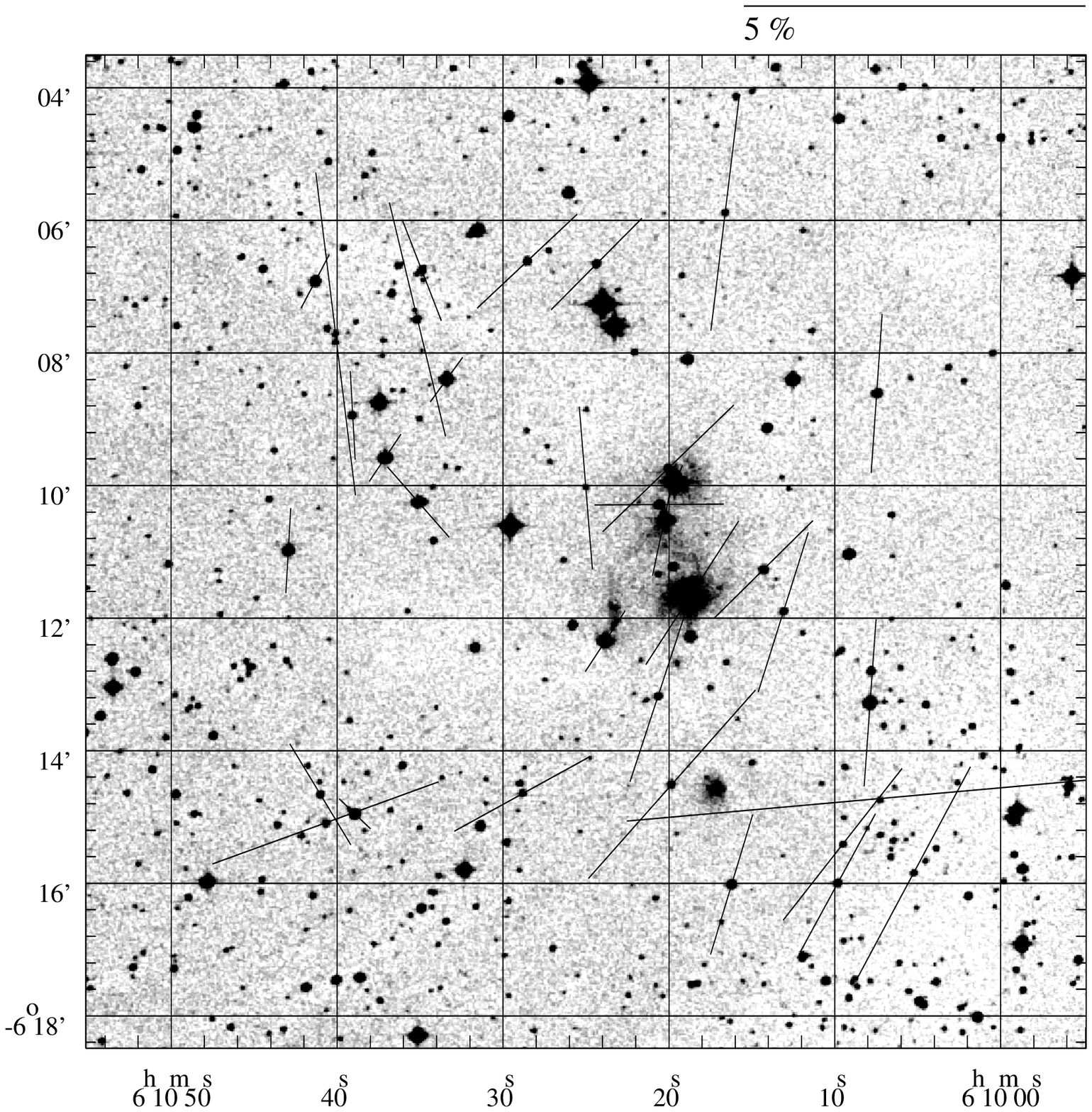}
\includegraphics[width=9cm,angle=0]{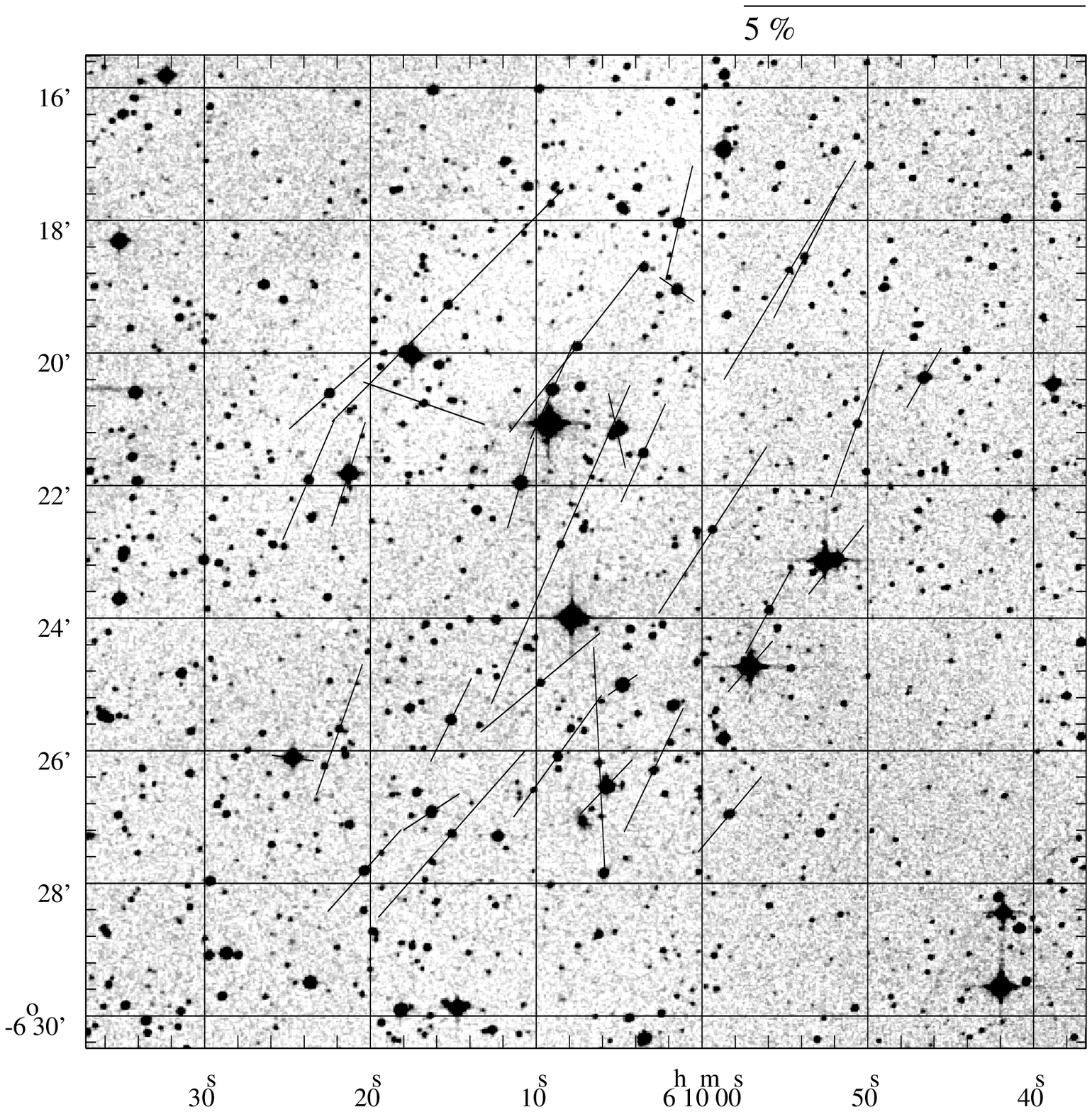}

\vspace{-2cm}
\includegraphics[width=6cm,angle=-90.]{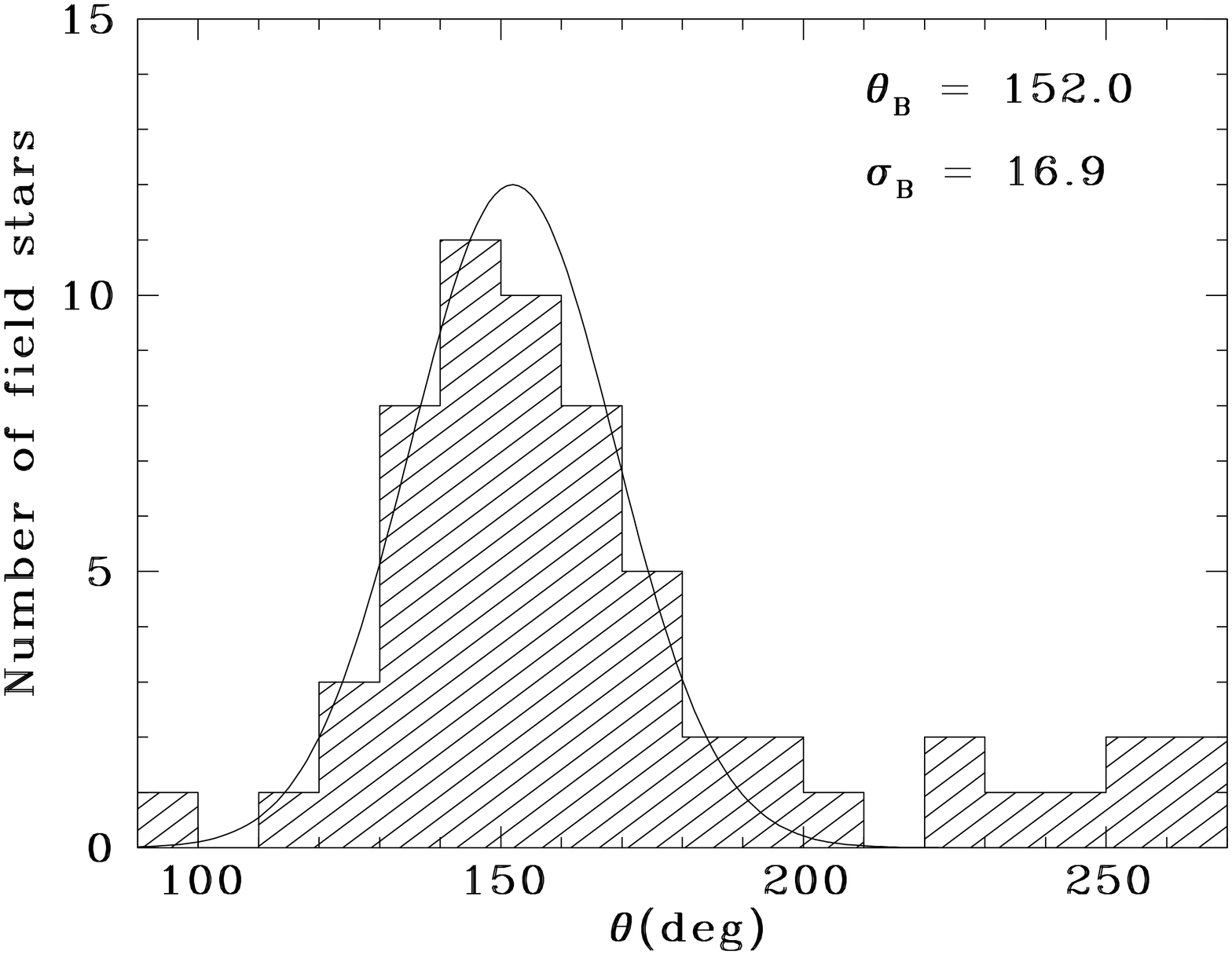}
\caption{Polarimetry of Fields~24A e 24B. Upper left panel: The observed polarization vectors of Field~24A overplotted on a DSS2 red image. The coordinates are B1950. Upper right panel: The same for Field~24B. Lower panel: The histogram of the position angles of the polarization, $\theta$, of Fields~24A e 24B. In the upper right corner, it is shown the average and the dispersion of the interstellar magnetic field used in the analysis. A Gaussian curve using these values is also depicted. }
\label{fig:field24a}
\end{figure}

\clearpage

\begin{figure}
\includegraphics[width=13cm,angle=0]{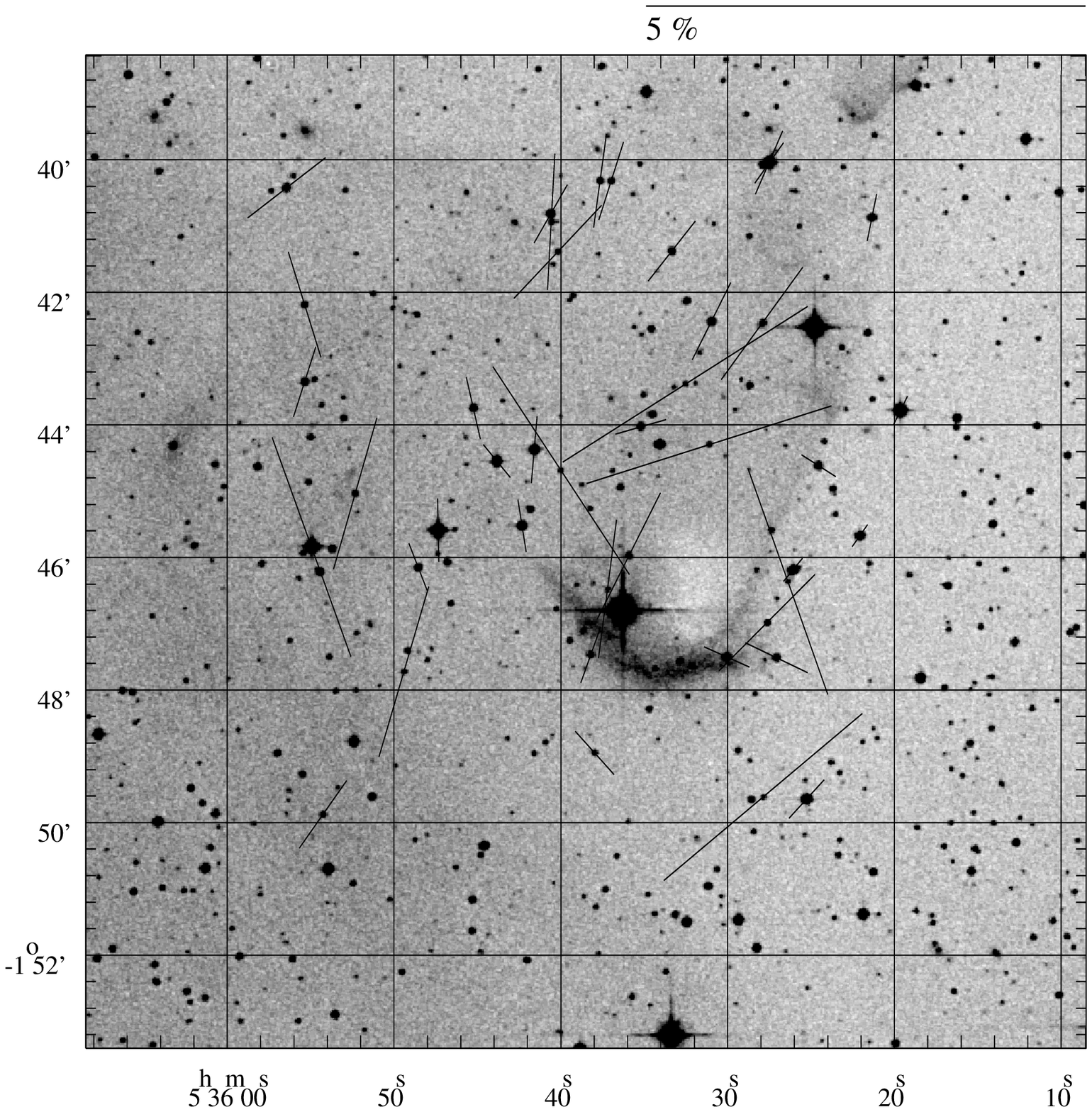}

\vspace{-5cm}
\includegraphics[width=6cm,angle=-90.]{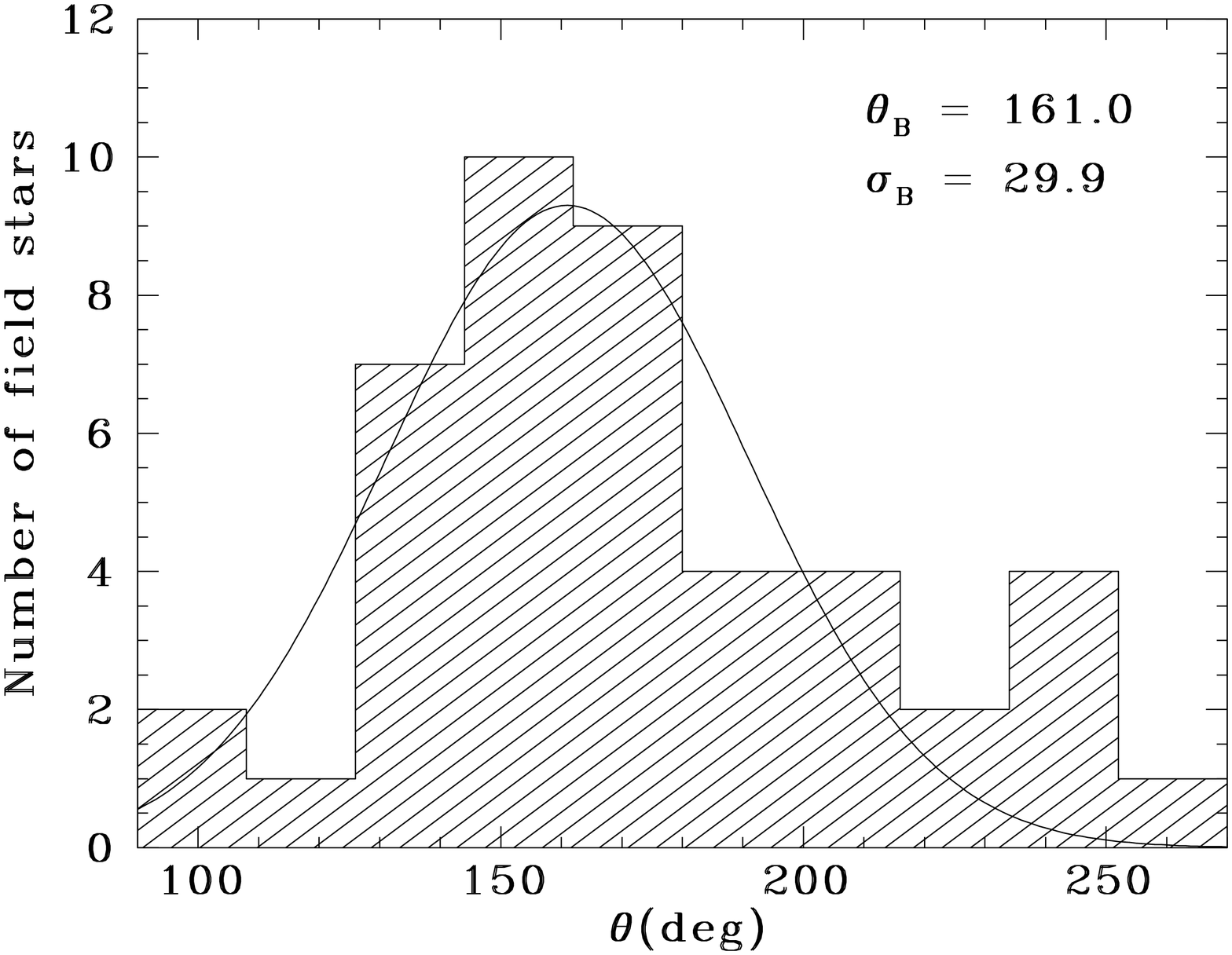}
\caption*{The same of Figure \ref{fig:hh139} for Field~25.}
\label{fig:field25}
\end{figure}

\clearpage

\begin{figure}
\includegraphics[width=13cm,angle=0]{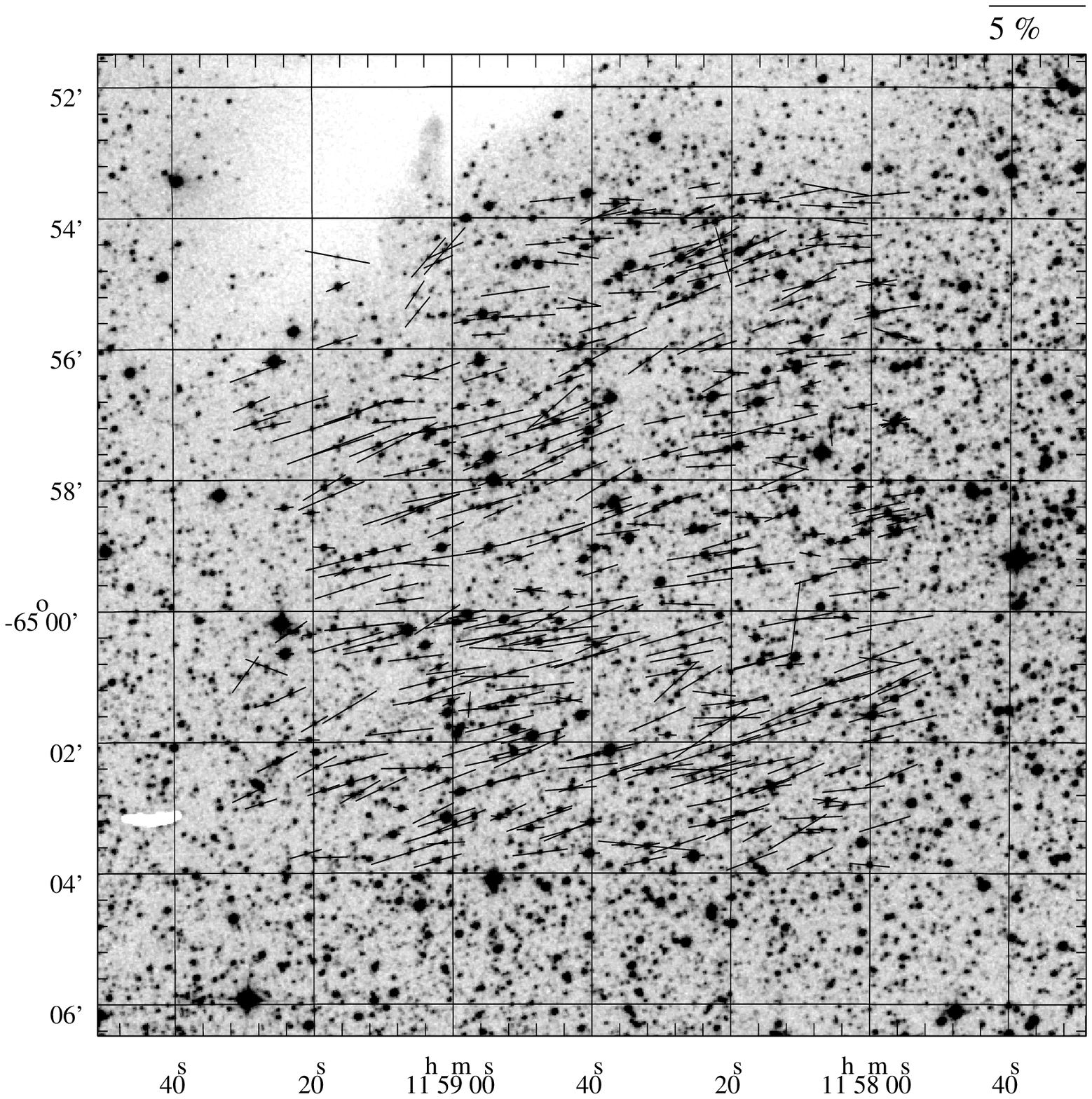}

\vspace{-5cm}
\includegraphics[width=6cm,angle=-90.]{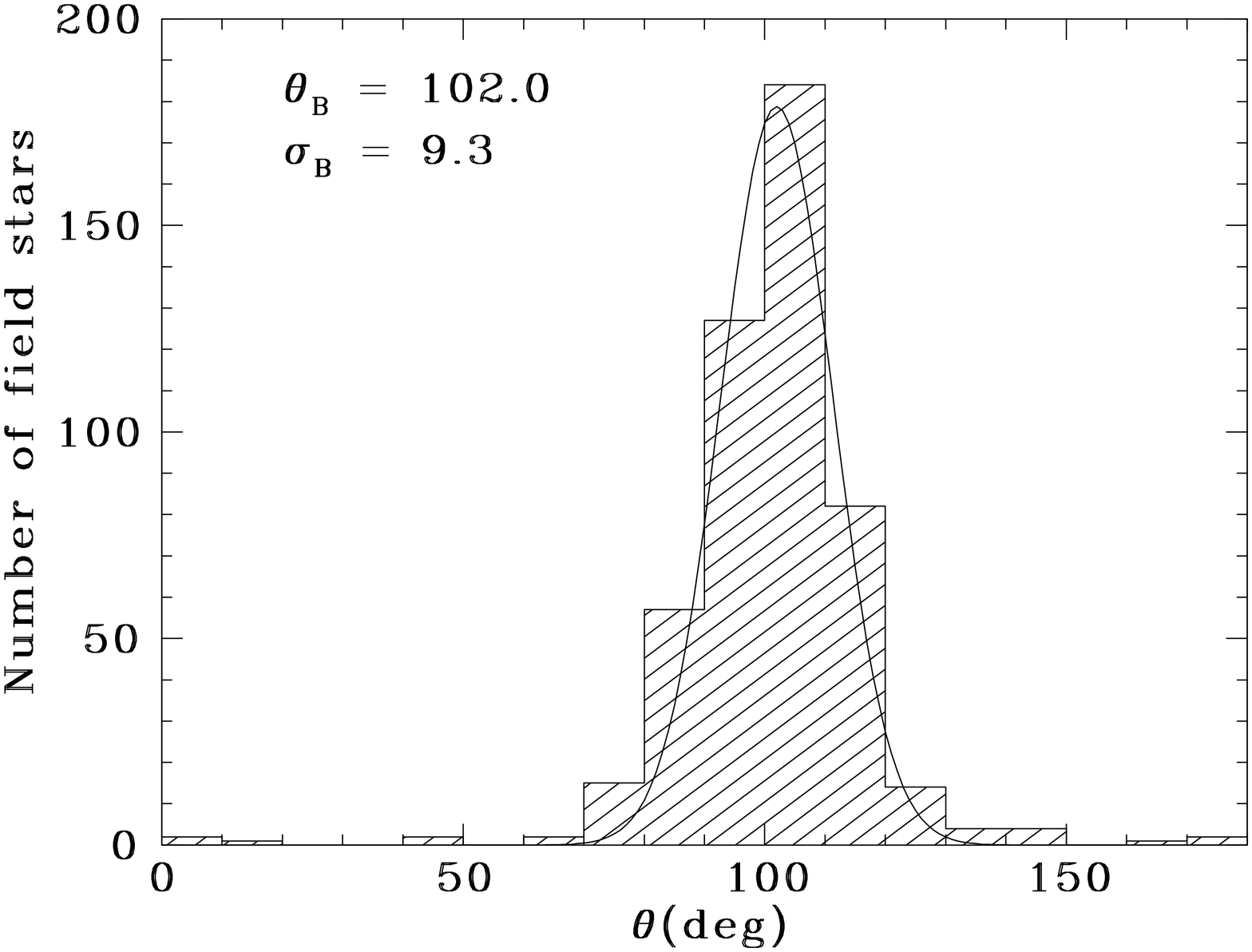}
\caption*{The same of Figure \ref{fig:hh139} for Field~26.}
\label{fig:field26}
\end{figure}

\clearpage

\begin{figure}
\includegraphics[width=13cm,angle=0]{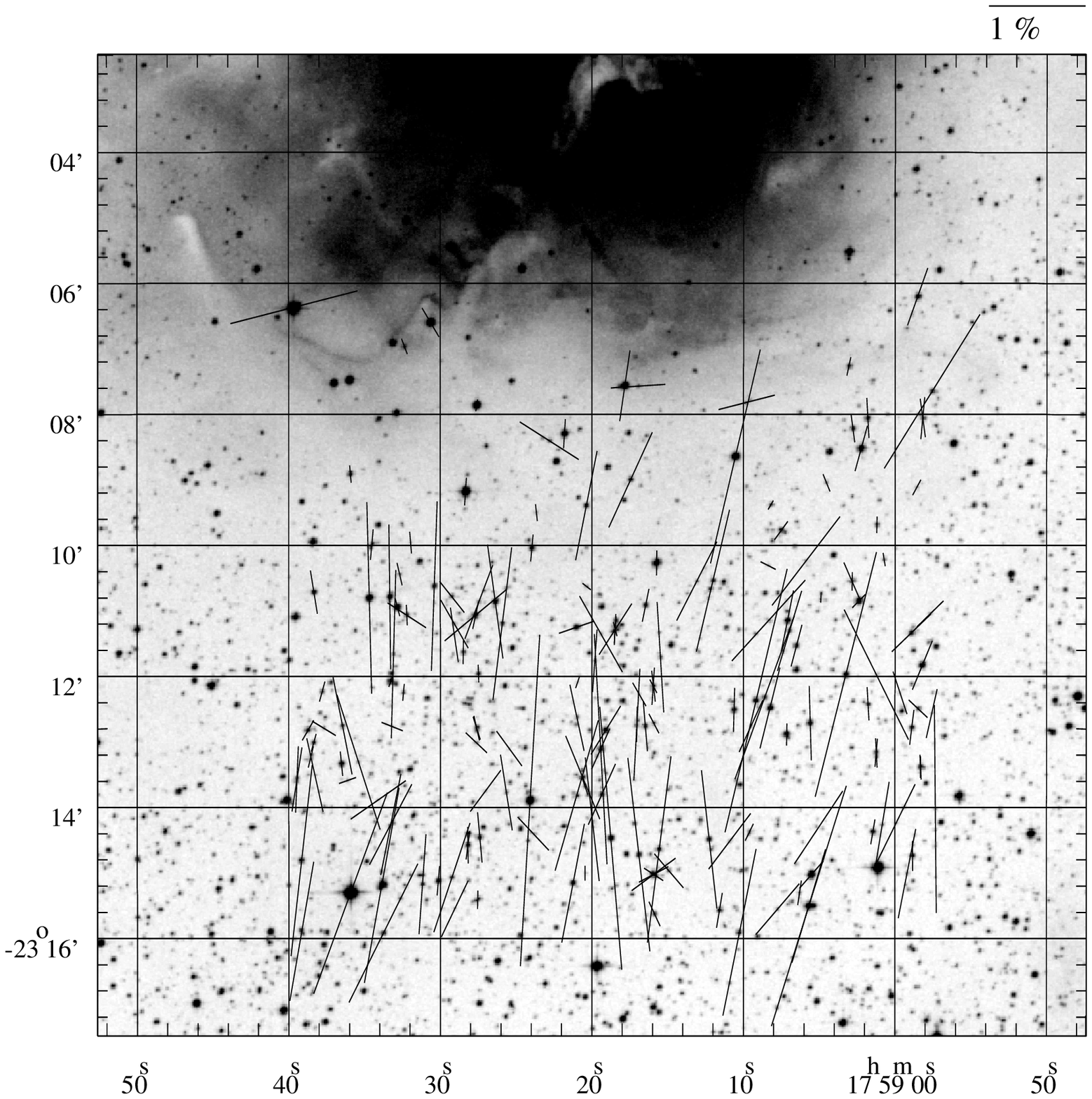}

\vspace{-5cm}
\includegraphics[width=6cm,angle=-90.]{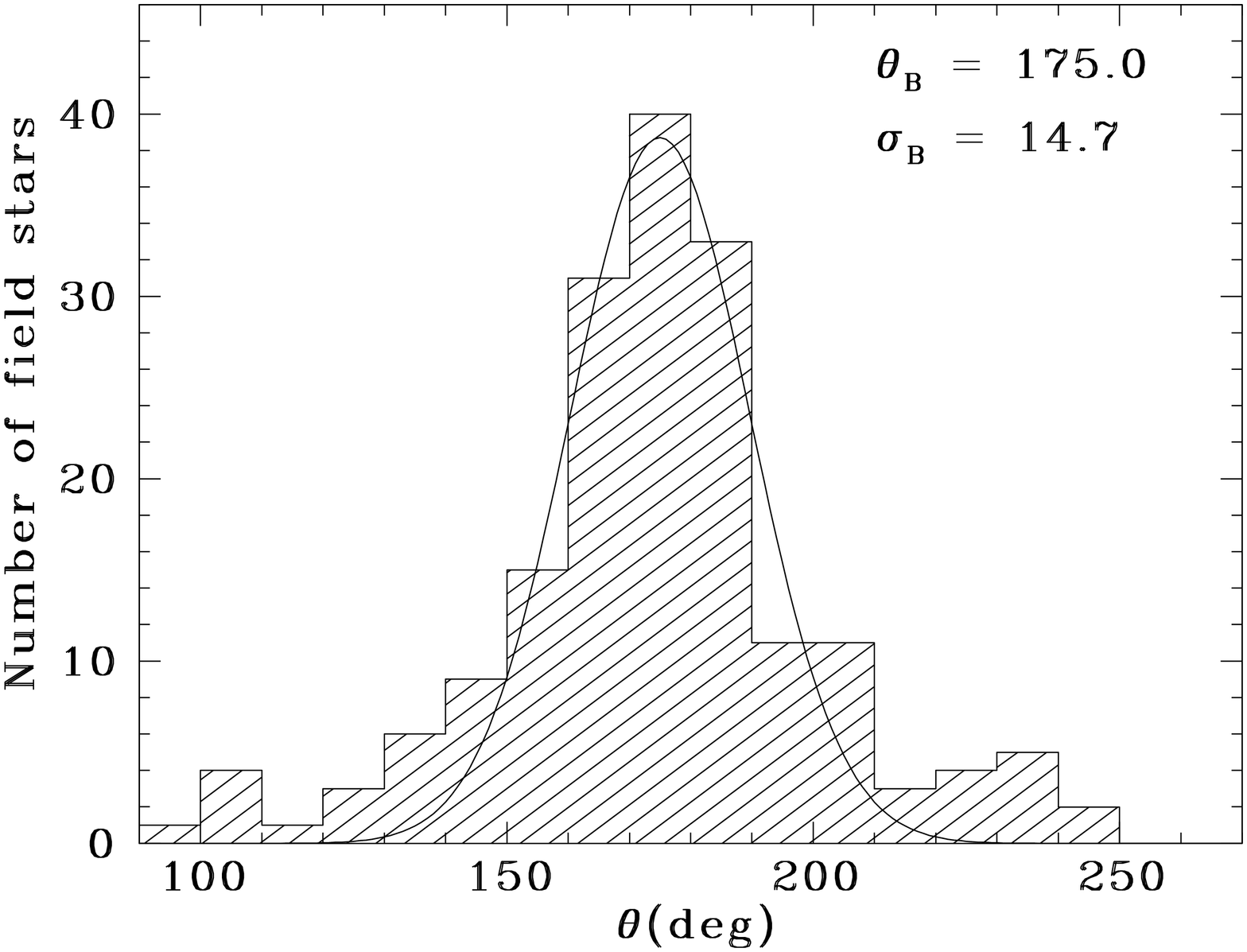}
\caption*{The same of Figure \ref{fig:hh139} for Field~27.}
\label{fig:field27}
\end{figure}

\clearpage

\begin{figure}
\includegraphics[width=13cm,angle=0]{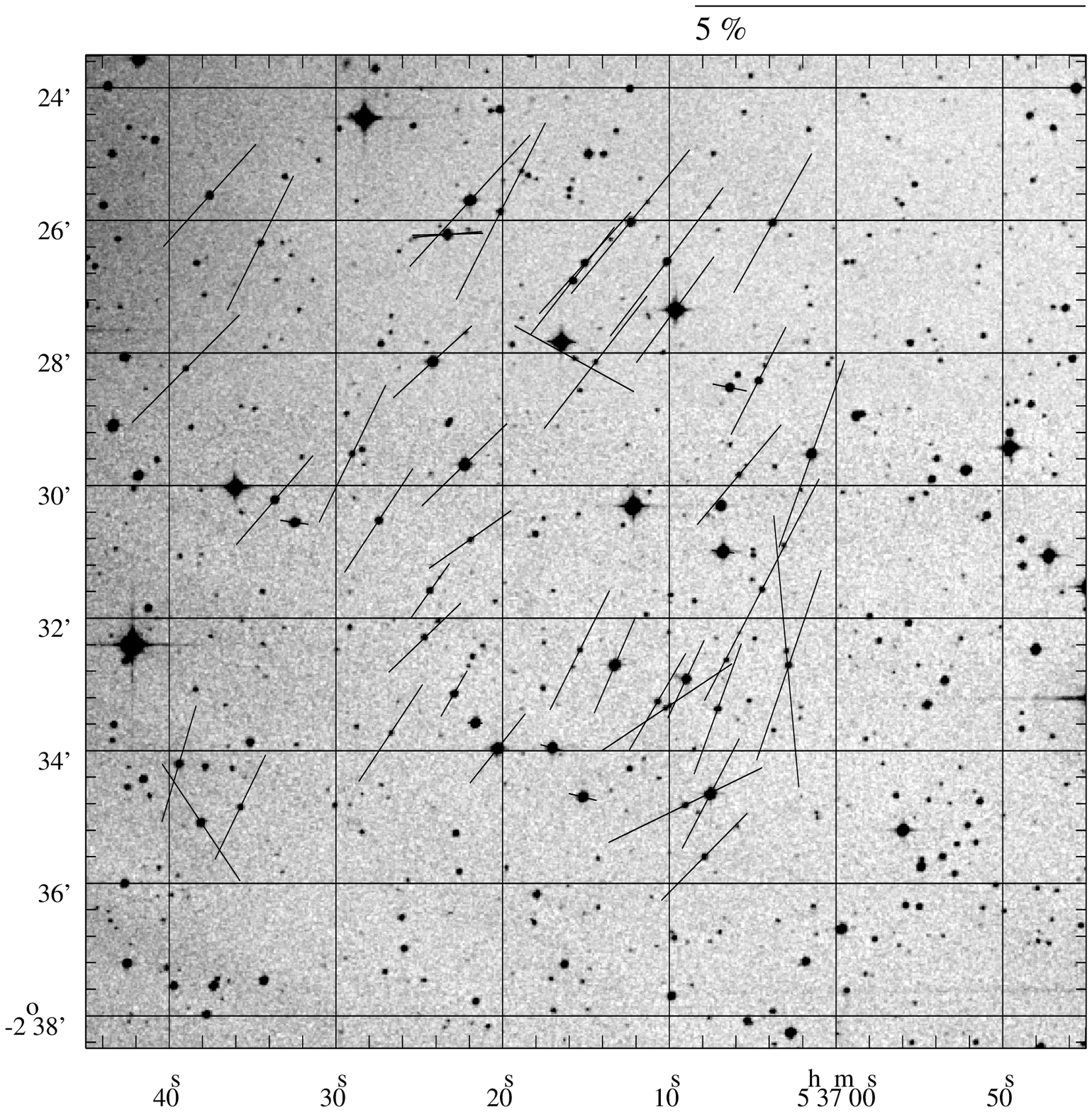}

\vspace{-5cm}
\includegraphics[width=6cm,angle=-90.]{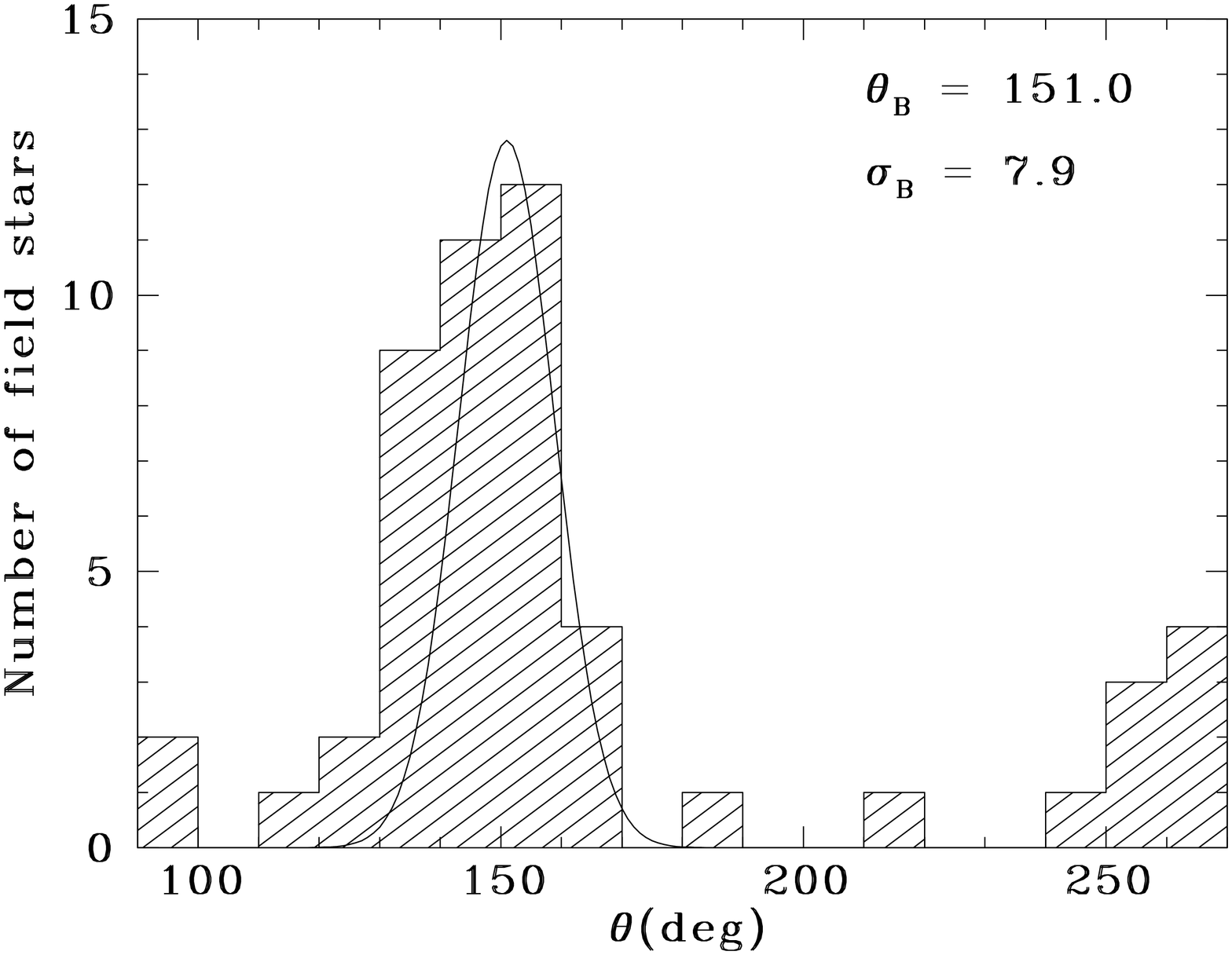}
\caption*{The same of Figure \ref{fig:hh139} for Field~28.}
\label{fig:field28}
\end{figure}

\end{document}